\documentclass[preprint,12pt]{elsarticle}

\usepackage{amsmath,amsfonts,amssymb,amsthm} 
\usepackage{bm} 

\usepackage{graphicx}    
\usepackage{float}       
\usepackage{tikz}        
\usepackage{pgfplots}    
\pgfplotsset{compat=1.18}

\usepackage[labelfont=bf]{caption} 
\usepackage{subcaption}            

\usepackage{booktabs}    
\usepackage{array}       
\usepackage{longtable}   
\usepackage{multirow}    
\usepackage{adjustbox}   
\usepackage[table,xcdraw]{xcolor} 
\usepackage{hhline}

\usepackage{parskip}     
\usepackage{enumitem}    
\usepackage{lscape}      
\usepackage{rotating}    
\usepackage{afterpage}   
\usepackage{pgffor}      
\usepackage{titlesec}    

\usepackage{algorithm}   
\usepackage{algorithmic} 

\usepackage{nomencl}     

\usepackage{url}         
\usepackage{soul}        
\usepackage{verbatim}    
\usepackage{hyperref}    

\titleformat{\subsection}{\normalfont\bfseries}{\thesubsection}{1em}{}
\titleformat{\subsubsection}{\normalfont\bfseries}{\thesubsubsection}{1em}{}
\titleformat{\paragraph}[runin]{\normalfont\bfseries}{\theparagraph}{1em}{}

\titlespacing*{\subsection}{0pt}{3.8ex plus 1ex minus .2ex}{1.5ex plus .2ex}
\titlespacing*{\subsubsection}{0pt}{3.2ex plus 1ex minus .2ex}{1.2ex plus .2ex}

\usepackage{multicol}
\usepackage[margin=2cm, nomarginpar,]{geometry}

\makenomenclature

\journal{Building and Environment}

\makeatletter
\def\ps@pprintTitle{%
  \let\@oddhead\@empty
  \let\@evenhead\@empty
  \def\@oddfoot{\footnotesize\itshape
    Preprint submitted to \ifx\@journal\@empty Elsevier\else\@journal\fi\hfill\today}%
  \let\@evenfoot\@oddfoot}
\makeatother  

\begin{document}
\begin{frontmatter}
\title{When does boundary-condition uncertainty outweigh thermal and vegetation effects in street canyon flow? A global sensitivity analysis of input uncertainty}

\author[BSC]{Albert Moreno Català}
\author[UPC,BSC]{Joan Calafell Sandiumenge}
\author[TUD]{Nikolaj T. Mücke}
\author[UPC,BSC]{Jan Mateu Armengol\corref{cor1}}
\ead{jan.mateu@upc.edu}
\cortext[cor1]{Corresponding author}

\affiliation[BSC]{organization={Barcelona Supercomputing Center}, city={Barcelona}, country={Spain}}
\affiliation[UPC]{organization={Department of Fluid Mechanics, Universitat Politècnica de Catalunya - BarcelonaTech (UPC)}, city={Barcelona}, country={Spain}}
\affiliation[TUD]{organization={Faculty of Civil Engineering and Geosciences, TU Delft},city={Delft}, country={Netherlands}}

\begin{abstract}
Computational Fluid Dynamics (CFD) is increasingly used to resolve wind, heat, and pollution patterns in cities with high fidelity but high computational cost. Thermal effects and vegetation are often neglected despite being important drivers, yet their influence may be masked by variability from other uncertain inputs. This work determines how precisely model parameters and boundary conditions must be known for these effects to govern output uncertainty in the velocity, temperature, and concentration fields of a two-dimensional street canyon. A global Sobol analysis of 24 parameters is made tractable by a surrogate trained on Reynolds-Averaged Navier--Stokes (RANS) simulations, combining principal component analysis (PCA) with multilayer perceptrons (MLPs). Two ventilation regimes assess the variance due to thermal and vegetation parameters as uncertainty in all other inputs is progressively reduced. At typical uncertainty levels, turbulence closure constants and inflow turbulence profiles govern variability in all three fields in both regimes. Under weak ventilation, solar forcing becomes the main source of temperature variability once the remaining uncertainties are halved, and of concentration variance once they are reduced to a quarter (about $\pm11\%$ for the dominant closure constant and $\pm12\%$ for the emission rate). In contrast, vegetation contributes significantly at typical uncertainty levels. Under strong ventilation, however, the canopy never overtakes the closure constants, even when the dominant one is known to $\pm5\%$. Spatially, both effects peak at road level and along canyon walls, which influence outdoor exposure and indoor infiltration, suggesting that high-resolution exposure studies should account for them.
\end{abstract}

\end{frontmatter}

\nomenclature{$W$}{Street canyon width [m]}
\nomenclature{$H$}{Building height [m]}
\nomenclature{$\nabla$}{Nabla operator}
\nomenclature{$\rho$}{Density [kg/$\text{m}^3$]}
\nomenclature{$\mathbf{u}$}{Velocity vector [m/s]}
\nomenclature{$p$}{Pressure [Pa]}
\nomenclature{$\mathbf{I}$}{Identity tensor}
\nomenclature{$\bm{\tau}$}{Viscous stress tensor [Pa]}
\nomenclature{$\bm{\tau}_t$}{Turbulent stress tensor [Pa]}
\nomenclature{$g$}{Gravity acceleration [m/s$^2$]}
\nomenclature{$\mathbf{S}_u$}{Momentum sink [Pa/m]}
\nomenclature{$D$}{Mass diffusivity [m$^2$/s]}
\nomenclature{$D_t$}{Turbulent mass diffusivity [m$^2$/s]}
\nomenclature{$c$}{Concentration [$\mu$mol/m$^3$]}
\nomenclature{$Q_c$}{Line pollutant source [$\mu$mol/(m·s)]}
\nomenclature{$\delta$}{Dirac delta function}
\nomenclature{$y_{_Q}$}{Pollutant source height [m]}
\nomenclature{$x_{_Q}$}{Pollutant source horizontal position [m]}
\nomenclature{$c_p$}{Specific heat capacity [J/(kg·K)]}
\nomenclature{$T$}{Temperature [K]}
\nomenclature{$\lambda$}{Thermal conductivity [W/(m·K)]}
\nomenclature{$\lambda_t$}{Turbulent thermal conductivity [W/(m·K)]}
\nomenclature{$Q_p$}{Line heat source [W/m]}
\nomenclature{$S_p$}{Volumetric heat flux [W/m$^3$]}
\nomenclature{$k$}{Turbulent Kinetic Energy [m$^2$/s$^2$]}
\nomenclature{$\varepsilon$}{Turbulent Dissipation Rate [m$^2$/s$^3$]}
\nomenclature{$\mu$}{Dynamic viscosity [Pa·s]}
\nomenclature{$\mu_t$}{Turbulent dynamic viscosity [Pa·s]}
\nomenclature{$\sigma_k$}{$k$\,--\,$\varepsilon$ model parameter}
\nomenclature{$P_k$}{Production of TKE [kg/(m·s$^3$)]}
\nomenclature{$S_k$}{TKE source [kg/(m·s$^3$)]}
\nomenclature{$\sigma_\varepsilon$}{$k$\,--\,$\varepsilon$ model parameter}
\nomenclature{$C_{\varepsilon_1}$}{$k$\,--\,$\varepsilon$ model parameter}
\nomenclature{$C_{\varepsilon_2}$}{$k$\,--\,$\varepsilon$ model parameter}
\nomenclature{$B$}{Wall function parameter}
\nomenclature{$f_\mu$}{Turbulent viscosity damping function}
\nomenclature{$f_\varepsilon$}{Turbulent dissipation damping function}
\nomenclature{$S_\varepsilon$}{TKE dissipation rate source [kg/(m·s$^4$)]}
\nomenclature{$C_\mu$}{$k$\,--\,$\varepsilon$ model parameter}
\nomenclature{$Sc_t$}{Turbulent Schmidt number}
\nomenclature{$Pr_t$}{Turbulent Prandtl number}
\nomenclature{$LAD$}{Leaf Area Density [1/m]}
\nomenclature{$C_d$}{Drag coefficient}
\nomenclature{$\beta_p$}{Vegetation model parameter}
\nomenclature{$\beta_d$}{Vegetation model parameter}
\nomenclature{$C_{\varepsilon_4}$}{Vegetation model parameter}
\nomenclature{$C_{\varepsilon_5}$}{Vegetation model parameter}
\nomenclature{$T_0$}{Reference temperature [$^\circ$C]}
\nomenclature{$C$}{Vegetation model factor [s$^{0.5}$m$^{-1}$]}
\nomenclature{$l$}{Characteristic leaf size [m]}
\nomenclature{$x$}{Horizontal coordinate [m]}
\nomenclature{$y$}{Vertical coordinate [m]}
\nomenclature{$U_\tau$}{Friction velocity [m/s]}
\nomenclature{$\kappa$}{von Kármán constant}
\nomenclature{$y_0$}{Aerodynamic roughness length [m]}
\nomenclature{$p_k$}{Inlet TKE constant}
\nomenclature{$p_\varepsilon$}{Inlet dissipation rate constant}
\nomenclature{$B_k$}{Ground level inlet concentration [$\mu$mol/m$^3$]}
\nomenclature{$p_{_B}$}{Inlet concentration factor}
\nomenclature{$\Gamma$}{Adiabatic lapse rate [K/m]}
\nomenclature{$P_s$}{Direct Normal Irradiance [W/m$^2$]}
\nomenclature{$I_0$}{Solar constant [W/m$^2$]}
\nomenclature{$\tau_a$}{Atmospheric optical depth}
\nomenclature{$\theta$}{Solar altitude [deg]}
\nomenclature{$a_s$}{Solar absorptivity}
\nomenclature{$s_p$}{Surface heat flux [W/m$^2$]}
\nomenclature{$\mathcal{H}$}{Heaviside function}
\nomenclature{$\xi$}{Local spatial coordinate [m]}
\nomenclature{$\boldsymbol{\theta}$}{Input parameter vector}
\nomenclature{$\mathcal{M}$}{Surrogate model}
\nomenclature{$\mathbf{w}$}{Latent weight vector}
\nomenclature{$\mathbf{\hat{w}}$}{Predicted latent weight vector}
\nomenclature{$\mathbf{f}$}{Ground-truth field vector}
\nomenclature{$\hat{\mathbf{f}}$}{Predicted field vector}
\nomenclature{$\mathbf{\bar{f}}$}{Mean field vector}
\nomenclature{$\boldsymbol{\phi}$}{Spatial basis function}

\vspace{-10pt}
\section{Introduction}
Air pollution is a major environmental threat to human health, with well-established links to respiratory and cardiovascular disease and elevated mortality~\cite{newby2015expert,al2020environmental,landrigan2017air}. Ultimately, chronic exposure reduces life expectancy and well-being, leading the World Health Organization (WHO) to identify it as the greatest environmental risk to public health. As of 2022, more than 99\% of the global population resided in locations where pollutant concentrations exceeded WHO-recommended thresholds~\cite{world2021global}. Exposure is particularly critical in cities, where dense traffic, industrial activity, and growing energy demands for space conditioning generate persistently high pollution levels~\cite{fenger1999urban}. Since the proportion of the world's population living in urban areas is projected to reach 68\% by 2050~\cite{united2019department}, an increasing share will be exposed to these elevated pollution levels.\par
Effectively mitigating these health risks requires a precise characterization of the spatial distribution of pollutant concentrations. Official monitoring networks are the primary source of air quality data, offering accurate measurements at high temporal resolution; however, the limited number of stations restricts their ability to capture spatial variability~\cite{santiago2013computational,rosa2026revealing}. Low-cost sensors and passive dosimeters partially address this limitation, as their reduced cost allows for denser spatial deployment; yet, they remain point-based and carry substantial measurement uncertainty, which constrains their reliability as a standalone solution. Satellite-based observations, in turn, offer extensive spatial coverage, but their coarse spatial resolution prevents them from capturing concentration patterns at the street scale.\par
Global and regional air quality models provide pollutant concentration estimates with full spatial coverage, both horizontally and vertically. Chemistry--transport models such as WRF-Chem~\cite{grell2005fully} and operational services such as the Copernicus Atmosphere Monitoring Service (CAMS)~\cite{peuch2022copernicus} are widely used for large-scale air quality forecasting; however, their spatial resolution remains too coarse for street-level applications. Alternative approaches targeting the urban scale typically couple regional models with Gaussian dispersion schemes~\cite{criado2023data}, but these rely on simplified assumptions that neglect how buildings alter local flow and street-level exposure~\cite{yazid2014review}.\par
Computational Fluid Dynamics (CFD) addresses this gap by directly resolving the governing flow equations, offering high spatial resolution and an explicit representation of the urban obstacles that the previous approaches could not capture~\cite{pullen2005comparison,leelHossy2014dispersion}. Fully resolving all scales of motion is, however, infeasible for city-sized domains, so urban CFD relies on turbulence modeling to lower the cost while retaining the fidelity required for the application. Reynolds-Averaged Navier--Stokes (RANS) and wall-modeled Large-Eddy Simulation (LES) are the two main methods used for urban CFD. LES dynamically resolves the larger turbulent scales directly, capturing much of the flow's unsteady behavior, whereas RANS models all turbulent scales statistically, trading some fidelity for a substantially lower computational demand. Nevertheless, even RANS simulations remain computationally heavy compared to simplified dispersion models, and their solutions are highly sensitive to the uncertainty inherited from their boundary conditions and emission inputs, meaning that a substantial computational cost does not necessarily translate into a correspondingly reliable output.\par
Compounding this issue, urban CFD studies frequently disregard thermal effects and vegetation, despite growing evidence that both substantially shape local flow and pollutant dispersion~\cite{xie2005impact,vranckx2015impact,nazarian2016realistic,shen2023new}. Whether the effects of thermal processes and vegetation canopies can be reliably identified depends not only on the uncertainty associated with these inputs, but also on the relative magnitude of the uncertainty in the other model inputs such as the boundary conditions, which may dominate the predicted variability and mask their effects. Global sensitivity analysis has been used to disentangle the effects of the different sources of uncertainty by quantifying their respective contributions to the variance of the model output and distinguishing inputs whose uncertainty must be accurately characterized from those with a comparatively minor influence on the predicted fields. To date, however, uncertainty quantification and sensitivity studies of urban CFD have addressed these inputs in isolation. Inflow boundary conditions, mainly wind speed, direction, and surface roughness, have received the most attention~\cite{garcia2014quantifying,garcia2017quantifying}, while the calibration of the $k$\,--\,$\varepsilon$ closure constants has been examined separately~\cite{guillas2014bayesian}. Related efforts have built reduced-order models to propagate atmospheric and source uncertainties at reduced cost~\cite{nony2023reduced,lumet2025uncertainty}, yet these too vary only a limited subset of inputs. Vegetation has entered such analyses only through the aerodynamic uncertainty of isolated plant canopies~\cite{giacomini2022framework}, not as part of an urban canyon. To our knowledge, no study has assessed these inputs jointly, leaving the interactions that only a combined analysis can capture unquantified.\par
This work addresses that gap through a global sensitivity analysis over a 24-dimensional parameter space that brings the thermal and vegetation parameters together with the turbulence, inflow, and emission uncertainties considered in previous studies. The Sobol method is adopted for this purpose, as it apportions the total output variance among the uncertain inputs, including their interactions. The influence of the thermal and vegetation effects is therefore quantified as the share of output variance associated with their respective parameters, that is, the predictive uncertainty incurred by not characterizing these effects precisely. Because this share is measured relative to the uncertainty carried by the other inputs, we progressively tighten the uncertainty of those inputs and determine how accurately they must be known before the thermal and vegetation contributions come to dominate the output variance. This yields a practical criterion for when the two effects must be represented explicitly and when, instead, the larger uncertainty in the remaining inputs governs the predicted fields.\par
However, accurately estimating Sobol indices over a 24-dimensional parameter space requires on the order of $1{\times}10^{5}$ model evaluations, so, due to the curse of dimensionality and the computational cost of each CFD simulation, performing this analysis directly on the CFD model is generally infeasible. To address this limitation, we develop a data-driven surrogate model that combines Principal Component Analysis (PCA) with a Multi-Layer Perceptron (MLP), following the approach we previously applied to the canonical thermal street canyon~\cite{moreira5492793assessment}. Here, the framework is extended to a vegetated canyon and to the joint prediction of the velocity, temperature, and concentration fields. A two-dimensional canyon with wind perpendicular to the street axis is adopted, as it represents the least-ventilated configuration and a standard benchmark for canyon dispersion; it also keeps the generation of the training database tractable, since a three-dimensional counterpart would be considerably more expensive to generate. Once trained, the surrogate predicts the velocity, temperature, and concentration fields orders of magnitude faster than the CFD solver. The analysis is carried out for two environmental regimes, a thermal-dominated scenario at low wind speeds and a vegetation-dominated scenario at higher wind speeds, to examine how the ranking of the input parameters shifts with the prevailing flow regime. For each field, sensitivity is reported both as point-wise spatial maps and as aggregated global indices, quantifying the relative importance of each governing parameter within the canyon. The aggregated indices follow the generalized formulation developed for multivariate and functional outputs~\cite{lamboni2011multivariate,gamboa2013sensitivity,xiao2017multivariate}, which to the best of our knowledge has not previously been applied to urban CFD, and which here requires no additional cost since the PCA basis is already available from the surrogate.\par
This paper is structured as follows. Section~\ref{sec:Methods} covers the generation of the ground-truth CFD database, the construction of the PCA+MLP surrogate, and the Sobol sensitivity analysis framework. Then, Section~\ref{sec:Results} first assesses the surrogate against the CFD reference data and presents the sensitivity analysis for the thermal and vegetation-dominated scenarios. Section~\ref{sec:Conclusion} summarizes the main conclusions.

\section{Methods}\label{sec:Methods}
\subsection{Ground truth database generation}
\subsubsection{Physical model}
\begin{figure}[h]
    \centering
    \includegraphics[scale=1.05, angle=-90]{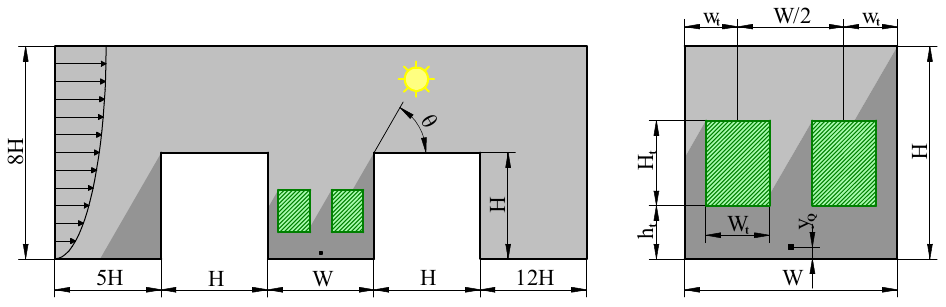}
    \caption{Representation of the two-dimensional computational domain (left) and a detailed view of the street canyon region (right). The full layout illustrates the domain extensions along with the solar elevation angle ($\theta$) driving thermal effects. The magnified view details the placement of the two tree crowns. \textit{Note:} dimensions are not to scale to enhance the visibility of the canyon elements and vegetation.}
    \label{fig:modelSetup}
\end{figure}

The physical configuration extends the thermal street canyon of Moreira \textit{et al}.~\cite{moreira5492793assessment}: a canonical two-dimensional canyon formed by two square buildings of height $H$ separated by a street of equal width ($W\,{=}\,H\,{=}\,20$\,m), with the flow imposed perpendicular to the street axis (Fig.~\ref{fig:modelSetup}). Vehicular emissions are represented by heat and pollutant line sources located at the street centerline at height $y_{_Q}$, with black carbon (BC) treated as a passive scalar~\cite{shrestha2010black, acp-23-1421-2023}. A detailed description of the baseline configuration is provided in~\cite{moreira5492793assessment}.\par
The main modification with respect to this reference setup is the inclusion of urban vegetation. Two tree crowns are placed within the canyon and modeled as rectangular porous zones, a standard approach in which the foliage is not geometrically resolved but represented through volumetric source terms in the governing equations~\cite{gromke2015cfd, buccolieri2018review}. In line with typical urban tree geometries~\cite{pretzsch2015crown,martin2022accuracy}, the crowns are parametrized by their width ($W_t\,{=}\,3W/10$), height ($H_t\,{=}\,2H/5$), base height ($h_t\,{=}\,H/4$), and horizontal distance from the tree centerline to the adjacent wall ($w_t\,{=}\,W/4$). Foliage density is characterized by the leaf area density ($LAD$), which quantifies the one-sided leaf area per unit crown volume and governs the strength of the vegetation--flow interaction, with $LAD\,{=}\,0$ recovering the tree-free canyon. The vegetation interacts with the flow, the thermal field, and pollutant transport through several competing mechanisms. Aerodynamically, the foliage exerts a drag force that decelerates the wind within and around the crowns, while enhancing turbulence production in the wake of the leaves and, at the same time, accelerating the dissipation of larger turbulent eddies. Thermally, the crowns cool the surrounding air through transpiration and intercept part of the incoming solar radiation, shading the underlying surfaces. These mechanisms have opposing consequences for air quality: reduced ventilation and turbulence attenuation favor pollutant accumulation within the canyon, whereas evaporative cooling and shading weaken the buoyancy-driven circulation that shapes the dispersion patterns. The net effect on street-level concentrations is therefore configuration-dependent, which motivates the explicit modeling of vegetation. 
Despite the geometric idealization, the coupling between buoyancy-induced recirculations and crown drag gives rise to complex flow structures within the canyon.

\subsubsection{Mathematical model}
The governing equations follow the formulation of Moreira \textit{et al}.~\cite{moreira5492793assessment}: the steady-state RANS equations for mass, momentum, species, and energy transport of a weakly compressible Newtonian fluid under the low-Mach-number approximation, with the density evaluated from the ideal gas law at constant reference pressure ($\rho=\rho(T)$)~\cite{paillere2000comparison}:
\setlength{\jot}{6pt}
\begin{gather}
    \nabla\cdot\left(\rho\mathbf{u}\right)=0\\
    \rho\left(\mathbf{u}\cdot\nabla\right)\mathbf{u}=\nabla\cdot\left[-p\mathbf{I}+\bm{\tau}+\bm{\tau}_t\right]+\rho\mathbf{g}+\mathbf{S}_u\\
    \mathbf{u}\cdot\nabla c + \nabla\cdot\left(-\left(D+D_t\right)\nabla c\right)=Q_c\,\delta(\mathbf{r}-\mathbf{r}_{_Q})\\
    \rho c_p\mathbf{u}\cdot\nabla T + \nabla\cdot\left(-\left(\lambda+\lambda_t\right)\nabla T\right)=Q_p\,\delta(\mathbf{r}-\mathbf{r}_{_Q})+S_p
\end{gather}
where $\mathbf{u},p,c,T$ correspond to the Favre-averaged velocity, pressure, concentration, and temperature, respectively. In the momentum equation, the identity tensor is denoted by $\mathbf{I}$, while $\bm{\tau}$ and $\bm{\tau}_t$ represent the Favre-averaged viscous and turbulent stress tensors. Gravitational acceleration is given by $\mathbf{g}=-g\mathbf{e}_y$. The molecular mass diffusivity is fixed at $D = 1{\times}10^{-6}\,\text{m}^2/\text{s}$, in line with reported diffusivities of black carbon in air~\cite{brilhac2000experimental}, and $D_t=\mu_t/(\rho\,Sc_t)$ is the turbulent mass diffusivity, where $\mu_t=\rho C_\mu\frac{k^2}{\varepsilon}f_\mu$ is the turbulent viscosity and $Sc_t$ is the turbulent Schmidt number. For the energy equation, $c_p$ is the specific heat capacity at constant pressure, whereas $\lambda$ and $\lambda_t=c_p\mu_t/Pr_t$ represent the molecular and turbulent thermal conductivities, where $Pr_t$ follows the Kays--Crawford model~\cite{Kays1994TurbulentPN}. $Q_c$ and $Q_p$ are the pollutant and heat line sources applied at the emission point $\mathbf{r}_{_Q}=(x_{_Q},y_{_Q})$, and $\mathbf{r}=(x,y)$ denotes the spatial position vector in the two-dimensional domain, with $x$ and $y$ representing the horizontal and vertical coordinates (origin located at the bottom-left corner). The terms $\mathbf{S}_u$ and $S_p$ are the vegetation momentum and energy sources defined below. The system is closed with the low-Reynolds-number $k$\,--\,$\varepsilon$ model, which introduces transport equations for the turbulent kinetic energy (TKE), $k$, and its dissipation rate, $\varepsilon$~\cite{abe1994new, ignat2000universal}:
\begin{gather}
    \rho\left(\mathbf{u}\cdot\nabla\right)k=\nabla\cdot\left[\left(\mu+\dfrac{\mu_t}{\sigma_k}\right)\nabla k\right]+P_k-\rho\varepsilon+S_k\\    \rho\left(\mathbf{u}\cdot\nabla\right)\varepsilon=\nabla\cdot\left[\left(\mu+\dfrac{\mu_t}{\sigma_\varepsilon}\right)\nabla \varepsilon\right]+C_{\varepsilon_1}\dfrac{\varepsilon}{k}P_k-C_{\varepsilon_2}\rho\dfrac{\varepsilon^2}{k}f_\varepsilon+S_\varepsilon
\end{gather}

The remaining closure relations (total stress tensor $\bm{\tau}+\bm{\tau}_t$, TKE production $P_k$, and near-wall damping functions $f_\mu$ and $f_\varepsilon$) are standard and detailed in~\cite{moreira5492793assessment, abe1994new}. The closure constants $C_{\varepsilon_1}$, $C_{\varepsilon_2}$, $C_\mu$, $\sigma_k$, $Sc_t$, together with the wall-function parameter $B$~\cite{moreira5492793assessment}, constitute the turbulence-related uncertain inputs of this study (further discussed in Table~\ref{tab:parameters}). The model constant $\sigma_\varepsilon$ is computed according to~\cite{richards2011appropriate} and is therefore not treated as an independent uncertain parameter.\par
Vegetation effects are incorporated into the governing equations through additional terms that depend on the leaf area density $\left(LAD\right)$ and drag coefficient $\left(C_d\right)$. $LAD$ represents the one-sided leaf area per unit canopy volume and characterizes the aerodynamic interaction between the vegetation and the surrounding flow field, while $C_d$ is fixed at 0.2 following common practice in urban flow simulations~\cite{gromke2015cfd, buccolieri2018review}.\par
The resulting aerodynamic resistance is modeled through a drag force $\left(\mathbf{S}_u\right)$ added to the momentum equation~\cite{buccolieri2018review, amorim2013cfd}:
\begin{equation}\label{eq:vegDrag}
    S_{u_i}=-\rho LADC_d Uu_i
\end{equation}
Vegetation also modifies the turbulent structure of the flow, which is accounted for through additional terms in the transport equations for turbulent kinetic energy and its dissipation rate~\cite{buccolieri2018review, amorim2013cfd}:
\begin{equation}\label{eq:vegK}
    S_k=\rho LADC_d\left(\beta_p U^3-\beta_d Uk\right)
\end{equation}
\begin{equation}\label{eq:vegEps}
    S_\varepsilon=\rho LADC_d\left(C_{\varepsilon_4}\beta_p\dfrac{\varepsilon}{k}U^3-C_{\varepsilon_5}\beta_d U\varepsilon\right)
\end{equation}
Here, $\beta_p$ accounts for turbulence production induced by foliage wake effects and is set to 1 following standard canopy formulations~\cite{katul2004one}, while $\beta_d$ governs turbulence dissipation associated with vegetation--turbulence interactions. $C_{\varepsilon_4}$ and $C_{\varepsilon_5}$ are model constants, and $U$ denotes the velocity magnitude, defined as $U=\left(\sum_i u_i^2\right)^{1/2}$.\par
Transpirational effects are incorporated into the energy equation through a term that depends on $LAD$, wind speed, and the local temperature difference between foliage and air~\cite{bruse1998simulating,yang2015cfd,defraeye2017transpirative}:
\begin{equation}\label{eq:vegEnergy}
    S_p=2\rho LADc_p\dfrac{T_0-T}{C\sqrt{l/U}}
\end{equation}
Here, the denominator models the leaf boundary layer resistance, relating the heat flux to the local flow conditions and leaf geometry, where $C$ is an empirical constant that depends on the vegetation type and $l$ denotes the characteristic leaf diameter.

\subsubsection{Boundary conditions}
\paragraph{Left boundary: Inlet}\hfill\par
The inlet conditions follow the formulation of Moreira \textit{et al}.~\cite{moreira5492793assessment}. A fully developed atmospheric boundary layer is imposed normal to the inflow face, with the mean velocity given by the logarithmic law for rough surfaces and the turbulent quantities prescribed accordingly:
\begin{equation}
    u(y)=\dfrac{U_{\tau}}{\kappa}\ln\left(\dfrac{y+y_0}{y_0}\right)
    \hspace{30pt}
    k=p_k\dfrac{U_{\tau}^2}{\sqrt{C_\mu}}
    \hspace{30pt}
    \varepsilon(y)=p_\varepsilon\dfrac{U_{\tau}^3}{\kappa\left(y+y_0\right)}
\end{equation}
where $U_\tau$ is the friction velocity, $\kappa$ the von Kármán constant, $y_0$ the aerodynamic roughness length, and $p_k$ and $p_\varepsilon$ scaling factors that tune the inflow turbulence levels.\par
Ambient pollution and thermal stratification are represented by linear vertical profiles~\cite{stull2012introduction}:
\begin{equation}
    c(y)=B_k\left(1+p_{_B}\dfrac{y}{8H}\right)
    \hspace{30pt}
    T(y)=T_0-\Gamma y
\end{equation}
where $B_k$ is the background concentration at ground level, $p_{_B}$ controls its vertical variation, $T_0=20^{\circ}$C is the reference temperature, and $\Gamma=9.8^{\circ}$C/km is the dry adiabatic lapse rate. The seven inflow and background parameters ($U_\tau$, $\kappa$, $y_0$, $p_k$, $p_\varepsilon$, $B_k$, $p_{_B}$) are treated as uncertain inputs of this study (Table~\ref{tab:parameters}).

\paragraph{Lower boundary: walls and solar forcing}\hfill\par
A no-slip condition is enforced at all solid boundaries. For species transport, a zero-flux condition is applied, so that surfaces act as impermeable and non-depositing walls. Pollutant deposition on the foliage is likewise neglected, since the deposition velocities of fine particles such as BC are small and their impact on canyon concentrations is secondary compared to the aerodynamic effects of the crowns~\cite{buccolieri2018review}.\par
Solar forcing is modeled by applying heat fluxes to all sunlit surfaces, including the ground, building walls, and the exterior of tree crowns. In contrast to the Dirichlet approach (a prescribed wall temperature) of Moreira \textit{et al}.~\cite{moreira5492793assessment}, this Neumann flux formulation links the surface heating directly to two uncertain environmental parameters (Table~\ref{tab:parameters}): the atmospheric optical depth ($\tau_a$) and the solar incidence angle ($\theta$). First, the Direct Normal Irradiance (DNI) reaching the surface is obtained from the Bouguer--Beer--Lambert law~\cite{ogunjobi2004influence}:
\begin{equation}
    P_s=I_0e^{-\tau_a/\sin\theta}
\end{equation}
where $I_0=1361$~W/m$^2$ is the extraterrestrial solar irradiance and $\theta\in\left[0^\circ,180^\circ\right]$ sweeps the solar path from sunrise ($\theta=0^\circ$) to sunset ($\theta=180^\circ$), with $\theta=90^\circ$ corresponding to the zenith. The irradiance is then projected onto the horizontal and vertical surfaces and scaled by the solar absorptivity $a_s$:
\begin{equation}
    s_{p,h} = P_s a_s \sin\theta \quad\qquad s_{p,v} = P_s a_s \left|\cos\theta\right|
\end{equation}
where $a_s$ is set to 0.8 for the building walls and ground~\cite{qin2011modeling} and to 0.5 for the tree crowns. Finally, mutual shading among buildings and trees is accounted for by modulating the absorbed flux at each surface point. Denoting by $s_{p}^{*}\in\{s_{p,h},\,s_{p,v}\}$ the unshaded flux corresponding to the orientation of the surface, the net heat flux along a generic surface with local coordinate $\xi$ ($\xi = x$ on horizontal surfaces and $\xi = y$ on vertical ones) reads:
\begin{equation}\label{eq:shading}
    s_{p}(\xi,\theta) = s_{p}^{*}(\theta)\prod_{n=1}^{N}\mathcal{H}_n\left(\xi,\theta\right)
\end{equation}

where $N$ is the number of shadow boundaries intersecting the solar path towards that surface. Each factor $\mathcal{H}_n$ is a smoothed Heaviside function varying continuously between 0 (fully shaded) and 1 (fully sunlit):
\begin{equation}
    \mathcal{H}_n\left(\xi,\theta\right)=\dfrac{1}{2}+\dfrac{\sigma_n}{2}\tanh\left[m\left(\xi-\xi_{s,n}\left(\theta\right)\right)\right]
\end{equation}

where $\xi_{s,n}(\theta)$ is the position of the $n$-th shadow edge, determined by trigonometry, and $\sigma_n=\pm1$ selects the sunlit side of the edge: $\sigma_n=+1$ when the surface is illuminated for $\xi>\xi_{s,n}$, and $\sigma_n=-1$ otherwise. The steepness $m=70$~m$^{-1}$ smears the shadow edge over a few centimeters, so that the product in Eq.~\ref{eq:shading} equals one where the surface is fully exposed to the sun and vanishes wherever at least one shadow covers it, while avoiding a discontinuous heat flux on the numerical solver. This formulation generalizes the smoothed shadow transition of Moreira \textit{et al}.~\cite{moreira5492793assessment} to heat fluxes with multiple shadow boundaries cast by buildings and vegetation, providing an accurate and inexpensive description of shading for the present two-dimensional geometry; more general shadow-casting and radiative transfer algorithms can be found in open-source tools such as  SOLWEIG~\cite{lindberg2008solweig} or PALM~\cite{krc2021radiative}.

\paragraph{Upper and right boundaries}\hfill\par
The upper boundary is treated as a symmetry plane, allowing tangential flow while imposing zero normal velocity and zero normal gradients for all transported quantities, whereas at the outlet a zero-gradient condition is applied to all variables, assuming fully developed flow.

\newpage
\subsubsection{Numerical setup}
The governing equations are discretized using the finite element method (FEM) implemented in COMSOL Multiphysics v6.3~\cite{comsol2025}, employing first-order Lagrange elements with streamline upwind Petrov--Galerkin (SUPG) stabilization~\cite{zienkiewicz2005finite}.
Since the database generation requires thousands of unattended simulations spanning widely different flow regimes, this low-order stabilized formulation is chosen to prioritize convergence robustness over formal accuracy.
The steady-state solution of each case is considered converged when the mean residual of the velocity and turbulent kinetic energy fields within the street canyon satisfies $(\mathcal{R}(U)+\mathcal{R}(k))/2\leq1{\times}10^{-3}$. A total of 8192 CFD simulations were performed, of which $N_{\text{samp}}=4746$ completed successfully (58\% of all runs). Although the range of each individual parameter was selected based on physical considerations, the independent sampling of all parameters may result in physically extreme configurations, such as low-velocity cases dominated by buoyancy-driven Rayleigh--Bénard-like instabilities~\cite{berge1984rayleigh} or inconsistent closure constants $(C_{\varepsilon_1}\geq C_{\varepsilon_2})$, which contribute to the observed non-convergence rate.\par 
The computational domain spans $8H\,{\times}\,20H$ and is discretized with a structured mesh of quadrilateral elements and refined within the street canyon using square elements down to $0.125\,$m side length, resulting in a total of 88,880 mesh elements (around $6.4{\times}10^5$ degrees of freedom). Each simulation requires approximately 270~seconds on 16 CPU cores of a dual-socket AMD EPYC 7713 node, corresponding to roughly $1{\times}10^4$ core-hours for the full database, which is generated automatically through COMSOL batch runs on an in-house HPC cluster. The mesh and domain independence studies, together with the experimental validation of the RANS model and of the vegetation parameterization, are reported in \ref{app:modelValidation}.\par

\subsubsection{Distributions of input parameters}
The database is constructed by considering 24 uncertain input parameters expected to influence the CFD solution, organized into seven groups: the RANS closure, the inflow velocity, turbulence, and concentration profiles, the traffic emissions, the solar forcing, and the vegetation (Table~\ref{tab:parameters}). The uncertainty characterization extends the framework of Moreira \textit{et al}.~\cite{moreira5492793assessment}, from which the distributions of the shared parameters are adopted, to the solar and vegetation parameters introduced in this study, whose ranges are derived from the referenced literature. All parameters are treated as mutually independent, as required for the variance-based Sobol decomposition to admit its standard interpretation and consistent with their role as free inputs in the governing equations and boundary conditions. In particular, the leaf area density spans from $LAD=0$, recovering the tree-free canyon, to densely foliated crowns, so that the database covers both vegetated and non-vegetated configurations. The atmospheric optical depth range is chosen to span conditions from relatively clear ($\tau_a\approx0.15$) to optically thick atmospheric conditions ($\tau_a\approx3$), consistent with optical depth values reported from ground-based remote sensing measurements~\cite{sassen2001midlatitude,lolli2023climatological}.\par
For clarity in the presentation of the results, a representative parameter is highlighted for each group in Table~\ref{tab:parameters}, selected \textit{a posteriori} as the parameter exhibiting the highest total-order Sobol index within its group (Section~\ref{sec:sensitivityAnalysis}). The discussion is articulated around these representatives, while the sensitivity indices of the complete parameter set are reported in \ref{app:AdditRes}, where the remaining parameters exhibit weaker and/or qualitatively similar sensitivity patterns.\par
\begin{table}[h]
\centering
\begin{adjustbox}{max width=\textwidth}
\renewcommand{\arraystretch}{1.2} 
\begin{tabular}{|c|c|c|c|c|c|}
\hline
\rowcolor[HTML]{E0E0E0}
Group &
  \multicolumn{1}{c|}{Parameter ($\theta_i$)} & \multicolumn{1}{c|}{Description} & \multicolumn{1}{c|}{Unit} & \multicolumn{1}{c|}{Distribution} & \multicolumn{1}{c|}{Ref.} \\ \hline
\multirow{6}{*}{\shortstack[c]{RANS\\closure}}
    & $Sc_t$ & Turbulent Schmidt number & - & $U(0.2, 1)$ & \cite{moreira5492793assessment, bazdidi2015turbulent} \\ \hhline{|~|-----}
    & $C_{\varepsilon_1}$ & \multirow{4}{*}{$k$\,--\,$\varepsilon$ model parameter} & - & $N(1.4,0.1)$ & \multirow{5}{*}{\cite{moreira5492793assessment, guillas2014bayesian}} \\ \hhline{|~|-|~|--|~|}
    & \cellcolor[HTML]{DAE8FC}$C_{\varepsilon_2}$ &  & - & $N(2.1,0.3)$ & \\ \hhline{|~|-|~|--|~|}
    & $C_\mu$ &  & - & $\mathcal{L}(\ln(0.025),0.35)$ & \\ \hhline{|~|-|~|--|~|}
    & $\sigma_k$ &  & - & $\mathcal{L}(\ln(0.6),0.2)$ & \\ \hhline{|~|----|~|}
    & $B$ & Wall function parameter & - & $U(4.4,5.6)$ & \\ \hline
\multirow{3}{*}{\shortstack[c]{Inflow\\velocity}}
    & \cellcolor[HTML]{DAE8FC}$U_\tau$ & Friction velocity & m/s & $U(0.15,1)$ & -  \\ \hhline{|~|-----}
    & $\kappa$ & von Kármán constant & - & $U(0.3,0.6)$ & \cite{moreira5492793assessment, guillas2014bayesian} \\ \hhline{|~|-----}
    & $y_0$ & Aerodynamic roughness length & m & $U(0.15,2.8)$ & \cite{moreira5492793assessment, burian2003morphological} \\  \hline
\multirow{2}{*}{\shortstack[c]{Inflow\\turbulence}}
    & \cellcolor[HTML]{DAE8FC}$p_k$ & TKE factor  & - & $\mathcal{L}(0,0.4)$ & - \\ \hhline{|~|-----}
    & $p_\varepsilon$ & Dissipation rate factor & - & $\mathcal{L}(0,0.4)$ & - \\ \hline
\multirow{2}{*}{\shortstack[c]{Inflow\\concentration}}
    & \cellcolor[HTML]{DAE8FC}$B_k$ & Background conc. constant & $\mu$mol/m$^3$ & $U(0,0.65)$ & \cite{moreira5492793assessment, reche2011variability} \\ \hhline{|~|-----}
    & $p_{_B}$ & Background conc. linear factor & - & $U(-0.8,0.8)$  & - \\ \hline
\multirow{3}{*}{Emission}
    & \cellcolor[HTML]{DAE8FC}$Q_c$ & Emission rate & $\mu$mol/(m·s) & $U(0.01,10)$ & \cite{moreira5492793assessment, smit2010validation, jevzek2018traffic} \\ \hhline{|~|-----}
    & $y_{_Q}$ & Emission source height & m & $U(0,0.75)$ & - \\ \hhline{|~|-----}
    & $Q_p$ & Emission heat release & kW/m & $\mathcal{B}(1.7,2.9,0,80)$ & \cite{moreira5492793assessment, jin2016sensitivity, torres2024performance} \\ \hline
\multirow{2}{*}{\shortstack[c]{Solar\\forcing}}
    & $\theta$ & Solar incidence angle & deg & $U(0,180)$ & - \\ \hhline{|~|-----}
    & \cellcolor[HTML]{DAE8FC}$\tau_a$ & Atmospheric optical depth & - & $\mathcal{B}(1,1.2,0.15,3)$ & \cite{sassen2001midlatitude,lolli2023climatological} \\ \hline
\multirow{6}{*}{Vegetation}
    & \cellcolor[HTML]{DAE8FC}$LAD$ & Leaf Area Density & 1/m & $U(0,2)$ & \cite{buccolieri2018review,gao2023estimating} \\ \hhline{|~|--|---}
    & $\beta_d$ & \multirow{3}{*}{Vegetation model parameter} & - & $U(3,6.5)$ & \multirow{3}{*}{\cite{buccolieri2018review}} \\ \hhline{|~|-|~|--|~|} 
    & $C_{\varepsilon_4}$ &  & - & $N(1.5,0.15)$ & \\ \hhline{|~|-|~|--|~|}
    & $C_{\varepsilon_5}$ &  & - & $N(1.2,0.2)$ & \\ \hhline{|~|--|---}
    & $C$ & Vegetation model factor & s$^{0.5}$m$^{-1}$ & $N(143.5,18)$ & \multirow{2}{*}{\cite{bruse1998simulating}} \\ \hhline{|~|----|~|}
    & $l$ & Characteristic leaf size & m & $U(0.02,0.5)$ & \\ \hline
\end{tabular}
\end{adjustbox}
\caption{Statistical distributions used for generating the ground-truth database by sampling input parameters ($\theta_i$) from prescribed probability distributions. $U(a,b)$ denotes a uniform distribution between minimum $a$ and maximum $b$; $N(\mu,\sigma)$ and $\mathcal{L}(\mu,\sigma)$ represent normal and log-normal distributions with mean $\mu$ and standard deviation $\sigma$. $\mathcal{B}(\alpha,\beta,a,b)$ indicates a bounded Beta distribution with shape parameters $\alpha$ and $\beta$, scaled onto the continuous interval $[a,b]$. Highlighted cells mark the representative parameter of each group.}
\label{tab:parameters}
\end{table}

The 24-dimensional input space is sampled using a quasi-Monte Carlo, low-discrepancy Sobol sequence~\cite{niederreiter1992random}. A total of $2^{13}\,{=}\,8192$ parameter combinations are generated with the \texttt{SciPy} library~\cite{2020SciPy-NMeth}, and each sample is used to run an independent CFD simulation to form the high-fidelity database.

\subsection{Surrogate model}
To bypass the high computational cost of the CFD simulations required to perform the proposed high-dimensional sensitivity analysis, we develop data-driven surrogate models, substantially increasing the number of available model evaluations without resorting to additional CFD simulations. The surrogate $\mathcal{M}$ is trained on the ground-truth CFD database and maps the vector of input parameters $\boldsymbol{\theta}\in\mathbb{R}^{24}$ defining a CFD simulation to the resulting velocity, temperature, and concentration fields. It combines Principal Component Analysis (PCA) ---also referred to as Proper Orthogonal Decomposition (POD)--- and Multi-Layer Perceptrons (MLP): PCA projects the fields onto low-dimensional latent spaces, and the MLPs are trained to map the input parameters to the corresponding principal component coefficients, from which the full spatial fields are reconstructed.\par
To alleviate the training cost, the data samples are obtained by linearly interpolating the high-resolution CFD mesh data onto a uniform $128\,{\times}\,128$ Cartesian grid covering $x\in[67.5,195.0]\,\text{m}$ and $y\in[0.0,31.5]\,\text{m}$, i.e., the street canyon and its near wake, where the complex wind, temperature, and concentration patterns develop. This resolution follows Moreira \textit{et al}.~\cite{moreira5492793assessment}, who reported an interpolation error of approximately 2\% relative to a high-resolution reference grid. Nodes lying within the buildings are discarded, resulting in $N_p=13{,}264$ effective grid points.\par
\begin{figure}[h!]
	\centering
	\includegraphics[scale=1.44]{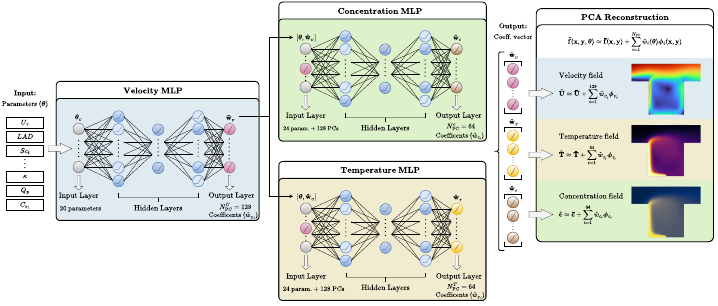}
	\caption{Schematic of the sequential PCA+MLP surrogate framework. The velocity MLP maps the reduced parameter set ($\boldsymbol{\theta}_{_U}$) to the velocity PCA coefficients ($\mathbf{\hat{w}}_{_U}$), which are concatenated with the full parameter set to form the augmented inputs $[\boldsymbol{\theta},\mathbf{\hat{w}}_{_U}]$ of the temperature and concentration MLPs. The physical fields ($\mathbf{\hat{U}}$, $\mathbf{\hat{T}}$, and $\mathbf{\hat{c}}$) are reconstructed via linear combination with their respective pre-computed PCA basis functions ($\boldsymbol{\phi}$).}
	\label{fig:MLP_PCA}
\end{figure}

Rather than predicting all fields independently, the proposed framework follows a sequential approach (Fig.~\ref{fig:MLP_PCA}) in which the predicted velocity PCA coefficients act as an intermediate physical descriptor for the scalar fields, a strategy whose benefits are quantified in Section~\ref{sec:trainEval}.\par
For the velocity field, and after preliminary tests, a reduced parameter set $\boldsymbol{\theta}_{_U}\in\mathbb{R}^{20}$ is adopted by omitting four parameters with no influence on the flow dynamics. Since the pollutant has no feedback on the velocity field, the three parameters governing the mass emission and the background concentration ($Q_c$, $B_k$, and $p_{_B}$) are excluded. Additionally, the wall-function parameter $B$ is discarded, as its influence on the flow within the canyon proved negligible over the sampled range. To capture the directional flow dynamics rather than merely the velocity magnitude, the surrogate targets both velocity components simultaneously, concatenating the discrete fields over the $N_p$ grid points into a unified vector $\mathbf{U}=[\mathbf{u},\mathbf{v}]\in\mathbb{R}^{2N_p}$. The surrogate models for the velocity, temperature, and concentration fields are then:
\begin{equation}
    \hat{\mathbf{U}}=\mathcal{M}_{_U}(\boldsymbol{\theta}_{_U}),\qquad
    \hat{\mathbf{T}}=\mathcal{M}_{_T}(\boldsymbol{\theta}),\qquad
    \hat{\mathbf{c}}=\mathcal{M}_{c}(\boldsymbol{\theta})
\end{equation}
where $\boldsymbol{\theta}_{_U}\subset\boldsymbol{\theta}$, so all three depend only on the input parameters. Their explicit form is given in Section~\ref{sec:MLP}.\par

\subsubsection{Principal Component Analysis}
PCA is applied independently to each field, so that velocity, temperature, and concentration end up with their own mean field, spatial basis, and reduced representation. The principal components are constructed using exclusively the training subset to ensure a strict separation of data. As a result, the PCA basis functions are defined only by the variance within the training set, and are then used to project both validation and test datasets. Prior to the PCA projection, the velocity components are normalized by the friction velocity $U_\tau$ of each simulation, yielding a dimensionless velocity field, while the temperature and concentration fields are normalized via min--max scaling to the range $[0,1]$, with the minimum and maximum values likewise computed on the training set. Mathematically, any physical field snapshot $\mathbf{f}(\mathbf{x},\mathbf{y},\boldsymbol{\theta})$ can be approximated by its mean field and a linear combination of $N_{_\text{PC}}$ orthogonal spatial modes:
\begin{equation}\label{eq:PCA}
    \mathbf{f}\left(\mathbf{x},\mathbf{y},\boldsymbol{\theta}\right)\approx\mathbf{\bar{f}}\left(\mathbf{x},\mathbf{y}\right)+\sum^{N_{_\text{PC}}}_{i=1}w_i\left(\boldsymbol{\theta}\right)\boldsymbol{\phi}_i\left(\mathbf{x},\mathbf{y}\right)
\end{equation}
In this framework, $\mathbf{\bar{f}}\left(\mathbf{x},\mathbf{y}\right)$ is the spatial mean field of the training dataset, $\boldsymbol{\phi}_i\left(\mathbf{x},\mathbf{y}\right)$ denotes the spatial basis functions that capture the directions of maximum variance of each field, obtained as the right singular vectors of the training data matrix via singular value decomposition (SVD), and $w_i\left(\boldsymbol{\theta}\right)$ represent the scalar coefficients corresponding to the specific set of input parameters $\boldsymbol{\theta}$, with each mode having an associated variance $\lambda_i$. These coefficients form the target latent weight vector $\mathbf{w} = [w_1, w_2, \dots, w_{N_{_\text{PC}}}]$, which the MLPs are trained to approximate as $\mathbf{\hat{w}}$. The map from $\mathbf{w}$ back to the physical field in Eq.~\ref{eq:PCA} is denoted $\text{PCA}^{-1}$.\par
Through this linear transformation, PCA compresses the dimensionality of each spatial field from its uniform grid size $N_p$ (or $2N_p$ for velocity) down to its respective dimension $N_{_\text{PC}}$. By retaining a reduced number of modes $N_{_\text{PC}}\,{\ll}\,N_p$ with the highest associated variance, the high-dimensional physical problem is projected into a low-dimensional latent space that captures most of the field's variability. In this work, $N_{_\text{PC}}^{U}=128$ and $N_{_\text{PC}}^{T}=N_{_\text{PC}}^{c}=64$, a choice discussed in \ref{app:nPC}.

\subsubsection{Multi-Layer Perceptron}\label{sec:MLP}
Multi-Layer Perceptrons map the input parameters to their respective latent spaces derived from PCA. The velocity MLP maps the reduced parameter set to the velocity principal component coefficients:
\begin{equation}
    \mathbf{\hat{w}}_{_U} = \text{MLP}_{_U}(\boldsymbol{\theta}_{_U}), \quad \mathbf{\hat{w}}_{_U} \in \mathbb{R}^{N_{_\text{PC}}^{U}}
\end{equation}
The predicted velocity coefficients are appended to the full parameter set, $[\boldsymbol{\theta},\mathbf{\hat{w}}_{_U}]\in\mathbb{R}^{24+N_{_\text{PC}}^{U}}$, to serve as augmented inputs for the two scalar MLPs:
\begin{equation}
    \mathbf{\hat{w}}_{_T} = \text{MLP}_{_T}([\boldsymbol{\theta}, \mathbf{\hat{w}}_{_U}]), \quad \mathbf{\hat{w}}_{_T} \in \mathbb{R}^{N_{_\text{PC}}^{T}}
\end{equation}
\begin{equation}
    \mathbf{\hat{w}}_c = \text{MLP}_c([\boldsymbol{\theta}, \mathbf{\hat{w}}_{_U}]), \quad \mathbf{\hat{w}}_c \in \mathbb{R}^{N_{_\text{PC}}^{c}}
\end{equation}
Once the coefficients ($\mathbf{\hat{w}}_{_U},\mathbf{\hat{w}}_{_T},\mathbf{\hat{w}}_c$) are predicted by the MLPs, the spatial fields of velocity ($\mathbf{\hat{u}},\mathbf{\hat{v}}$), temperature ($\mathbf{\hat{T}}$), and concentration ($\mathbf{\hat{c}}$) are reconstructed using the pre-computed PCA basis vectors:
\begin{equation}
    \hat{\mathbf{U}}=\text{PCA}^{-1}_{_U}(\mathbf{\hat{w}}_{_U}),\qquad
    \hat{\mathbf{T}}=\text{PCA}^{-1}_{_T}(\mathbf{\hat{w}}_{_T}),\qquad
    \hat{\mathbf{c}}=\text{PCA}^{-1}_{c}(\mathbf{\hat{w}}_{c})
\end{equation}
To account for the varying target dimensions of each field, different network architectures are used for velocity, temperature, and concentration predictions. The velocity model incorporates five hidden layers comprising 64, 128, 256, 512, and 256 neurons. In contrast, the temperature and concentration fields share an identical four-hidden-layer structure with 512, 256, 512, and 1024 neurons. In all networks, the hidden layers employ the Sigmoid Linear Unit (SiLU) activation function~\cite{elfwing2018sigmoid, farea2025learning}, which provides a smooth, continuously differentiable alternative to traditional rectified units. To prevent overfitting and enhance generalization, dropout is applied to selected hidden layers with rates up to 0.10.

\subsubsection{Training and evaluation}\label{sec:trainEval}
The $N_{\text{samp}}$ converged simulations forming the database are randomly partitioned into training (80\%), validation (10\%), and test (10\%) sets. The training set is used to optimize the MLP weights and biases, the validation set monitors training to prevent overfitting by evaluating generalization capabilities during training time, and the test set is held out to compare performance among different model configurations or architectures.\par
The three MLPs are trained separately in sequence: the velocity model is trained first, after which its PCA coefficients become available as inputs to the temperature and concentration models. To preserve the physical consistency between the flow and scalar fields, the temperature and concentration models use the ground-truth velocity principal component coefficients ($\mathbf{w}_{_U}$) as inputs during training, whereas the predicted coefficients from the velocity MLP ($\mathbf{\hat{w}}_{_U}$) are employed during validation and testing. This prevents the networks from learning inconsistent mappings caused by pairing imperfect velocity predictions with the true scalar targets.\par
The networks are trained using batches of 64 samples and optimized with the AdamW algorithm~\cite{loshchilov2017decoupled}, applying a decoupled weight decay of 0.1 for the velocity network and 0.5 for the scalar models. The velocity network is trained by minimizing the smooth $L_1$ loss function~\cite{girshick2015fast}, which behaves quadratically for small errors and linearly for large errors. This makes it more robust to outliers than an $L_2$-based loss, which is relevant for the velocity field due to localized structures (boundary layers, wakes, recirculation) that produce large-gradient regions. This loss choice yielded a lower prediction error than the mean relative $L_2$ loss used to train the scalar models. The initial learning rate is set to $5{\times}10^{-4}$, $1{\times}10^{-3}$, and $1{\times}10^{-4}$ for the velocity, temperature, and concentration models, respectively. During training, it is adjusted using a ReduceLROnPlateau scheduler, which halves the learning rate after 50 epochs without improvement in the validation loss. Training is regulated through early stopping with a patience of 100 epochs, restoring the model weights corresponding to the minimum validation loss. All networks are implemented using the \texttt{PyTorch} library~\cite{paszke2019pytorch}.\par
Surrogate accuracy is quantified through the mean relative $L_2$ error. For a single simulation $i$, the relative reconstruction error is:
\begin{equation}\label{eq:errorL2}
    \epsilon_i = \frac{\lVert \mathbf{f}_i - \hat{\mathbf{f}}_i \rVert_2}{\lVert \mathbf{f}_i \rVert_2} = \sqrt{\frac{\sum_{j=1}^{N_p} (f_{ij} - \hat{f}_{ij})^2}{\sum_{j=1}^{N_p} f_{ij}^2}}
\end{equation}
where $\mathbf{f}_i$ is the ground-truth field vector and $\hat{\mathbf{f}}_i$ its surrogate prediction. For velocity, $\mathbf{f}_i$ is the concatenated vector $\mathbf{U}$, while the temperature is evaluated as the anomaly $\mathbf{T}'=\mathbf{T}-T_0$, so that the metric measures the error on the thermal response rather than being masked by the absolute temperature baseline. Note that, being normalized by the field magnitude, $\epsilon_i$ can become large for near-zero fields. The overall error reported throughout this work is the mean over the test set:
\begin{equation}\label{eq:errorL2_global}
    \epsilon = \frac{1}{N_{\text{test}}} \sum_{i=1}^{N_{\text{test}}} \epsilon_i
\end{equation}
With this metric, the sequential approach reduces the mean relative $L_2$ error of the concentration from 10.0\% to 8.7\% with respect to predicting all fields independently. For temperature, it performs comparably to the independent baseline (13.3\% versus 12.9\%) while ensuring physical consistency with the underlying advective flow.

\subsection{Global sensitivity analysis}
\subsubsection{Sobol indices}
To quantify the influence of the input parameters on each field, a variance-based global sensitivity analysis is performed using the Sobol method~\cite{sobol2001global, saltelli2002making, saltelli2010variance}. This approach decomposes the total variance of the model output into contributions associated with individual inputs and their interactions. The sum of all sensitivity indices equals unity:
\begin{equation}
    \sum_{i=1}^dS_i+\sum_{i=1}^d\sum_{j>i}^dS_{ij}+\dots+S_{1,2,\dots,d}=1
\end{equation}
where $d$ is the number of uncertain parameters, and $S_{ij}$ represents the sensitivity index of the interaction between parameters $i$ and $j$.\par
The first-order sensitivity index ($S_i$) measures the main effect of a parameter $X_i$ independently of any interactions:
\begin{equation}\label{eq:S_i}
    S_i=\dfrac{\mathbb{V}[\mathbb{E}(Y{\mid}X_i)]}{\mathbb{V}(Y)}=\dfrac{\mathbb{V}_i}{\mathbb{V}(Y)}
\end{equation}
where $Y$ is the model output (or quantity of interest), and $\mathbb{E}$ and $\mathbb{V}$ denote the mathematical expectation and variance operators, respectively. The total-effect index ($S_{T_i}$) quantifies the contribution of a parameter, including both its first-order effect and all higher-order interactions with other variables:
\begin{equation}\label{eq:S_Ti}
    S_{T_i}=\dfrac{\mathbb{E}[\mathbb{V}(Y{\mid}X_{\sim i})]}{\mathbb{V}(Y)}=1-\dfrac{\mathbb{V}[\mathbb{E}(Y{\mid}X_{\sim i})]}{\mathbb{V}(Y)}
\end{equation}
where $X_{\sim i}$ denotes the set of all input parameters except $X_i$. Since the model output is a spatial field rather than a scalar, Eqs.~\ref{eq:S_i} and~\ref{eq:S_Ti} can be evaluated independently at each grid point, yielding a sensitivity map $S_i(x,y)$ (and $S_{T_i}(x,y)$) for every parameter. These maps resolve \textit{where} within the canyon each parameter governs the output variability. However, comparing full spatial fields across the $d$ parameters is impractical, which motivates a scalar summary of each parameter's overall influence.\par
To this end, generalized first-order ($G_i$) and total-effect ($G_{T_i}$) indices are defined. Since the PCA basis is already available from the surrogate construction, the mode-wise Sobol indices $S_{i,j}$ and $S_{T_{i,j}}$ can be computed directly from the principal component coefficients and aggregated by weighting each mode by its explained variance ratio~\cite{lamboni2011multivariate,gamboa2013sensitivity,xiao2017multivariate}. This formulation has been applied to dynamic and spatial outputs in other fields~\cite{perrin2021functional} and is adopted here for the two-dimensional canyon fields.
\begin{equation}
    G_i = \sum_{j=1}^{N_{_\text{PC}}}\dfrac{S_{i,j}\lambda_j}{\sum_{p=1}^{N_\text{PC}}\lambda_p}
    \hspace{30pt}
    G_{T_i} = \sum_{j=1}^{N_{_\text{PC}}}\dfrac{S_{T_{i,j}}\lambda_j}{\sum_{p=1}^{N_{_\text{PC}}}\lambda_p}
\end{equation}
where $\lambda_j$ denotes the eigenvalue associated with the $j$-th principal component, representing its contribution to the total variance of the model output, and $S_{i,j}$ and $S_{T_{i,j}}$ are the first-order and total-effect Sobol indices of parameter $i$ with respect to the $j$-th principal component. The weights are normalized over the $N_{_\text{PC}}$ retained components.\par
Owing to the orthonormality of the PCA basis, the generalized indices admit a direct physical interpretation: $G_i$ equals the spatial average of the local sensitivity map $S_i(x,y)$, weighted at each point by the local output variance,
\begin{equation}\label{eq:Gi_interp}
    G_i = \dfrac{\sum_{x,y}S_i(x,y)\,\mathbb{V}[Y(x,y)]}{\sum_{x,y}\mathbb{V}[Y(x,y)]}
\end{equation}
so that regions of the canyon where the field varies most contribute proportionally more to the global ranking. The local maps and the generalized indices therefore describe the same variance decomposition at two levels of aggregation, the former resolving \textit{where} a parameter acts and the latter providing a single scalar descriptor per parameter.

\subsubsection{Uncertainty scenarios and sampling}\label{sec:uncertScenarios}
The analysis is designed to quantify the share of output variance attributable to the thermal and vegetation parameters, that is, the predictive uncertainty incurred by not characterizing these two effects precisely. Since this share is measured relative to the uncertainty carried by the remaining inputs, it is not a fixed property of the canyon but depends on how well those inputs are known. Two environmental scenarios are therefore defined, each isolating one of the two effects. Within each scenario, the uncertainty of the remaining parameters is then progressively reduced, which determines how accurately those parameters must be known before the thermal and vegetation contributions come to govern the output variance.\par
To reduce the uncertainty of the parameters consistently across different distribution types, a relative measure $\gamma$ is defined for each. For uniform distributions, $\gamma$ denotes the relative half-width of the interval with respect to the mean, such that the bounds are given by $\mu(1\pm\gamma)$. For normal distributions, $\gamma$ is defined from the $3\sigma$ interval as $\gamma=3\sigma/\mu$. Similarly, for lognormal distributions, it is expressed as $\gamma=e^{3\sigma}-1$, corresponding to the multiplicative deviation derived from the $3\sigma$ interval in log-space. Starting from a baseline level ($\gamma_{_0}$), the uncertainty of the parameters not defining a given scenario is successively reduced to 50\%, 25\%, and 12.5\% of $\gamma_{_0}$.\par
\begin{figure}[h]
    \centering
    \includegraphics[scale=.95]{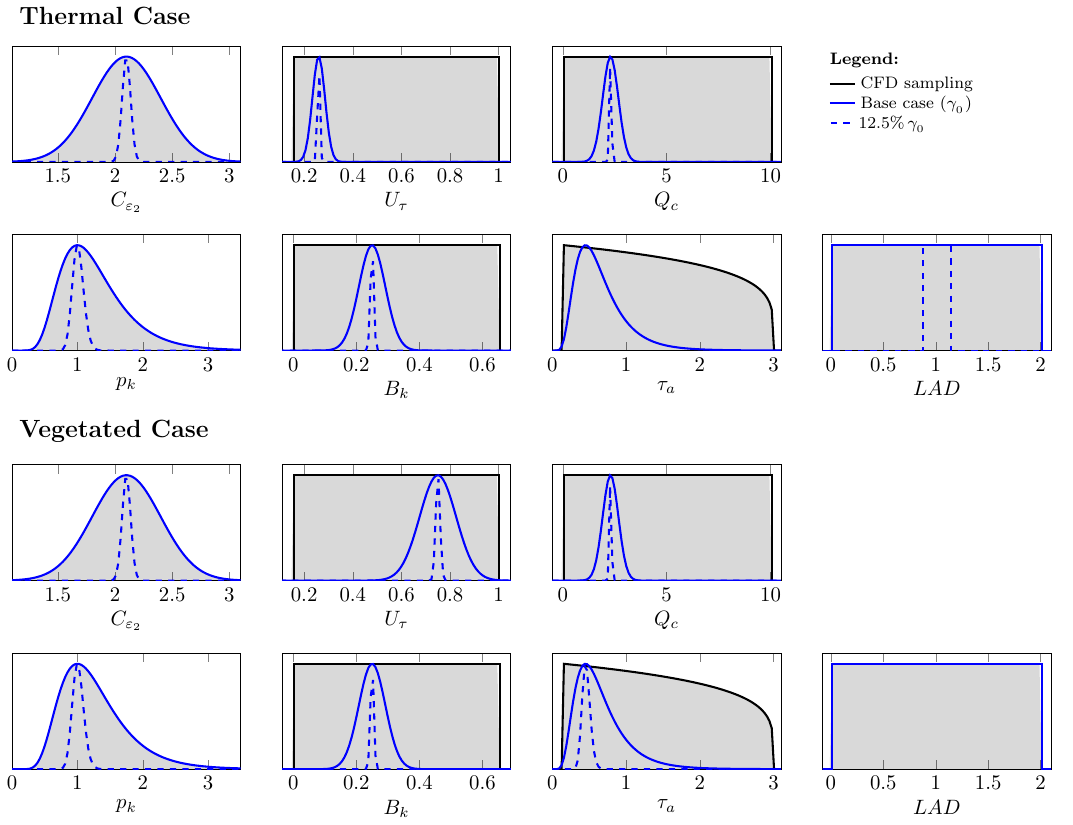}
    \caption{Probability density functions of the representative parameter of each group (Table~\ref{tab:parameters}) for the thermal (top) and vegetated (bottom) scenarios. Grey shaded areas bounded by solid black curves denote the original CFD sampling distributions used to generate the database; solid blue curves denote the baseline uncertainty level adopted for the Sobol sensitivity analysis ($\gamma_{_0}$); dashed blue curves denote the most reduced uncertainty level considered ($12.5\%\,\gamma_{_0}$). Note that $\tau_a$ is held at $\gamma_{_0}$ throughout the thermal scenario, and $LAD$ is held at $\gamma_{_0}$ throughout the vegetated scenario, as these parameters define their respective scenario and are therefore not reduced.}
    \label{fig:uncertScen}
\end{figure}

Both scenarios are constructed within the parameter space of the CFD database (Table~\ref{tab:parameters}), so as to avoid extrapolating the surrogate beyond the region covered by the training database. Within that space, the mean values are chosen to represent plausible urban conditions under which each effect is expected to be influential, and the parameters defining the corresponding effect are held at $\gamma_{_0}$ throughout, since reducing them would suppress the very contribution being measured. Figure~\ref{fig:uncertScen} compares the original CFD sampling distributions with the baseline and most reduced levels for the representative parameter of each group; the complete set is reported in \ref{app:statDistr}.\par
\textbf{Thermal scenario.} The friction velocity is centered at $U_\tau=0.26$\,m/s, corresponding to a weakly ventilated canyon in which buoyancy governs the recirculation, and the solar incidence angle is constrained to $\theta=125^\circ\pm8\%$, within the zone of maximum thermal impact. The atmospheric optical depth is centered at $\tau_a=0.45$, representative of a moderately attenuating atmosphere~\cite{gil2023geometrical}. Together these correspond to a calm, clear-sky situation in which solar forcing is prone to dominate the canyon dynamics. The parameters governing the thermal response ($\tau_a$, $\theta$, $Q_p$, and $y_{_Q}$) are held at $\gamma_{_0}$, while the uncertainty of all remaining parameters is progressively reduced.\par
\textbf{Vegetated scenario.} The friction velocity is centered at $U_\tau=0.75$\,m/s, so that the aerodynamic interaction between the flow and the canopy, rather than buoyancy, drives the canyon dynamics. The leaf area density is centered at $LAD=1.0$\,m$^{-1}$, corresponding to a densely foliated crown. The parameters governing the vegetation response ($LAD$, $\beta_d$, $C_{\varepsilon_4}$, $C_{\varepsilon_5}$, $C$, and $l$) are held at $\gamma_{_0}$, while the uncertainty of all remaining parameters is progressively reduced.\par

Since accurate estimation of Sobol indices requires thousands of model evaluations to achieve statistical convergence, performing a direct sensitivity analysis through CFD simulations would carry a computational cost several orders of magnitude beyond that of the training database. Therefore, the surrogate model is used to perform the evaluations, while the open-source \texttt{SALib} Python library~\cite{Iwanaga2022,Herman2017} is employed both to generate the samples via Saltelli's scheme~\cite{saltelli2002making,campolongo2011screening} and to compute the sensitivity indices. For each scenario and uncertainty level, a sample size of $N=8192$ is adopted within the 24-dimensional input space, resulting in  $N(d+2)=212{,}992$ model evaluations per case, with $d=24$ the number of uncertain parameters.

\section{Results}\label{sec:Results}
\subsection{Surrogate performance}\label{sec:surrogatePerf}
Evaluated on the held-out test set, the velocity, temperature, and concentration surrogates achieve mean relative $L_2$ errors of 2.4\%, 13.3\%, and 8.7\%, respectively, demonstrating robust generalization to unseen data. The following paragraphs examine how this error is distributed, first qualitatively across the predicted fields and then quantitatively across the parametric space.\par
\begin{figure}[h!]
	\centering
	\includegraphics[scale=.60]{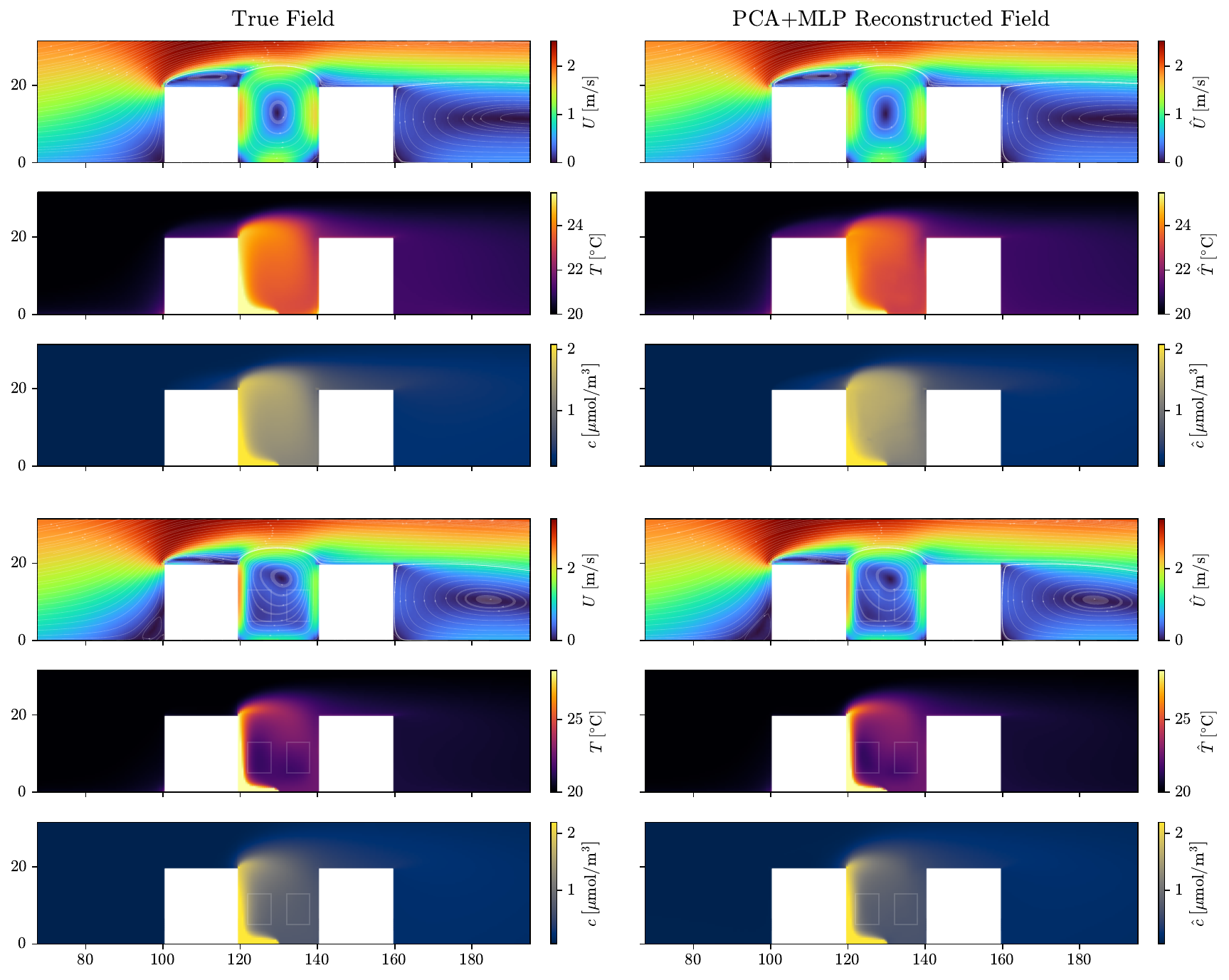}
	\caption{Comparison between ground truth CFD simulations (left column) and PCA+MLP surrogate model predictions (right column) for two samples from the test set. Rows 1--3 correspond to Sample 1, and rows 4--6 correspond to Sample 2. From top to bottom within each sample block, the panels illustrate velocity magnitude ($U$, $\hat{U}$), temperature ($T$, $\hat{T}$), and concentration ($c$, $\hat{c}$).}
	\label{fig:inferenceFields}
\end{figure}

Figure~\ref{fig:inferenceFields} compares the surrogate predictions with the CFD reference for two test cases matched at $\theta\approx125^\circ$ and $U_\tau\approx0.3\,\text{m/s}$, differing mainly in the presence of vegetation. In the tree-free canyon (rows 1--3) the flow organizes into a single recirculation centered in the cavity, producing diffuse temperature and concentration plumes. With crowns (rows 4--6), the main circulation is squeezed between the foliage and the building walls, displacing the vortex core upward, above the crowns, and slightly downstream. The resulting channeling along the facades~\cite{buccolieri2009aerodynamic,gromke2008dispersion} confines the temperature and concentration plumes into sharper structures near the source. In both cases the surrogate reproduces the flow topology, the position of the vortex core, and the extent of the thermal and concentration plumes, with no visible distortion of the predicted structures.\par

\begin{figure}[h!]
	\centering
	\includegraphics[scale=.51]{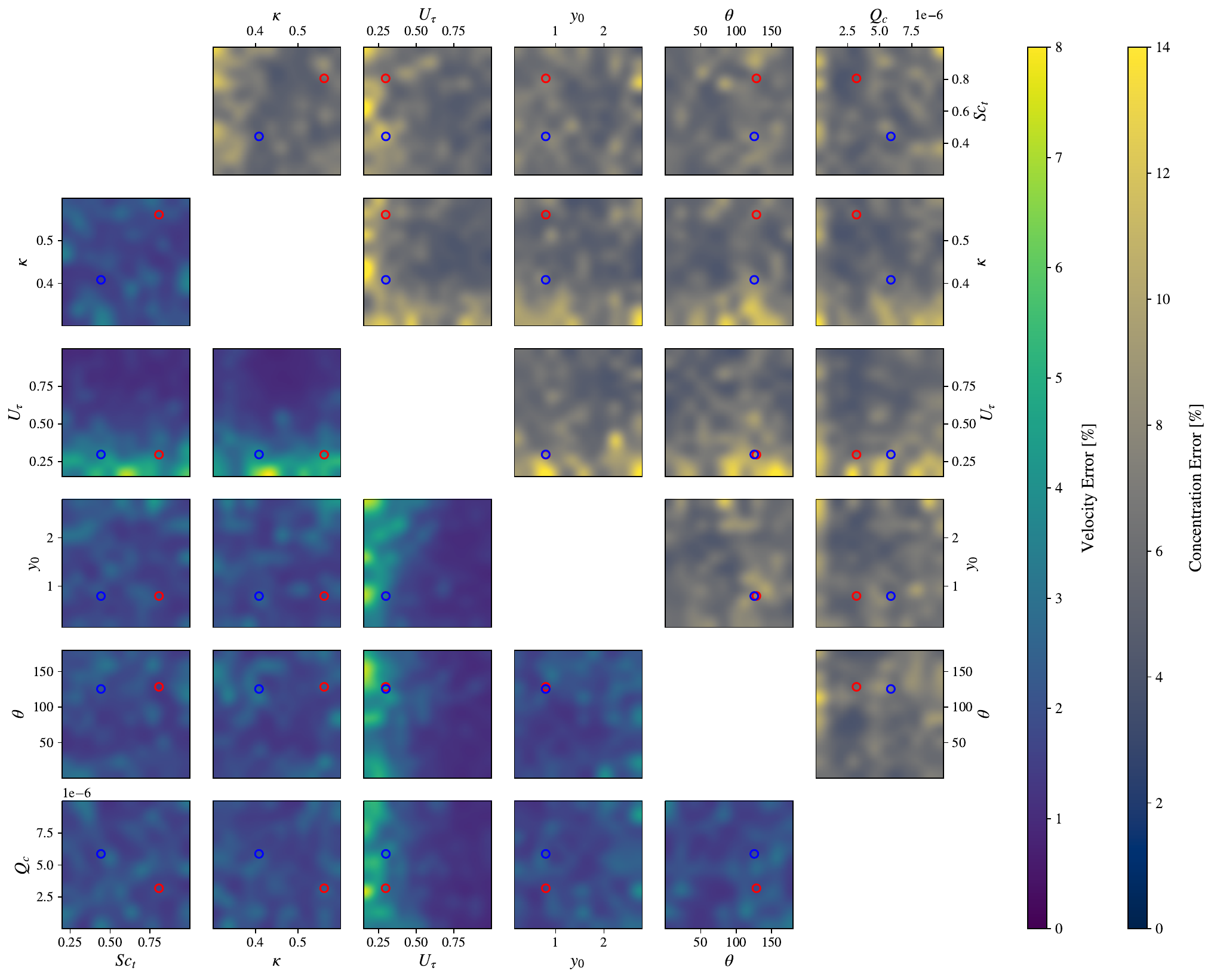}
	\caption{Error heatmap matrix illustrating the localized performance of the PCA+MLP surrogate models across pairwise parameter combinations. The lower triangle shows the mean relative $L_2$ error for the velocity field, whereas the upper triangle shows the corresponding error for the concentration field. Each map is computed by averaging the test errors within two-dimensional parameter bins. Red and blue circles mark the locations of the two representative test cases (Sample 1 and Sample 2) within the parametric space.}
	\label{fig:surrogateError}
\end{figure}

The qualitative agreement observed in these illustrative cases is quantified in Fig.~\ref{fig:surrogateError}, which reports the model error throughout the parametric space. The maps aggregate the errors of the complete test set, averaged within two-dimensional parameter bins. Six parameters sampled from uniform distributions are selected for the visualization, since their constant sampling density populates the bins evenly. The lower and upper triangles show the velocity and concentration errors, respectively, while the corresponding temperature maps are reported in~\ref{sec:tempModelError} (Fig.~\ref{fig:surrogateErrorTemp}).\par
Surrogate accuracy degrades where the canyon flow becomes strongly nonlinear. The largest errors for velocity predictions occur at low inflow velocities ($U_\tau\leq0.4\,\text{m/s}$), where buoyancy competes with mechanical forcing and the flow switches between topologies: the main canyon vortex weakens, splits into counter-rotating cells, or reverses. In this regime, neighboring parameter combinations no longer correspond to a smooth deformation of a single flow structure, and a linear PCA basis reconstructs such transitions poorly. This region also contains the highest proportion of non-converged CFD simulations, so fewer training samples are available precisely where the parameter-to-field mapping is most complex.\par
Since the concentration surrogate takes the predicted velocity coefficients as input, errors in the velocity surrogate propagate to the scalar transport model, contributing to the higher concentration errors observed in the same regions. Slightly higher concentration errors also appear for solar angles $\theta\in[100^\circ,150^\circ]$ and at low values of $\kappa$, although these variations are close to the bin-to-bin scatter of the maps and should be read as a tendency rather than a well-resolved effect. A possible explanation is that both parameters modify the canyon vortex morphology and the resulting pollutant plume: in the case of $\theta$, solar heating of the windward facade weakens the primary vortex and may induce vortex reversal~\cite{moreira5492793assessment}. The maps show no visible trend with respect to the remaining parameters. The temperature error follows a pattern similar to that of the concentration, with the largest errors at low $U_\tau$ and comparable tendencies at low $\kappa$ and around $\theta\approx150^\circ$, consistent with the shared dependence of both scalar fields on the predicted velocity.\par

Overall, the surrogate recovers the flow topology, the vortex position, and the extent of the thermal and concentration plumes, with the larger errors concentrated in the thermally dominated regime at low $U_\tau$. The Sobol analysis that follows does not rely on individual field predictions but on variance statistics accumulated over large ensembles of evaluations, and the sensitivity indices are reported together with bootstrap confidence intervals, so that the sampling uncertainty of the estimates can be assessed. The residual surrogate error nevertheless remains an additional source of uncertainty on the reported indices, particularly for the thermal scenario, which lies in the low-velocity region discussed above. This trade-off is what makes the analysis feasible, since the Sobol estimators require over $2{\times}10^5$ evaluations, which the surrogate produces in a fraction of the time needed by the CFD solver.\par

\subsection{Sensitivity analysis}\label{sec:sensitivityAnalysis}
\subsubsection{Reference mean and variance fields}
\begin{figure}[h!]
	\centering
	\includegraphics[scale=.585]{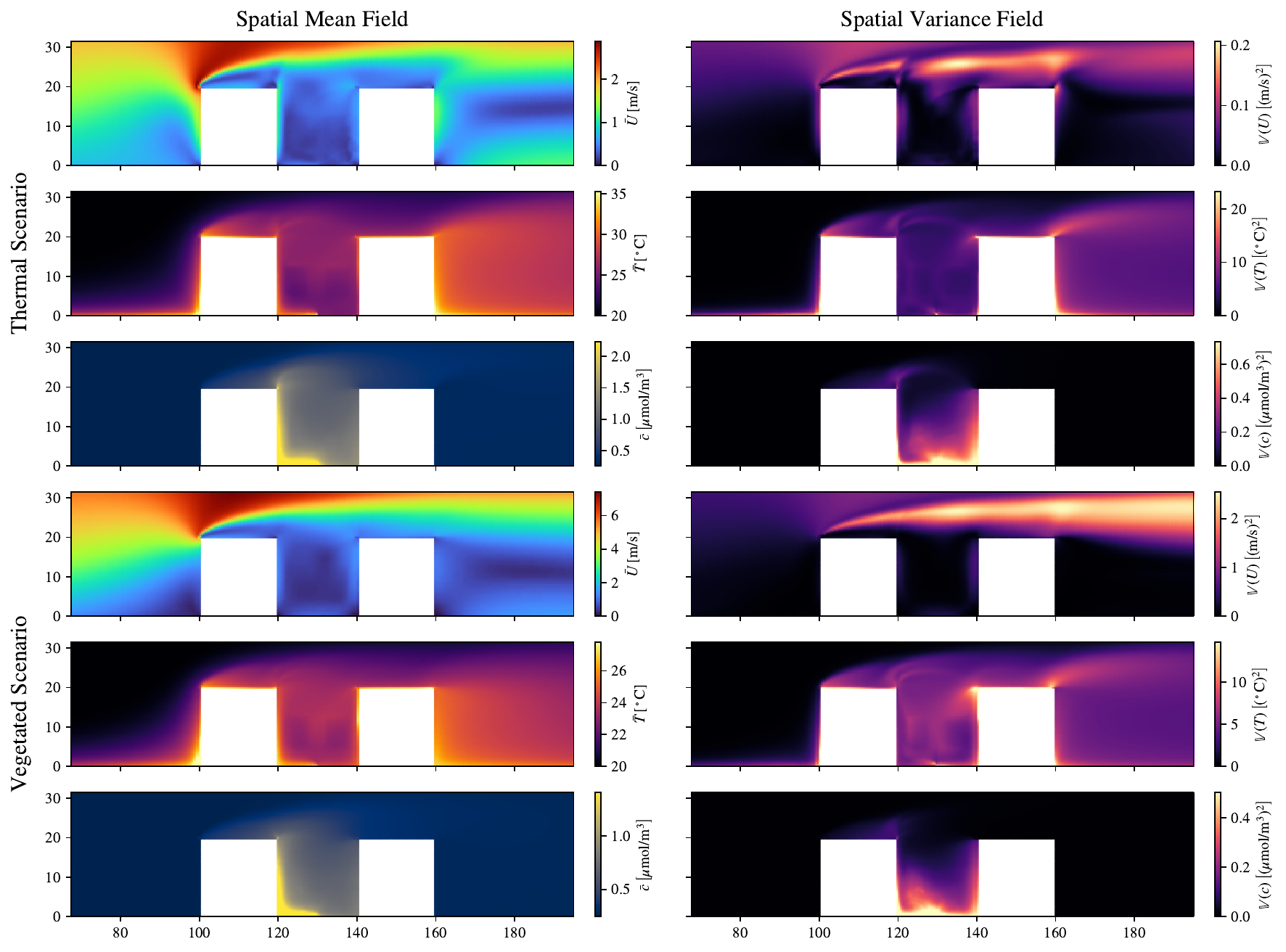}
	\caption{Spatial distribution of the mean (left) and variance (right) fields of velocity magnitude ($U$), temperature ($T$), and concentration ($c$), evaluated under the baseline distributions ($\gamma_{_0}$) of the thermal and vegetated scenarios. These absolute fields provide the reference scale needed to interpret the Sobol sensitivity indices discussed in the following sections.}
\label{fig:meanVarFields}
\end{figure}

Since the Sobol indices reported below are variance fractions, they identify which parameter dominates the local variability but not how large that variability is. Figure~\ref{fig:meanVarFields} therefore reports the absolute mean and variance fields of the three variables of study (velocity, temperature, and concentration) under both scenarios, evaluated at the baseline uncertainty level ($\gamma_{_0}$), providing the scale against which the indices are read; in regions where the variance is negligible, the indices carry correspondingly little practical weight.\par

The spatial patterns are broadly similar across scenarios, while the magnitudes differ according to the prescribed inflow. Velocity mean and variance are markedly larger in the vegetated scenario, with the variance being roughly one order of magnitude higher. Temperature mean and variance are instead higher in the thermal scenario, consistent with the dominant role of buoyancy under low-velocity conditions. The concentration variance is comparable between scenarios, whereas the mean is higher in the thermal case, since the stronger inflow of the vegetated scenario promotes greater pollutant dispersion.\par
Each field concentrates its variability in a distinct region: the velocity along the shear layer developing in the wake of the leeward building, extending further downstream; the temperature throughout the canyon and the downstream wake; and the concentration at road level and along both canyon walls, coinciding with the region of highest pedestrian exposure.\newpage

\subsubsection{Thermal scenario}
\begin{figure}[h!]
	\centering
	\includegraphics[scale=.49]{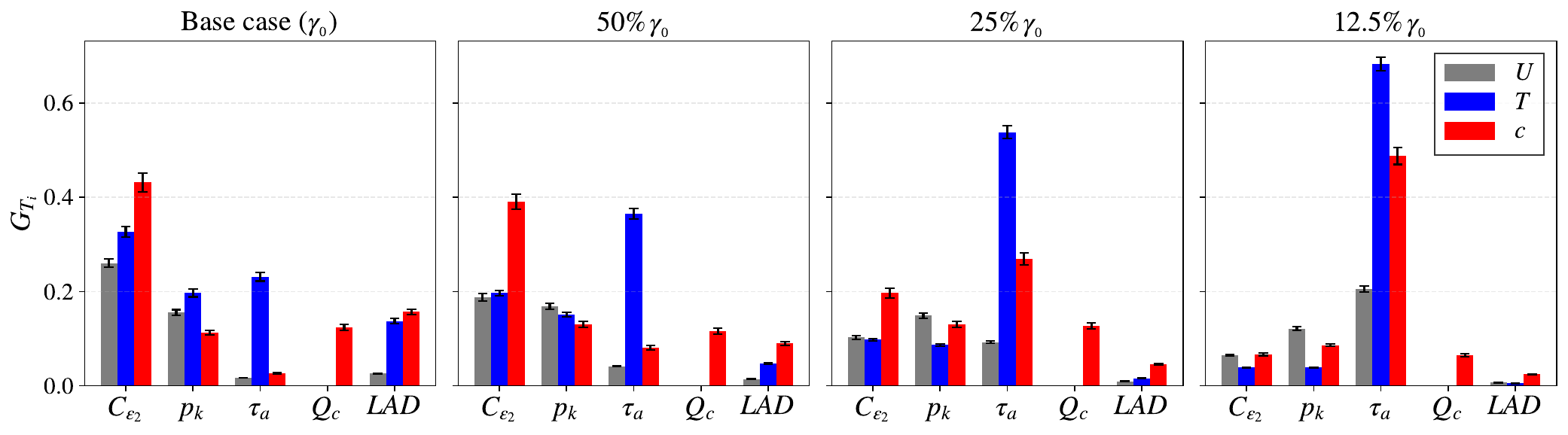}
	\caption{Variance-weighted total-effect ($G_{T_i}$) Sobol sensitivity indices for the velocity ($U$), temperature ($T$), and concentration ($c$) fields under the \textbf{thermal scenario}. The global indices are aggregated across the $N_{_\text{PC}}$ principal components, weighted by their respective explained variance, for different uncertainty levels: the base case, and reduced scenarios at $50\%$, $25\%$, and $12.5\%$ of the baseline uncertainty ($\gamma_{_0}$). Error bars represent the propagated 95\% confidence intervals from the Sobol analysis.}
	\label{fig:sBarTherm}
\end{figure}

Figure~\ref{fig:sBarTherm} reports the generalized total-effect indices for the representative parameter of each of the five groups: the RANS closure ($C_{\varepsilon_2}$), the inflow turbulence ($p_k$), the solar forcing ($\tau_a$), the traffic emission ($Q_c$), and the vegetation ($LAD$). The inflow velocity and background concentration representatives are omitted, as their global contribution is negligible at all uncertainty levels. At baseline uncertainty, the two turbulence groups jointly govern all three fields, while the solar forcing is confined to the temperature and remains essentially invisible in the velocity and concentration. Vegetation, in contrast, already ranks among the leading contributors to the concentration variance, exceeding the emission source and far outweighing the solar forcing, despite its comparatively minor role in the velocity field. This discrepancy is addressed by the spatial maps discussed below.\par
As the uncertainty of the non-thermal parameters is tightened, the solar forcing is the only group whose share grows, and it does so for all three fields. It overtakes the turbulence groups for temperature at $50\%\,\gamma_{_0}$, for concentration at $25\%\,\gamma_{_0}$, and for velocity at the most reduced level. The two turbulence groups decay at markedly different rates: the RANS closure contribution collapses, whereas the inflow turbulence profile proves comparatively persistent and overtakes it for both velocity and concentration. The emission source, in turn, retains an almost constant share of the concentration variance across all levels, so that tightening its characterization does not alter its relative weight. Taken together, these trends indicate that how precisely the solar forcing must be characterized is not a property of the configuration alone, but of how precisely the remaining inputs are known: while the turbulence parameters carry their baseline uncertainty, thermal effects contribute little to the predicted concentration field, whereas once they are constrained to about a quarter of that level ($25\%\,\gamma_{_0}$, corresponding to roughly $11\%$ for $C_{\varepsilon_2}$; Table~\ref{tab:paramSens}) the solar forcing becomes the dominant source of uncertainty.\par
\begin{figure}[h!]
	\centering
	\includegraphics[scale=.65]{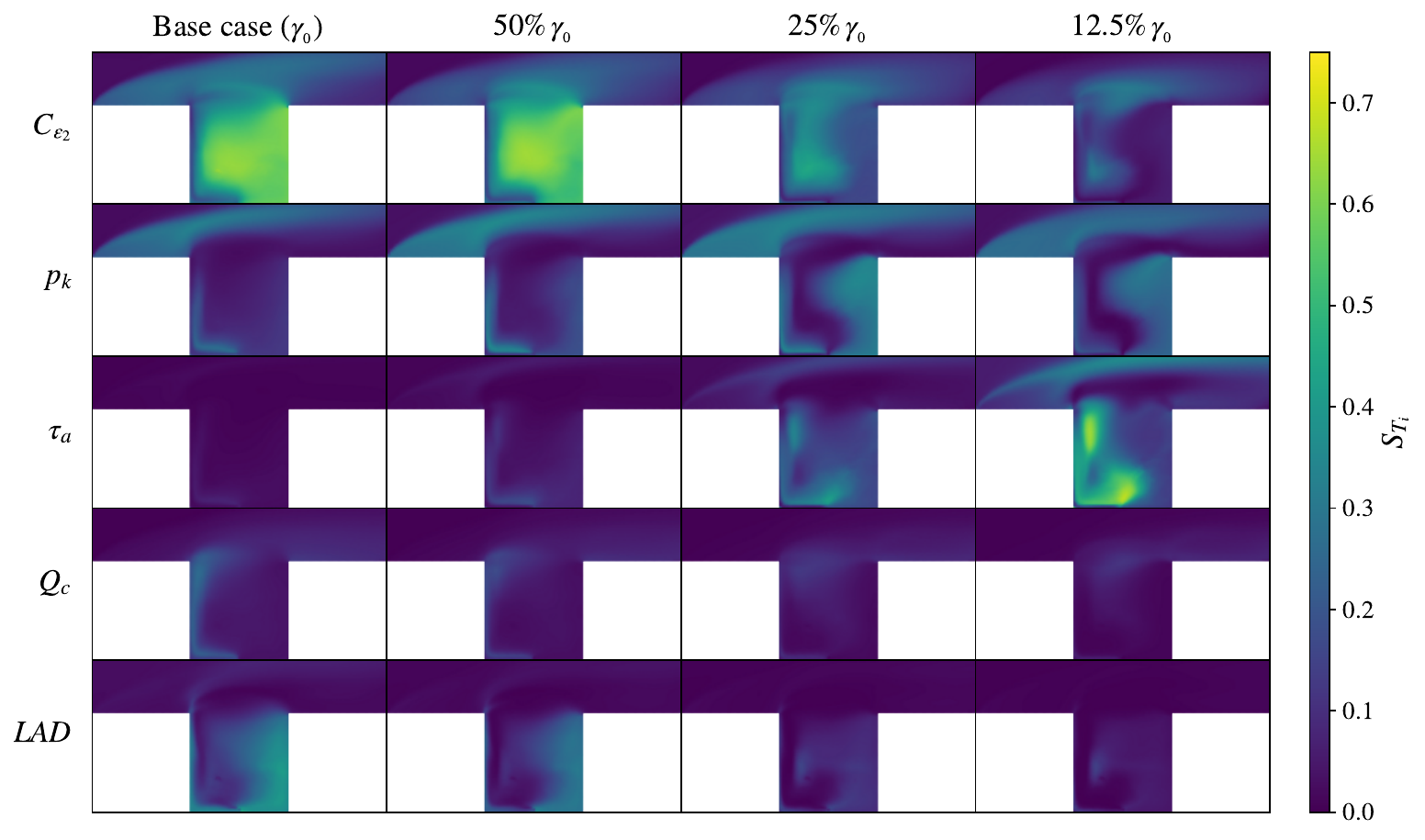} 
	\caption{Spatial distribution of the total-effect Sobol sensitivity indices ($S_{T_i}$) for the \textbf{concentration field} under the \textbf{thermal scenario}. The panels display the contours for five selected parameters ($C_{\varepsilon_2}, p_k, \tau_a, Q_c,LAD$) evaluated across different uncertainty levels: the base case, and reduced scenarios at $50\%$, $25\%$, and $12.5\%$ of the baseline uncertainty ($\gamma_{_0}$). Higher values denote regions where the concentration variance is predominantly driven by the uncertainty of the corresponding parameter.}
	\label{fig:sFieldThermC}
\end{figure}

Figure~\ref{fig:sFieldThermC} resolves these contributions spatially for the concentration field. The two turbulence groups dominate distinct regions: the RANS closure constant governs the canyon cavity itself, where the recirculating flow is well mixed, whereas the inflow turbulence profile governs the shear layer along the roofline and the adjacent leeward wall. This separation explains the persistence reported above, since the roof-level shear layer controls the exchange between the canyon and the overlying flow, and therefore pollutant removal, irrespective of how the interior is mixed. For practitioners, this distinguishes two different modeling efforts: calibrating closure constants improves the prediction of in-canyon mixing, while characterizing the approach flow improves the prediction of ventilation.\par
The solar forcing does not emerge uniformly as the remaining uncertainties are tightened. Instead, its contribution concentrates along the upper leeward wall and at road level, coinciding with the region of highest concentration variance identified in Fig.~\ref{fig:meanVarFields} and with the pedestrian breathing zone. Uncertainty in the solar forcing therefore does not degrade the solution uniformly, but specifically in the region where exposure is assessed.\par
The vegetation contribution follows the canyon vortex, forming a curved footprint through the cavity. Since $LAD$ is not among the parameters defining this scenario, its uncertainty is tightened along with the rest, and its footprint fades accordingly. The emission source, though equally tightened, remains confined to the immediate vicinity of the road and retains a near-constant share throughout.\newpage
The maps also explain the low global sensitivity of $LAD$ in the velocity field noted above: the crowns do influence the velocity within the canyon (Fig.~\ref{fig:ThermV1}), but the velocity variance is concentrated in the shear layer and the downstream wake, so this local contribution carries little weight when aggregated. Reporting the indices at both levels is thus necessary, since a parameter acting strongly in a low-variance region is not identifiable from the global ranking alone.\par
Outside the canyon, the concentration variance is dictated almost entirely by the background concentration, with the transition region governed by the inflow turbulence profiles and the turbulent Schmidt number; the complete 24-parameter maps are reported in \ref{app:AdditRes}.\par

\subsubsection{Vegetated scenario}
\begin{figure}[h!]
	\centering
	\includegraphics[scale=.49]{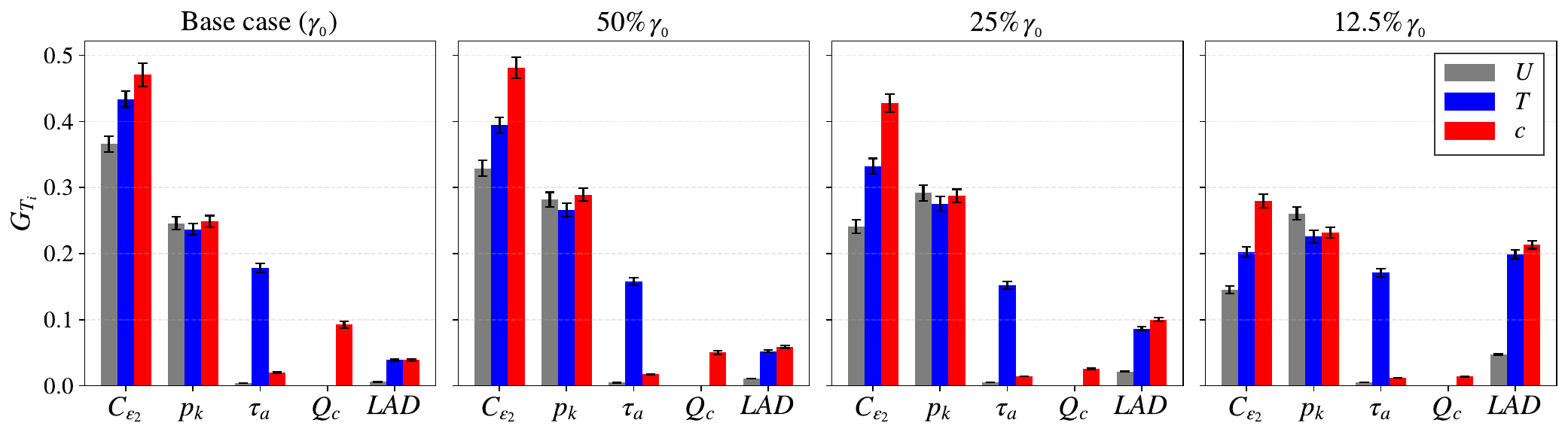}
	\caption{Variance-weighted total-effect ($G_{T_i}$) Sobol sensitivity indices for the velocity ($U$), temperature ($T$), and concentration ($c$) fields under the \textbf{vegetated scenario}. The global indices are aggregated across the $N_{_\text{PC}}$ principal components, weighted by their respective explained variance, for different uncertainty levels: the base case, and reduced scenarios at $50\%$, $25\%$, and $12.5\%$ of the baseline uncertainty ($\gamma_{_0}$). Error bars represent the propagated 95\% confidence intervals from the Sobol analysis.}
	\label{fig:sBarVeg}
\end{figure}

Figure~\ref{fig:sBarVeg} reports the generalized indices for the same five group representatives under the vegetated scenario. At baseline uncertainty the two turbulence groups govern all three fields more strongly than in the thermal case, the solar forcing is again confined to the temperature, and the vegetation contribution to the concentration variance is small, well below the emission source. This is a reversal with respect to the thermal scenario, where $LAD$ ranked among the leading contributors despite an identical distribution. Since the concentration variance is comparable between the two scenarios (Fig.~\ref{fig:meanVarFields}), the reduction is not a normalization effect but a genuine loss of influence: at the higher inflow velocity, pollutant removal is governed by turbulent exchange across the roof-level shear layer, and the canopy perturbs a circulation that no longer controls the concentration field. Vegetation modeling therefore matters most in weakly ventilated canyons, and its omission is least consequential precisely where ventilation is already strong.\par
As the uncertainty of the non-vegetation parameters is tightened, all of them are reduced together, so their mutual ranking is largely preserved and only the vegetation group grows. Its share increases monotonically for all three fields, and at the most reduced level it becomes comparable to the inflow turbulence profile for the concentration, yet it never overtakes the RANS closure. This contrasts with the thermal scenario, where the defining group had already taken the lead at $25\%\,\gamma_{_0}$. Vegetation effects thus remain entangled with turbulence-driven variability across the whole range considered, and even at the most reduced level the turbulence closure remains the dominant contributor. Characterizing the canopy more precisely therefore yields a smaller return under strong ventilation than characterizing the turbulence model.\par

\begin{figure}[h!]
	\centering
	\includegraphics[scale=.65]{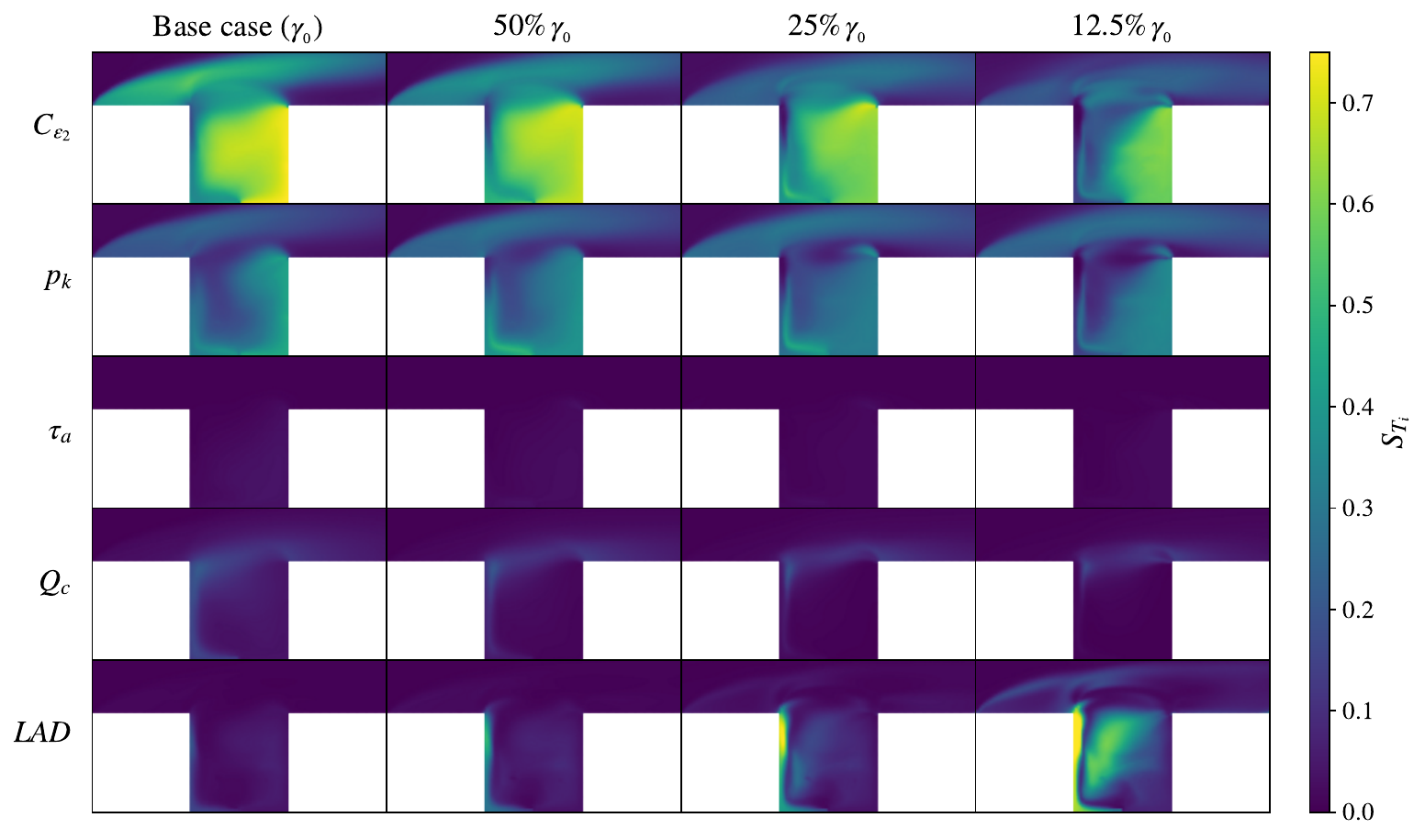}
	\caption{Spatial distribution of the total-effect Sobol sensitivity indices ($S_{T_i}$) for the \textbf{concentration field} under the \textbf{vegetated scenario}. The panels display the contours for five selected parameters ($C_{\varepsilon_2}, p_k, \tau_a, Q_c,LAD$) evaluated across different uncertainty levels: the base case, and reduced scenarios at $50\%$, $25\%$, and $12.5\%$ of the baseline uncertainty ($\gamma_{_0}$). Higher values denote regions where the concentration variance is predominantly driven by the uncertainty of the corresponding parameter.}
	\label{fig:sFieldVegC}
\end{figure}

The spatial maps in Fig.~\ref{fig:sFieldVegC} resolve these contributions for the concentration field. The solar forcing is absent at every level, whereas in the thermal scenario it was already faintly visible along the leeward wall at baseline and grew from there, so under strong ventilation it carries no practical weight for pollutant prediction. The turbulence groups retain the regional division identified previously, with the RANS closure constant occupying the cavity and the inflow profile the shear layer and the adjacent walls, while the emission source stays confined to road level.\par
Unlike the other groups, whose influence spreads over a region, the vegetation contribution concentrates in a narrow band along the leeward facade, where the crown sits closest to the wall, reaching local values far above its aggregated share. At $12.5\%\,\gamma_{_0}$ this band extends into the cavity along the vortex. The moderate global index therefore conceals a strongly localized effect: vegetation governs the concentration variance right at the facade, the plane through which windows exchange air with indoor spaces. Although indoor infiltration is outside the scope of this study, this suggests that canopy characterization may matter for near-facade concentration.\newpage

\subsubsection{Comparison between scenarios}
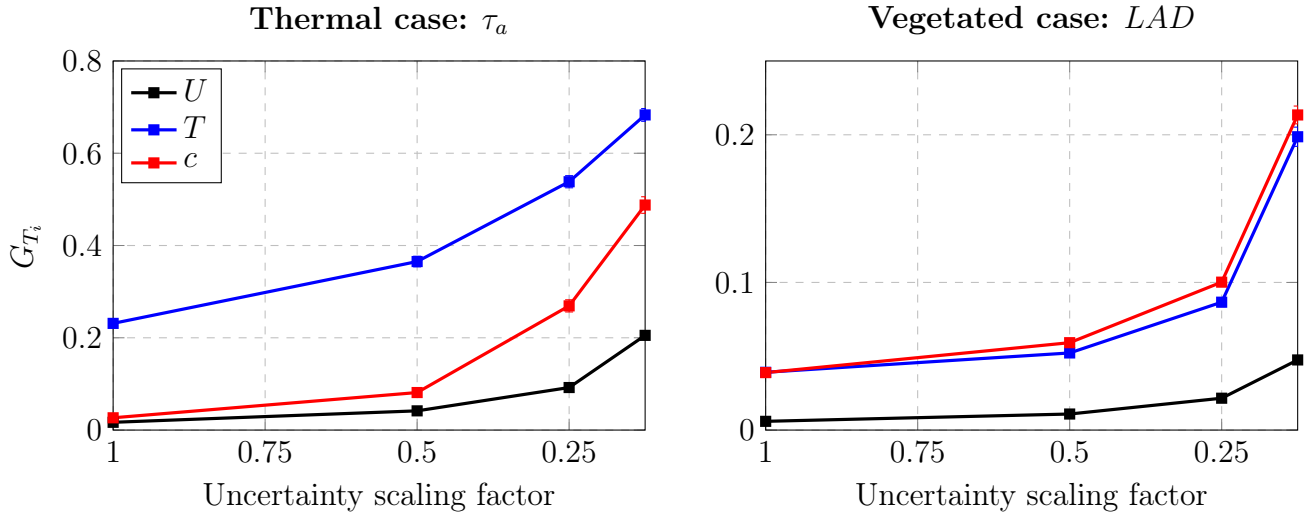
\begin{figure}[h]
    \centering
    \begin{subfigure}{0.49\textwidth}
        \centering
        \begin{tikzpicture}
        \begin{axis}[
            width=1\textwidth,
            height=.75\textwidth,
            ymin=0, ymax=0.8,
            xmin=0.125, xmax=1,
            x dir=reverse,
            xtick={1,0.75,0.5,0.25},
            xlabel={Uncertainty scaling factor},
            ylabel={$G_{T_i}$},
            title={\textbf{Thermal case:} $\tau_a$},
            xmajorgrids=true,
            ymajorgrids=true,
            grid style=dashed,
            legend style={at={(0.015,0.98)}, anchor=north west, draw=black},
            legend cell align={left},
        ]
            \addplot[
                black,
                mark=square*,
                mark size=1.5pt,
                line width=1.2pt,
                error bars/.cd,
                    y dir=both,
                    y explicit
            ] table[
                col sep=comma,
                x=factor,
                y=Therm_U_ST,
                y error=Therm_U_err
            ]{Data/ST_single.csv};
            \addlegendentry{$U$}
            \addplot[
                blue,
                mark=square*,
                mark size=1.5pt,
                line width=1.2pt,
                error bars/.cd,
                    y dir=both,
                    y explicit
            ] table[
                col sep=comma,
                x=factor,
                y=Therm_T_ST,
                y error=Therm_T_err
            ]{Data/ST_single.csv};
            \addlegendentry{$T$}
            \addplot[
                red,
                mark=square*,
                mark size=1.5pt,
                line width=1.2pt,
                error bars/.cd,
                    y dir=both,
                    y explicit
            ] table[
                col sep=comma,
                x=factor,
                y=Therm_c_ST,
                y error=Therm_c_err
            ]{Data/ST_single.csv};
            \addlegendentry{$c$}
        \end{axis}
        \end{tikzpicture}
    \end{subfigure}
    \hfill
    \begin{subfigure}{0.49\textwidth}
        \centering
        \begin{tikzpicture}
        \begin{axis}[
            width=1\textwidth,
            height=.75\textwidth,
            ymin=0, ymax=0.25,
            xmin=0.125, xmax=1,
            x dir=reverse,
            xtick={1,0.75,0.5,0.25},
            xlabel={Uncertainty scaling factor},
            title={\textbf{Vegetated case:} $LAD$},
            xmajorgrids=true,
            ymajorgrids=true,
            grid style=dashed,
            scaled y ticks=false,
            yticklabel style={
                /pgf/number format/fixed,
                /pgf/number format/precision=3
            },
        ]
            \addplot[
                black,
                mark=square*,
                mark size=1.5pt,
                line width=1.2pt,
                error bars/.cd,
                    y dir=both,
                    y explicit
            ] table[
                col sep=comma,
                x=factor,
                y=Veg_U_ST,
                y error=Veg_U_err
            ]{Data/ST_single.csv};
            \addplot[
                blue,
                mark=square*,
                mark size=1.5pt,
                line width=1.2pt,
                error bars/.cd,
                    y dir=both,
                    y explicit
            ] table[
                col sep=comma,
                x=factor,
                y=Veg_T_ST,
                y error=Veg_T_err
            ]{Data/ST_single.csv};
            \addplot[
                red,
                mark=square*,
                mark size=1.5pt,
                line width=1.2pt,
                error bars/.cd,
                    y dir=both,
                    y explicit
            ] table[
                col sep=comma,
                x=factor,
                y=Veg_c_ST,
                y error=Veg_c_err
            ]{Data/ST_single.csv};
        \end{axis}
        \end{tikzpicture}
    \end{subfigure}
    \caption{Variance-weighted total Sobol sensitivity indices ($G_{T_i}$) for the velocity ($U$), temperature ($T$), and concentration ($c$) fields plotted as a function of the uncertainty scaling factor: (a) thermal case characterizing the influence of atmospheric optical depth ($\tau_a$), and (b) vegetated case characterizing the influence of leaf area density ($LAD$). Note that the vertical axis range differs between the two panels. Error bars represent the propagated 95\% confidence intervals from the Sobol analysis.}
    \label{fig:sensitivitySingle}
\end{figure}

Figure~\ref{fig:sensitivitySingle} compares the generalized total-effect indices of the two defining parameters, $\tau_a$ in the thermal scenario and $LAD$ in the vegetated scenario, as the uncertainty of the remaining inputs is progressively reduced. Each is evaluated in the scenario constructed for it. The vegetated scenario was defined with a stronger inflow to emphasize the aerodynamic interaction between the flow and the canopy, yet at baseline uncertainty $LAD$ contributes roughly four times more to the concentration variance in the thermal scenario, where the inflow is weak, despite an identical distribution in both. This is because, under weak ventilation, pollutant removal depends on the in-canyon circulation that the crowns modify, whereas under strong ventilation it is controlled by the exchange across the roof-level shear layer, which the crowns barely affect. The comparison below therefore understates the influence of $LAD$.\par

The two share a common behavior. Both grow monotonically as the uncertainty of the remaining parameters is tightened, neither saturates within the range considered, and both remain weakest for the velocity field, consistent with the velocity variance being concentrated in the shear layer and the wake, where neither the crowns nor the heated surfaces act directly. The increase is also strongly non-uniform: in both scenarios, close to $90\%$ of the total gain in the concentration index occurs in the last two reduction steps. Halving the uncertainty of the remaining inputs once therefore changes little, and a substantial reduction is required before either effect emerges from the turbulence-driven variability. For practitioners this means that partial improvements in the characterization of the turbulence and inflow parameters yield little return in terms of resolving thermal or vegetation effects, and that the two are worth refining only once those inputs are well constrained.\par
The two effects nevertheless differ in two key respects. The first is their magnitude. At baseline, $\tau_a$ already makes an appreciable contribution to the temperature, whereas $LAD$ remains marginal for all three fields. At the lowest uncertainty level, the contribution of $\tau_a$ is roughly three times that of $LAD$ for temperature and twice as large for concentration. These differences partly reflect the wider baseline uncertainty assigned to $\tau_a$ (Table~\ref{tab:paramSens}) and the limited influence of $LAD$ under strong ventilation, as discussed above. The resulting ratios therefore depend not only on the physical influence of each parameter but also on the prescribed input distributions and the scenario considered. The second difference concerns how these effects are distributed across fields. $\tau_a$ acts primarily on temperature, influencing concentration indirectly through buoyancy-driven changes in circulation. Accordingly, its contribution to temperature consistently exceeds its contribution to concentration. By contrast, $LAD$ enters the momentum, turbulence, and energy equations directly, resulting in comparable contributions to both scalar fields.

\subsection{Limitations}\label{sec:limitations}
The two-dimensional configuration adopted here is what makes the analysis tractable, since generating a database of several thousand converged CFD solutions, dense enough to train a surrogate over a 24-parameter space, would carry a computational cost orders of magnitude higher in three dimensions. Two restrictions follow. The flow is imposed perpendicular to the street axis, so inflow angle effects cannot be assessed and would plausibly introduce an uncertainty larger than those considered here. The two scenarios also represent one choice among many: that the canopy contributes markedly less under strong than under weak ventilation was not anticipated, indicating that the flow regime conditions the parameter hierarchy more strongly than expected, so other combinations of inflow and environmental conditions may add nuance to the reported thresholds. In addition, the thermal scenario lies in the low-velocity region where the CFD convergence rate is lowest and the surrogate error highest (Section~\ref{sec:surrogatePerf}), so its thresholds carry the largest uncertainty.

\section{Conclusion}\label{sec:Conclusion}
This work assessed the relative contribution of 24 uncertain parameters, spanning turbulence closure, inflow conditions, pollutant emissions, solar forcing, and vegetation, to the velocity, thermal, and dispersion fields of a two-dimensional street canyon. A PCA+MLP surrogate emulating the CFD solution made a global Sobol analysis over the full parameter space tractable, and generalized variance-weighted indices were used to rank the parameters while retaining the spatially resolved sensitivity maps.\par
At the literature-based uncertainty levels of the base case ($\gamma_{_0}$), the turbulence parameters govern the velocity, temperature, and concentration fields in both the weakly ventilated and the strongly ventilated regime. Under weak ventilation the canopy already contributes to the concentration variance more than the emission rate itself, whereas under strong ventilation its contribution is marginal. The solar forcing, in contrast, contributes appreciably only to the temperature variance in both regimes, its influence on velocity and concentration being masked.\par

This ranking shifts only once the remaining inputs are substantially better characterized. Expressed in absolute terms, the $25\%\,\gamma_{_0}$ level corresponds to knowing the RANS closure constants to about $\pm11\%$, the emission rate to about $\pm12\%$, and the inflow turbulence profiles to about $\pm58\%$. The solar forcing overtakes the turbulence groups for the predicted concentration under weak ventilation at that point, despite the inflow profiles remaining loosely constrained. Vegetation never reaches an equivalent position under strong ventilation: even at $12.5\%\,\gamma_{_0}$, closure constants constrained to about $\pm5\%$, it remains below the closure contribution. Since roughly $90\%$ of the increase occurs over the last two reduction steps, tightening $C_{\varepsilon_2}$ from about $\pm43\%$ to $\pm21\%$ leaves the concentration ranking essentially unchanged; a reduction to about $\pm11\%$ is needed before the solar forcing emerges under weak ventilation, whereas under strong ventilation even $\pm5\%$ is not sufficient for the canopy to dominate.\par

The spatially resolved indices reveal a division of labor that the global ranking conceals. The closure constants govern mixing within the canyon cavity, whereas the inflow turbulence profiles govern the roof-level shear layer and hence the exchange between the canyon and the overlying flow. Calibrating model constants and characterizing the approach flow therefore improve different aspects of the prediction, and are not interchangeable as a modeling effort.\par

The thresholds reported here were obtained for a two-dimensional canyon under perpendicular flow and two ventilation regimes (Section~\ref{sec:limitations}); establishing the parameter hierarchy in this canonical setting is a first step that can inform the design of future studies in more realistic geometries.\par

A final result deserves emphasis for its practical implications. Although the global indices rank thermal and vegetation effects below the turbulence parameters in most of the cases examined, wherever these effects do contribute they do so at road level and along the canyon walls, shaping the fine-scale spatial structure of the concentration field precisely where exposure is determined. These are the locations sampled by pedestrians on the sidewalk, by occupants of vehicles at road level, and by the air drawn into buildings through facade openings. Studies addressing pedestrian or road-user exposure, or outdoor--indoor pollutant exchange, in configurations that include street canyons should therefore account for thermal and vegetation effects even when a global sensitivity screening suggests they are secondary.\par

\section*{CRediT authorship contribution statement}
\textbf{Albert Moreno Català:} Conceptualization, Methodology, Software, Investigation, Formal analysis, Data Curation, Visualization, Writing - Original Draft, Writing - Review \& Editing.
\textbf{Joan Calafell Sandiumenge:} Writing - Review \& Editing.
\textbf{Nikolaj T. Mücke:} Writing - Review \& Editing.
\textbf{Jan Mateu Armengol:} Conceptualization, Methodology, Resources, Supervision, Project administration, Funding acquisition, Writing - Original Draft, Writing - Review \& Editing.

\section*{Declaration of generative AI and AI-assisted technologies in the manuscript preparation process}
During the preparation of this work the authors used Claude in order to improve the structure, readability, and language of the manuscript. After using this tool, the authors reviewed and edited the content as needed and take full responsibility for the content of the published article.

\section*{Declaration of competing interest}
The authors declare that they have no known competing financial interests or personal relationships that could have appeared to influence the work reported in this paper.

\section*{Acknowledgements}
This work has received funding from the European Union’s Horizon Europe Research and Innovation Programme under the project UrbanAIR (grant agreement No. 101188131).
\section*{Data availability}
The datasets and code associated with this study are openly available in Zenodo at \cite{zenodo}.

\clearpage
\bibliographystyle{unsrt}
\bibliography{references}
\newpage

\appendix
\counterwithin*{figure}{section}
\renewcommand{\thefigure}{\Alph{section}.\arabic{figure}}
\counterwithin*{table}{section}
\renewcommand{\thetable}{\Alph{section}.\arabic{table}}

\section{Validation of the numerical solution}
\label{app:modelValidation}
\subsection{Mesh and domain independence}\label{sec:MeshDomConv}
Mesh and domain independence studies are conducted under isothermal conditions for the street canyon configuration proposed in Section~\ref{sec:Methods}. Convergence is assessed by evaluating the turbulent kinetic energy ($k$) profiles along the horizontal and vertical centerlines of the canyon. Five mesh levels, ranging from very coarse to very fine, are evaluated (consisting of 51,080, 54,080, 67,880, 88,880, and 106,880 elements, respectively). As shown in Fig.~\ref{fig:meshDomain}a, the solution converges with the fine grid. Furthermore, sensitivity to the domain height ($6H,\,8H,\,11H$, and $13H$) is tested to minimize top boundary interference (Fig.~\ref{fig:meshDomain}b). The profiles stabilize for domain heights of $8H$ and above. Consequently, the fine mesh and a domain height of $8H$ are selected for all further simulations, as illustrated in Fig.~\ref{fig:mesh}.
\begin{figure}[h]
    \centering
    \hspace{-20pt}
    \begin{subfigure}{0.48\textwidth}
        \centering
        \begin{tikzpicture}
        \begin{axis} [
            name=plotMesh,
            width=1.05\textwidth,
            height=.8\textwidth,
            xlabel={($x,y$)/$H$ [\,-\,]},
            ylabel={$k$ [\,m$^2$/s$^2$\,]},
            tick label style={font=\footnotesize},
            label style={font=\small},
            xmin=0, xmax=1,
            ymin=0, ymax=1.3,
            ymajorgrids=true,
            xmajorgrids=true,
            grid style=dashed,
            legend style={at={(0.015,0.98)}, anchor=north west, draw=black, font=\footnotesize}, 
            legend cell align={left},
            title={\textbf{Mesh Convergence}},
        ]
        
        \addplot[smooth, line width=1.2pt, color=black]
            table [header=false, x index=0, y index=1] {Data/CFDConv/MeshConv/Horizontal_elm_0.1.txt};
        \addlegendentry{Very fine}
        
        \addplot[smooth, line width=1.2pt, color=black, dotted, forget plot]
            table [header=false, x index=0, y index=1] {Data/CFDConv/MeshConv/Vertical_elm_0.1.txt};
        
        \addplot[smooth, line width=1.2pt, color=blue]
            table [header=false, x index=0, y index=1] {Data/CFDConv/MeshConv/Horizontal_elm_0.125.txt};
        \addlegendentry{Fine}
        
        \addplot[smooth, line width=1.2pt, color=blue, dotted, forget plot]
            table [header=false, x index=0, y index=1] {Data/CFDConv/MeshConv/Vertical_elm_0.125.txt};
        
        \addplot[smooth, line width=1.2pt, color=teal]
            table [header=false, x index=0, y index=1] {Data/CFDConv/MeshConv/Horizontal_elm_0.2.txt};
        \addlegendentry{Intermediate}
        
        \addplot[smooth, line width=1.2pt, color=teal, dotted, forget plot]
            table [header=false, x index=0, y index=1] {Data/CFDConv/MeshConv/Vertical_elm_0.2.txt};
        
        \addplot[smooth, line width=1.2pt, color=orange]
            table [header=false, x index=0, y index=1] {Data/CFDConv/MeshConv/Horizontal_elm_0.5.txt};
        \addlegendentry{Coarse}
        
        \addplot[smooth, line width=1.2pt, color=orange, dotted, forget plot]
            table [header=false, x index=0, y index=1] {Data/CFDConv/MeshConv/Vertical_elm_0.5.txt};
        
        \addplot[smooth, line width=1.2pt, color=red]
            table [header=false, x index=0, y index=1] {Data/CFDConv/MeshConv/Horizontal_elm_1.txt};
        \addlegendentry{Very coarse}
    
        \addplot[smooth, line width=1.2pt, color=red, dotted, forget plot]
            table [header=false, x index=0, y index=1] {Data/CFDConv/MeshConv/Vertical_elm_1.txt};
        
        \end{axis}
        \end{tikzpicture}
    \end{subfigure}
    \hspace{5pt}
    \begin{subfigure}{0.48\textwidth}
        \centering
        \begin{tikzpicture}
        \begin{axis} [
            name=plotDomain,
            width=1.05\textwidth,
            height=.8\textwidth,
            xlabel={($x,y$)/$H$ [\,-\,]},
            ylabel={$k$ [\,m$^2$/s$^2$\,]},
            yticklabel pos=right,
            tick label style={font=\footnotesize},
            label style={font=\small},
            xmin=0, xmax=1,
            ymin=0, ymax=1.3,
            ymajorgrids=true,
            xmajorgrids=true,
            grid style=dashed,
            legend style={at={(0.015,0.98)}, anchor=north west, draw=black, font=\footnotesize}, 
            legend cell align={left},
            title={\textbf{Domain Convergence}},
        ]
        
        \addplot[smooth, line width=1.2pt, color=black]
            table [header=false, x index=0, y index=1] {Data/CFDConv/Dh/Horizontal_Dh_12.txt};
        \addlegendentry{$\text{Domain height}=13H$}
        
        \addplot[smooth, line width=1.2pt, color=black, dotted, forget plot]
            table [header=false, x index=0, y index=1] {Data/CFDConv/Dh/Vertical_Dh_12.txt};
        
        \addplot[smooth, line width=1.2pt, color=blue]
            table [header=false, x index=0, y index=1] {Data/CFDConv/Dh/Horizontal_Dh_10.txt};
        \addlegendentry{$\text{Domain height}=11H$}
        
        \addplot[smooth, line width=1.2pt, color=blue, dotted, forget plot]
            table [header=false, x index=0, y index=1] {Data/CFDConv/Dh/Vertical_Dh_10.txt};
        
        \addplot[smooth, line width=1.2pt, color=teal]
            table [header=false, x index=0, y index=1] {Data/CFDConv/Dh/Horizontal_Dh_7.txt};
        \addlegendentry{$\text{Domain height}=8H$}
        
        \addplot[smooth, line width=1.2pt, color=teal, dotted, forget plot]
            table [header=false, x index=0, y index=1] {Data/CFDConv/Dh/Vertical_Dh_7.txt};
        
        \addplot[smooth, line width=1.2pt, color=red]
            table [header=false, x index=0, y index=1] {Data/CFDConv/Dh/Horizontal_Dh_5.txt};
        \addlegendentry{$\text{Domain height}=6H$}
    
        \addplot[smooth, line width=1.2pt, color=red, dotted, forget plot]
            table [header=false, x index=0, y index=1] {Data/CFDConv/Dh/Vertical_Dh_5.txt};
        
        \end{axis}
        \end{tikzpicture}
    \end{subfigure}
    \caption{Independence studies for the numerical setup: (a) mesh convergence study evaluated across five grid sizes (from very coarse to very fine); and (b) domain convergence study for different domain heights. Profiles correspond to the turbulent kinetic energy sampled along the horizontal (solid lines) and vertical (dotted lines) centerlines of the square canyon.}
    \label{fig:meshDomain}
\end{figure}
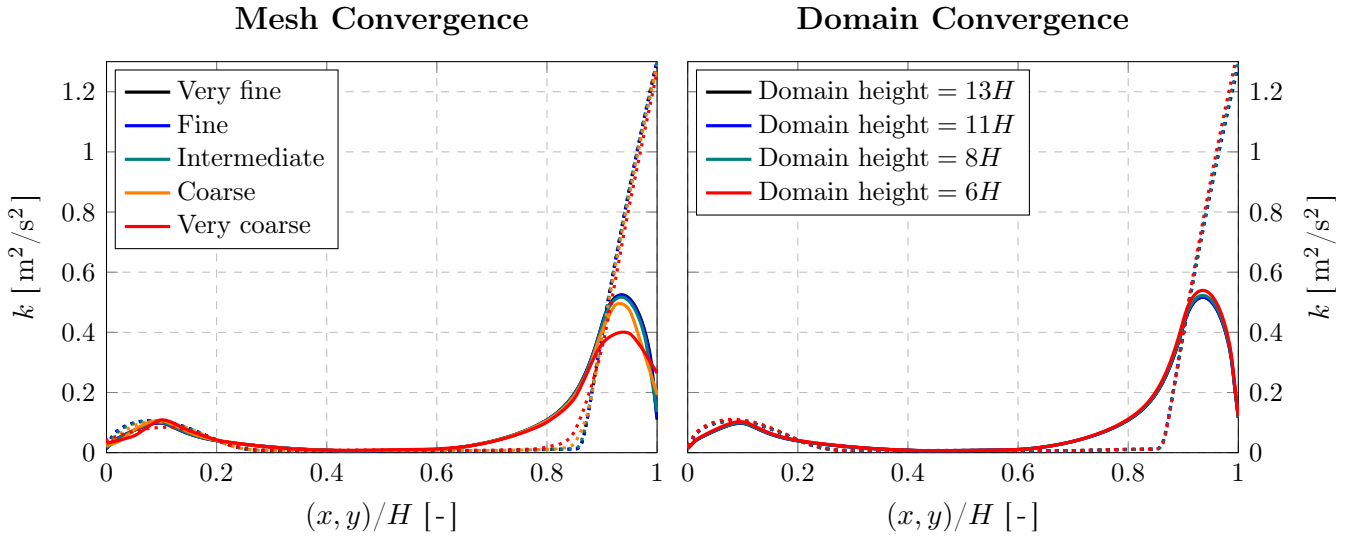
\begin{figure}[ht]
    \centering
    \begin{subfigure}[b]{0.48\textwidth}
        \centering
        \includegraphics[width=\textwidth]{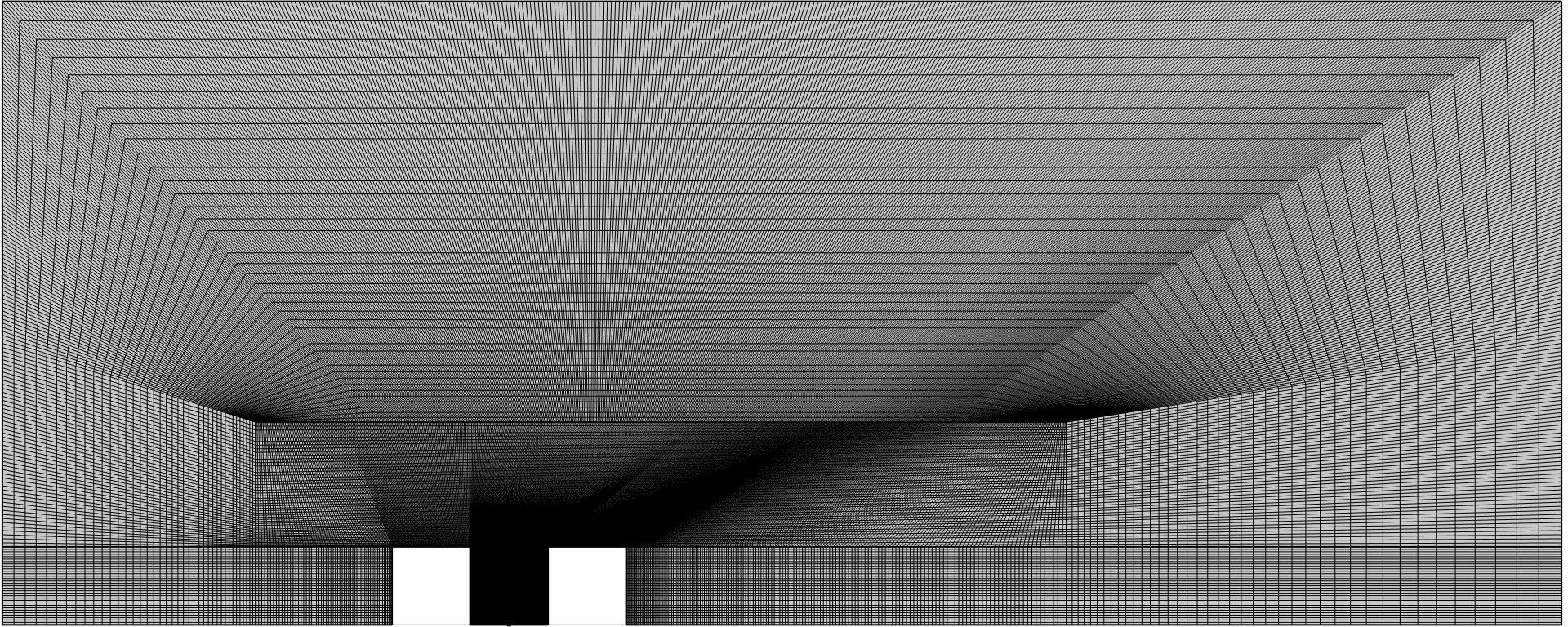}
        \caption{Full domain}
    \end{subfigure}
    \hfill
    \begin{subfigure}[b]{0.48\textwidth}
        \centering
        \includegraphics[width=\textwidth]{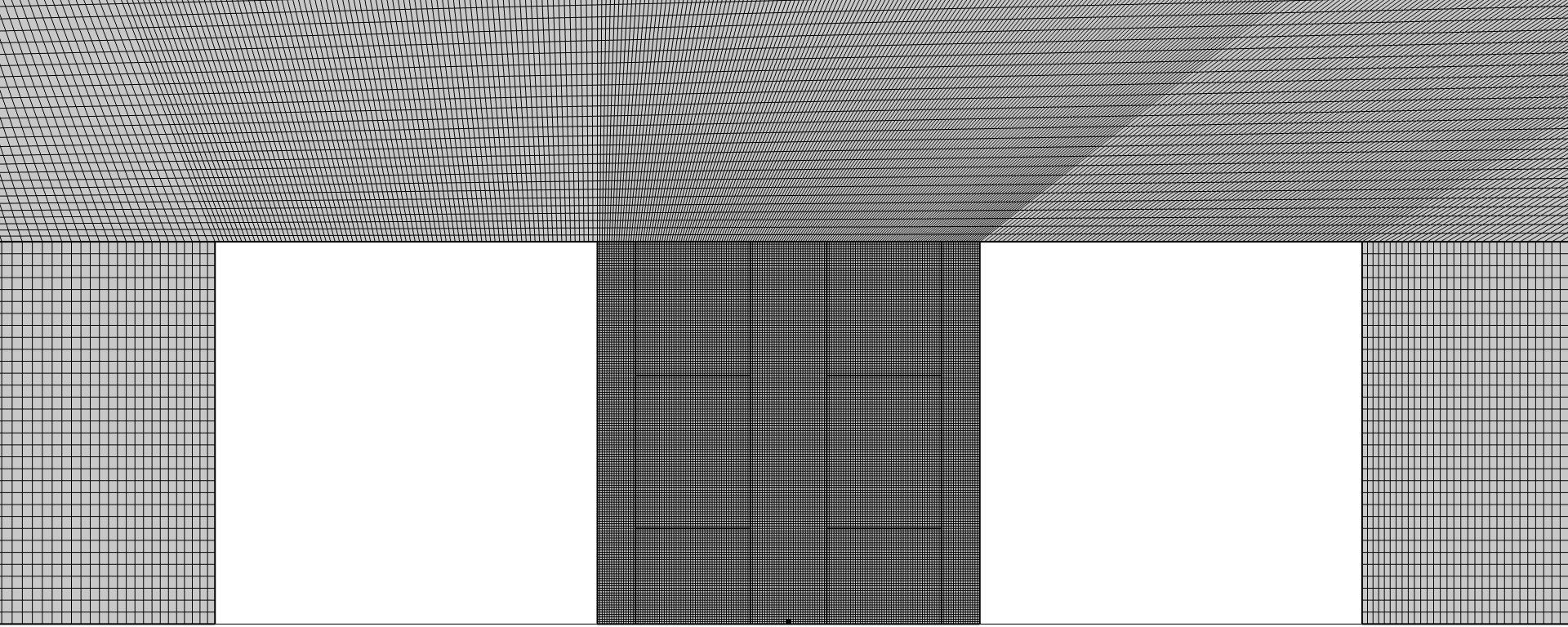}
        \caption{Detailed view}
    \end{subfigure}  
    \caption{Computational mesh: (a) full computational domain; (b) enlarged view of the street canyon cavity.}
    \label{fig:mesh}
\end{figure}

\clearpage
\subsection{RANS model validation}\label{sec:RANSval}
\begin{figure}[h]
    \centering
    \begin{subfigure}{0.48\textwidth}
        \centering
        \begin{tikzpicture}
        \begin{axis}[
            width=1.01\textwidth,
            height=.76\textwidth,
            xtick={0,.2,.4,.6,.8,1},
            ymin=10, ymax=100,
            xmin=0, xmax=1,
            ylabel={$C\mbox{*}$ [\,-\,]},
            xlabel={$y/H$ [\,-\,]},
            ymajorgrids=true,
            xmajorgrids=true,
            grid style=dashed,
            legend style={at={(0.986,0.98)}, anchor=north east, draw=black, font=\footnotesize},
            legend cell align={left},
            title={\textbf{Concentration profile}},
        ]
            \addplot+[
                only marks, mark=square*, black, mark options={black, scale=0.75},
                smooth,
                error bars/.cd, y fixed, y dir=both, y explicit
            ] table [x=x, y=y,y error=error, col sep=comma, only marks] {
                x, y, error
                0.9666666666,41.958041958041946,6.29
                0.6666666666,52.04081632653062,7.81
                0.3333333333,68.62244897959184,10.29
                0.1666666666,79.8469387755102,11.98
            };
            \addplot [line width=1.5pt, color=blue, mark=none] table [y=y, x=x, col sep=tab, mark=none]{Data/RANSVal/line_data1.txt};
            \legend{Measurements \cite{meroney1996study}, Low-Reynolds $k$\,--\,$\varepsilon$}
        \end{axis}
        \end{tikzpicture}
    \end{subfigure}
    \hfill
    \begin{subfigure}{0.48\textwidth}
        \centering
        \begin{tikzpicture}
        \begin{axis} [
            width=1.01\textwidth,
            height=.76\textwidth,
            title={\textbf{Temperature profile}},
            ymin=30, ymax=60,
            xmin=0, xmax=1,
            xlabel={$x/H$ [\,-\,]},
            ylabel={$T$ $[\,^\circ$C\,]},
            xmajorgrids=true,
            ymajorgrids=true,
            grid style=dashed,
            yticklabel pos=right,
            legend style={at={(0.986,0.98)}, anchor=north east, draw=black, font=\footnotesize},
            legend cell align={left},
        ]
            \addplot [line width=1.5pt, color=black, mark=none] table [col sep=comma, x=Xm, y=Ym, mark=none]{Data/RANSVal/0.95Bib.csv};
            \addplot [line width=1.5pt, color=red, mark=none, dashed] table [col sep=comma, x=Xn, y=Yn, mark=none, dashed]{Data/RANSVal/0.95Bib.csv};
            \addplot [line width=1.5pt, color=violet, mark=none, dashed] table [col sep=comma, x=Xl, y=Yl, mark=none]{Data/RANSVal/0.95Bib.csv};
            \addplot [line width=1.5pt, color=blue, mark=none] table [x expr=\thisrow{x}/0.2, y=y, mark=none]{Data/RANSVal/0.95.txt};

            \legend{Measurements \cite{allegrini2014buoyant}, NEWF \cite{allegrini2014buoyant}, LRNM \cite{allegrini2014buoyant}, Low-Reynolds $k$\,--\,$\varepsilon$}
        \end{axis}
        \end{tikzpicture}
    \end{subfigure}
    \caption{Validation of the implemented low-Reynolds-number RANS $k$\,--\,$\varepsilon$ model against experimental and numerical literature benchmarks: (a) vertical profile of non-dimensional concentration along the canyon leeward wall compared with the wind tunnel data from Meroney \textit{et al}.~\cite{meroney1996study}; and (b) horizontal temperature profile sampled at height $y/H = 0.95$ across the canyon width compared with wind tunnel measurements and alternative numerical models (NEWF and LRNM) reported by Allegrini \textit{et al}.~\cite{allegrini2014buoyant}.}
    \label{fig:RANSval}
\end{figure}
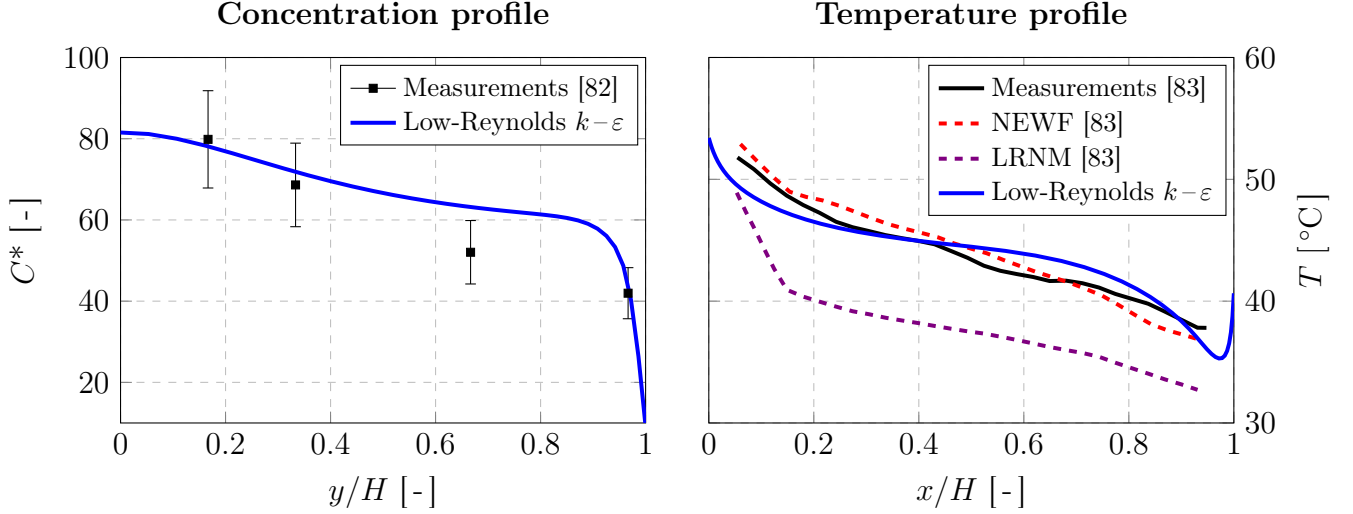

The low-Reynolds-number $k$\,--\,$\varepsilon$ model is validated against the isothermal 2D street canyon experiment of Meroney \textit{et al}.~\cite{meroney1996study}. As shown in Fig.~\ref{fig:RANSval}a, the vertical profile of non-dimensional concentration along the leeward wall is in good agreement with the experimental data. The model is also validated under buoyant conditions using the heated street canyon configuration of Allegrini \textit{et al}.~\cite{allegrini2014buoyant}. Figure~\ref{fig:RANSval}b compares the horizontal temperature profile at $y/H=0.95$ with both the experimental wind tunnel measurements and the alternative RANS formulations (NEWF and LRNM) reported in their work, indicating that the model reliably reproduces non-isothermal fields.

\subsection{Vegetation model validation}\label{sec:vegVal}
\begin{figure}[H]
    \centering
    \begin{tikzpicture}
    \begin{axis}[name=plotA,
        width=.5\textwidth, height=.25\textwidth,
        title={\textbf{z = 1.25H}},
        title style={at={(axis description cs:0.5,0.85)}, anchor=north},
        ymin=0, ymax=6, xmin=-5, xmax=19,
        ylabel={$U$ [\,m/s\,]},
        xticklabels={},
        ymajorgrids=true, grid style=dashed,
        scaled ticks=false,
    ]
        \addplot [only marks, mark=diamond*, mark size=2pt, color=black] 
            table [col sep=comma]{Data/VegValidation/valWT1.25.csv};
        \addplot [line width=1.5pt, color=red, mark=none]
            table [col sep=comma]{Data/VegValidation/valKE1.25.csv};
        \addplot [line width=1.5pt, color=blue, mark=none] 
            table [col sep=space]{Data/VegValidation/valLR1.25.csv};
    \end{axis}
    
    \begin{axis}[name=plotB, at=(plotA.outer east), anchor=outer west, xshift=4pt,
        width=.5\textwidth, height=.25\textwidth,
        title={\textbf{z = 1.25H}},
        title style={at={(axis description cs:0.5,0.85)}, anchor=north, xshift=48pt},
        ymin=0, ymax=3, xmin=-5, xmax=19,
        ylabel={$k$ [\,m$^2$/s$^2$\,]},
        xticklabels={},
        legend pos=north west, legend cell align={left}, legend style={font=\scriptsize, xshift=-3.6pt},
        ymajorgrids=true, grid style=dashed,
        scaled ticks=false,
        yticklabel pos=right,
    ]
        \addplot [only marks, mark=diamond*, mark size=2pt, color=black]
            table [col sep=comma]{Data/VegValidation/valWTk1.25.csv};
        \addplot [line width=1.5pt, color=red, mark=none]
            table [col sep=comma]{Data/VegValidation/valKEk1.25.csv};
        \addplot [line width=1.5pt, color=blue, mark=none]
            table [col sep=space]{Data/VegValidation/valLRk1.25.csv};
        \legend{Measurements \cite{liang2006improved}, Standard $k$\,--\,$\varepsilon$ \cite{liang2006improved}, Low-Reynolds $k$\,--\,$\varepsilon$}
    \end{axis}
    
    \begin{axis}[name=plotC, at={(plotA.south west)}, anchor=north west, yshift=-10pt,
        width=.5\textwidth, height=.25\textwidth,
        title={\textbf{z = 0.75H}},
        title style={at={(axis description cs:0.5,0.85)}, anchor=north},
        ymin=0, ymax=6, xmin=-5, xmax=19,
        ylabel={$U$ [\,m/s\,]},
        xticklabels={},
        ytick={0, 2, 4},
        ymajorgrids=true, grid style=dashed,
        scaled ticks=false,
    ]
        \addplot [only marks, mark=diamond*, mark size=2pt, color=black] 
            table [col sep=comma]{Data/VegValidation/valWT0.75.csv};
        \addplot [line width=1.5pt, color=red, mark=none]
            table [col sep=comma]{Data/VegValidation/valKE0.75.csv};
        \addplot [line width=1.5pt, color=blue, mark=none] 
            table [col sep=space]{Data/VegValidation/valLR0.75.csv};
    \end{axis}
    
    \begin{axis}[name=plotD, at=(plotC.outer east), anchor=outer west, xshift=4pt,
        width=.5\textwidth, height=.25\textwidth,
        title={\textbf{z = 0.75H}},
        title style={at={(axis description cs:0.5,0.85)}, anchor=north},
        ymin=0, ymax=3, xmin=-5, xmax=19,
        ylabel={$k$ [\,m$^2$/s$^2$\,]},
        xticklabels={},
        ymajorgrids=true, grid style=dashed,
        scaled ticks=false,
        ytick={0, 1, 2},
        yticklabel pos=right,
    ]
        \addplot [only marks, mark=diamond*, mark size=2pt, color=black] 
            table [col sep=comma]{Data/VegValidation/valWTk0.75.csv};
        \addplot [line width=1.5pt, color=red, mark=none]
            table [col sep=comma]{Data/VegValidation/valKEk0.75.csv};
        \addplot [line width=1.5pt, color=blue, mark=none] 
            table [col sep=space]{Data/VegValidation/valLRk0.75.csv};
    \end{axis}
    
    \begin{axis}[name=plotE, at={(plotC.south west)}, anchor=north west, yshift=-10pt,
        width=.5\textwidth, height=.25\textwidth,
        title={\textbf{z = 0.25H}},
        title style={at={(axis description cs:0.5,0.85)}, anchor=north},
        ymin=0, ymax=6, xmin=-5, xmax=19,
        xlabel={$x/H$ [\,-\,]}, ylabel={$U$ [\,m/s\,]},
        ymajorgrids=true, grid style=dashed,
        ytick={0, 2, 4},
        legend pos=north east, legend cell align={left}, legend style={font=\tiny, xshift=3.25pt},
        scaled ticks=false,
        xticklabel style={/pgf/number format/.cd,fixed,precision=3},
    ]
        \addplot [only marks, mark=diamond*, mark size=2pt, color=black] 
            table [col sep=comma]{Data/VegValidation/valWT0.25.csv};
        \addplot [line width=1.5pt, color=red, mark=none]
            table [col sep=comma]{Data/VegValidation/valKE0.25.csv};
        \addplot [line width=1.5pt, color=blue, mark=none] 
            table [col sep=space]{Data/VegValidation/valLR0.25.csv};
    \end{axis}
    
    \begin{axis}[name=plotF, at=(plotD.south west), anchor=north west, yshift=-10pt,
        width=.5\textwidth, height=.25\textwidth,
        title={\textbf{z = 0.25H}},
        title style={at={(axis description cs:0.5,0.85)}, anchor=north},
        ymin=0, ymax=3, xmin=-5, xmax=19,
        xlabel={$x/H$ [\,-\,]}, ylabel={$k$ [\,m$^2$/s$^2$\,]},
        ymajorgrids=true, grid style=dashed,
        scaled ticks=false,
        ytick={0, 1, 2},
        xticklabel style={/pgf/number format/.cd,fixed,precision=3},
        yticklabel pos=right,
    ]
        \addplot [only marks, mark=diamond*, mark size=2pt, color=black] 
            table [col sep=comma]{Data/VegValidation/valWTk0.25.csv};
        \addplot [line width=1.5pt, color=red, mark=none]
            table [col sep=comma]{Data/VegValidation/valKEk0.25.csv};
        \addplot [line width=1.5pt, color=blue, mark=none] 
            table [col sep=space]{Data/VegValidation/valLRk0.25.csv};
    \end{axis}
    \end{tikzpicture}
    \caption{Validation of the vegetation model. Comparison of horizontal profiles for velocity magnitude $(U)$ and turbulent kinetic energy $(k)$ at different heights ($z/H=1.25,0.75,0.25$) across the domain. The proposed low-Reynolds-number $k$\,--\,$\varepsilon$ model is validated against the wind tunnel measurements and standard $k$\,--\,$\varepsilon$ RANS simulations reported by Liang \textit{et al}.~\cite{liang2006improved}.}
    \label{fig:vegValidation}
\end{figure}
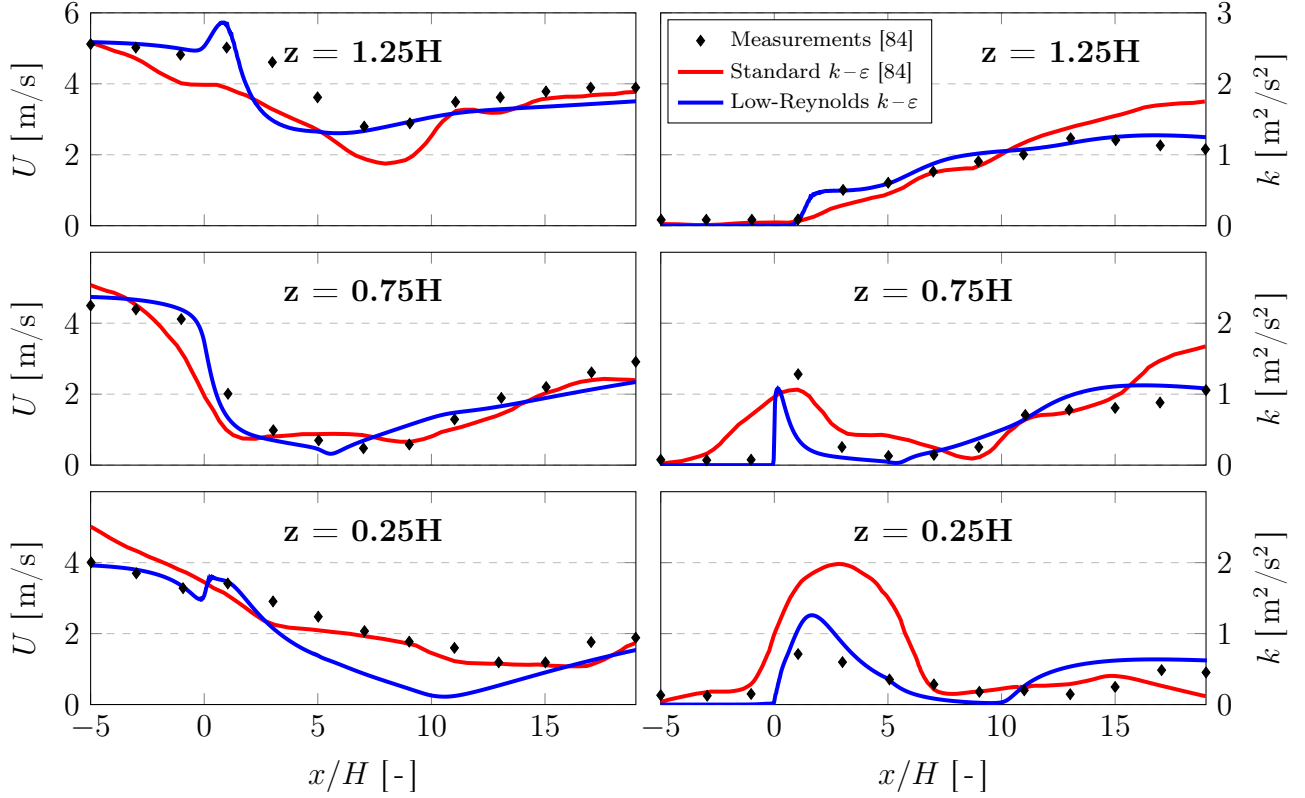

The vegetation parameterization is assessed using the wind tunnel data of a forest stand reported by Liang \textit{et al}.~\cite{liang2006improved}. In this configuration, $LAD$ and drag coefficient vary as a function of height to capture the distinct aerodynamic behaviors of the trunk zone and the upper canopy layer. Figure~\ref{fig:vegValidation} compares the simulated horizontal profiles of velocity magnitude ($U$) and turbulent kinetic energy ($k$), against both the experimental data and the standard $k$\,--\,$\varepsilon$ simulations at normalized heights of $z/H = 0.25, 0.75,$ and $1.25$. These results confirm the capability of the model to simulate vegetation-flow interactions.

\section{Selection of Principal Components}\label{app:nPC}
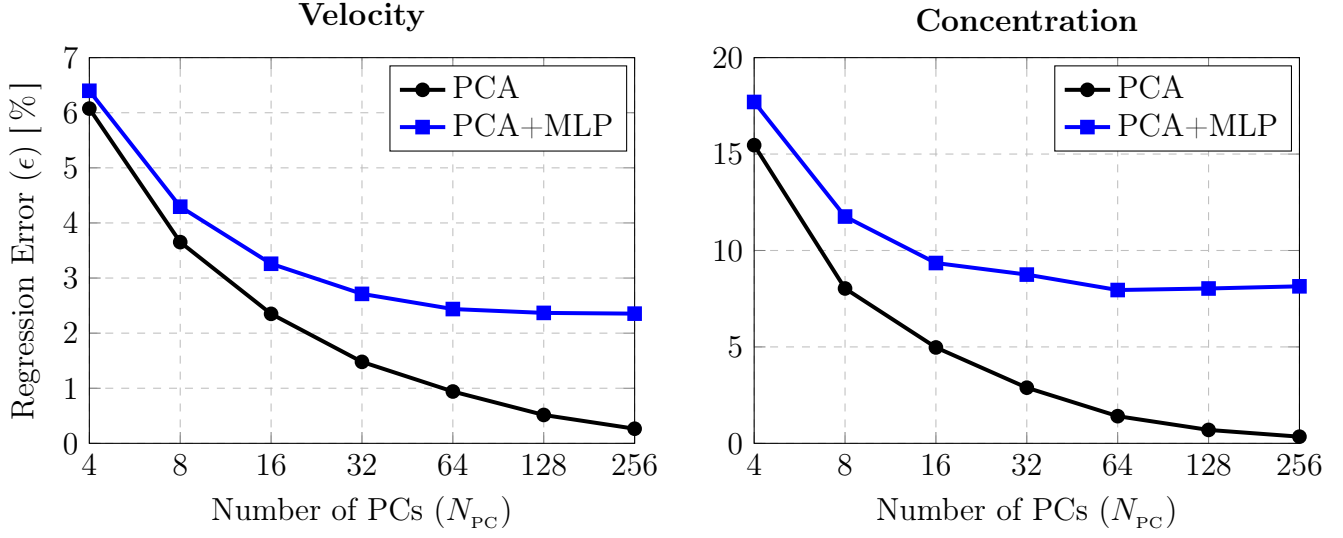
\begin{figure}[h]
    \centering
    \begin{tikzpicture}
    \begin{axis}[
        width=.5\textwidth,
        height=.38\textwidth,
        xlabel={Number of PCs ($N_{_\text{PC}}$)},
        ylabel={Regression Error $\left(\epsilon\right)$ [\,\%\,]},
        xmin=4, xmax=256,
        ymin=0, ymax=7,
        xmode=log, log basis x={2},
        xtick={4,8,16,32,64,128,256},
        log ticks with fixed point,
        ytick={0,1,2,3,4,5,6,7},
        ymajorgrids=true,
        xmajorgrids=true,
        grid style=dashed,
        legend style={xshift=1.3pt},
        legend cell align={left},
        title={\textbf{Velocity}},
    ]
    \addplot[line width=1.5pt, color=black, mark=*]
        table [x index=0, y index=1, col sep=comma] {Data/SurrogateError/ErrorVsPC_vel.csv};
    \addlegendentry{PCA}
    \addplot[line width=1.5pt, color=blue, mark=square*]
        table [x index=0, y index=2, col sep=comma] {Data/SurrogateError/ErrorVsPC_vel.csv};
    \addlegendentry{PCA+MLP}
    \end{axis}
    \begin{axis}[
        width=.5\textwidth,
        height=.38\textwidth,
        at={(0.5\textwidth,0)},
        anchor=south west,
        xlabel={Number of PCs ($N_{_\text{PC}}$)},
        xmin=4, xmax=256,
        ymin=0, ymax=20,
        xmode=log, log basis x={2},
        xtick={4,8,16,32,64,128,256},
        log ticks with fixed point,
        ytick={0,5,10,15,20,25},
        ymajorgrids=true,
        xmajorgrids=true,
        grid style=dashed,
        legend style={xshift=1.3pt},
        legend cell align={left},
        title={\textbf{Concentration}},
    ]
    \addplot[line width=1.5pt, color=black, mark=*]
        table [x index=0, y index=1, col sep=comma] {Data/SurrogateError/ErrorVsPC_con.csv};
    \addlegendentry{PCA}
    \addplot[line width=1.5pt, color=blue, mark=square*]
        table [x index=0, y index=2, col sep=comma] {Data/SurrogateError/ErrorVsPC_con.csv};
    \addlegendentry{PCA+MLP}
    \end{axis}
    \end{tikzpicture}
    \caption{Convergence of the relative $L_2$ test error as a function of the number of principal components ($N_{_\text{PC}}$) for the velocity and concentration fields. The plots compare the theoretical PCA truncation error (black circles) against the total regression error of the combined PCA+MLP surrogate architecture (blue squares), evaluated using 476 test samples.}
    \label{fig:MLPvsPCA}
\end{figure}

The convergence of the surrogate model error is presented in Fig.~\ref{fig:MLPvsPCA}. It is observed that while the theoretical PCA truncation error continues to decrease as more principal components are included, the mean test regression error of the PCA+MLP architecture reaches a performance floor.\par
For the velocity field, the reconstruction error stabilizes at approximately 2.4\% beyond $N_{_\text{PC}}=128$. In contrast, the concentration field reaches its minimum error of approximately 8.7\% at $N_{_\text{PC}}=64$. Notably, incorporating higher-order modes beyond this threshold for the scalar fields leads to a slight degradation in predictive performance.\par
This discrepancy suggests that the mapping between the 24-dimensional input parameter space and the higher-order PCA coefficients becomes increasingly complex and sensitive to non-linear flow features and numerical noise. Beyond these thresholds, the additional modes capture negligible variance and represent high-frequency fluctuations that do not correlate with the input parameters. Consequently, including these higher-order modes does not improve the predictive capability of the MLP, as the model reaches a regression floor where the signal-to-noise ratio is too low for effective learning.
\newpage
The higher error floor observed for the concentration field compared to the velocity field is likely associated with differences in the governing transport processes. While the velocity field is smoothed by momentum diffusion, the pollutant concentration exhibits localized plumes characterized by sharp spatial gradients, particularly near the emission source. Linear dimensionality reduction techniques, such as PCA, are less suited to representing these steep features. Furthermore, since scalar transport is coupled to the flow field, prediction errors in the velocity field may propagate into the scalar solution. These effects, together with the nonlinear interactions associated with buoyancy and aerodynamic vegetation effects, contribute to the increased complexity of the pollutant dispersion regression problem addressed by the MLP.

\newpage
\section{Statistical distributions}\label{app:statDistr}
Table~\ref{tab:paramSens} reports the complete set of statistical distributions and associated uncertainty levels corresponding to the base case ($\gamma_{_0}$) used to generate the thermal and vegetated uncertainty scenarios presented in Section~\ref{sec:uncertScenarios}; the reduced-uncertainty levels (50\%, 25\%, and 12.5\% of $\gamma_{_0}$) are obtained by scaling these distributions as described therein.
\begin{table}[h]
\centering
\begin{adjustbox}{max width=\textwidth}
\renewcommand{\arraystretch}{1.2}
\begin{tabular}{|c|c|c|c|c|c|c|}
\hline
\rowcolor[HTML]{E0E0E0}
Group &
  \multicolumn{1}{c|}{Parameter ($\theta_i$)} & \multicolumn{1}{c|}{Distribution} & \multicolumn{1}{c|}{Mean} & \multicolumn{1}{c|}{Uncertainty ($\gamma_{_0}$)} & \multicolumn{1}{c|}{Unit} & \multicolumn{1}{c|}{Ref.} \\ \hline
\multirow{6}{*}{\shortstack[c]{RANS\\closure}}
    & $Sc_t$ & $U(0.2, 1)$ & 0.6 & $\pm67\%$ & - & Table~\ref{tab:parameters} \\ \cline{2-7}
    & $C_{\varepsilon_1}$ & $N(1.4,0.1)$ & 1.4 & $\pm21\%$ & - & Table~\ref{tab:parameters} \\ \cline{2-7}
    & $C_{\varepsilon_2}$ & $N(2.1,0.3)$ & 2.1 & $\pm43\%$ & - & Table~\ref{tab:parameters} \\ \cline{2-7}
    & $C_\mu$ & $\mathcal{L}(\ln(0.025),0.35)$ & 0.025 & $\pm185\%$ & - & Table~\ref{tab:parameters} \\ \cline{2-7}
    & $\sigma_k$ & $\mathcal{L}(\ln(0.6),0.2)$ & 0.6 & $\pm82\%$ & - & Table~\ref{tab:parameters} \\ \cline{2-7}
    & $B$ & $U(4.4,5.6)$ & 5.0 & $\pm12\%$ & - & Table~\ref{tab:parameters} \\ \hline

\multirow{4}{*}{\shortstack[c]{Inflow\\velocity}}
    & \multirow{2}{*}{$U_\tau$} & $N(0.26,0.026)$\textsuperscript{a} & 0.26 & \multirow{2}{*}{$\pm30\%$} & \multirow{2}{*}{m/s} & \multirow{2}{*}{\cite{andreas1992uncertainty}} \\ \cline{3-4}
    &  & $N(0.75,0.075)$\textsuperscript{b} & 0.75 &  &  &  \\ \cline{2-7}
    & $\kappa$ & $U(0.36,0.44)$ & 0.4 & $\pm10\%$ & - & \cite{segalini2013uncertainty} \\ \cline{2-7} 
    & $y_0$ & $\mathcal{L}(\ln(1.5),0.167)$ & 1.5 & $\pm65\%$ & m & \cite{fang1992aerodynamic} \\ \hline
\multirow{2}{*}{\shortstack[c]{Inflow\\turbulence}}
    & $p_k$ & $\mathcal{L}(0,0.4)$ & 1.0 & $\pm232\%$ & - & Table~\ref{tab:parameters} \\ \cline{2-7}
    & $p_\varepsilon$ & $\mathcal{L}(0,0.4)$ & 1.0 & $\pm232\%$ & - & Table~\ref{tab:parameters} \\ \hline
\multirow{2}{*}{\shortstack[c]{Inflow\\concentration}}
    & $B_k$ & $N(0.25,0.042)$ & 0.25 & $\pm50\%$ & $\mu$mol/m$^3$ & - \\ \cline{2-7}
    & $p_{_B}$ & $U(-0.8,0.8)$ & 0.0 & $\pm0.8$\textsuperscript{c} & - & Table~\ref{tab:parameters} \\ \hline
\multirow{3}{*}{Emission}
    & $Q_c$ & $N(2.3,0.38)$ & 2.3 & $\pm50\%$ & $\mu$mol/(m·s) & \cite{lumbreras2009computation} \\ \cline{2-7}
    & $y_{_Q}$ & $N(0.35,0.035)$ & 0.35 & $\pm30\%$ & m & - \\ \cline{2-7}
    & $Q_p$ & $N(4,0.4)$ & 4.0 & $\pm30\%$ & kW/m &\cite{qin1993dispersion,li2019case} \\ \hline
\multirow{2}{*}{\shortstack[c]{Solar\\forcing}}
    & $\theta$ & $U(115,135)$ & 125 & $\pm8\%$ & deg & - \\ \cline{2-7}
    & $\tau_a$ & $\mathcal{L}(\ln(0.45),0.5)$ & 0.45 & $\pm348\%$ & - & \cite{gil2023geometrical} \\ \hline
\multirow{6}{*}{Vegetation}
    & $LAD$ & $U(0,2)$ & 1.0 & $\pm100\%$ & 1/m & Table~\ref{tab:parameters} \\ \cline{2-7}
    & $\beta_d$ & $U(3,6.5)$ & 4.75 & $\pm37\%$ & - & Table~\ref{tab:parameters} \\ \cline{2-7}
    & $C_{\varepsilon_4}$ & $N(1.5,0.15)$ & 1.5 & $\pm30\%$ & - & Table~\ref{tab:parameters} \\ \cline{2-7}
    & $C_{\varepsilon_5}$ & $N(1.2,0.2)$ & 1.2 & $\pm50\%$ & - & Table~\ref{tab:parameters} \\ \cline{2-7}
    & $C$ & $N(143.5,18)$ & 143.5 & $\pm38\%$ & s$^{0.5}$m$^{-1}$ & Table~\ref{tab:parameters} \\ \cline{2-7}
    & $l$ & $U(0.02,0.5)$ & 0.26 & $\pm92\%$ & m & Table~\ref{tab:parameters} \\ \hline
\end{tabular}
\end{adjustbox}
\caption{Statistical distributions and uncertainty ranges of the input parameters, corresponding to the base case ($\gamma_{_0}$) of the Sobol sensitivity analysis. Parameters referencing Table~\ref{tab:parameters} retain the distributions used to generate the CFD database. \textsuperscript{a}~Thermal scenario. \textsuperscript{b}~Vegetated scenario. \textsuperscript{c}~Since $p_{_B}$ is centered at zero, $\gamma_{_0}$ is reported as the absolute interval half-width rather than a percentage.}
\label{tab:paramSens}
\end{table}

\newpage
\section{Temperature Model Error}\label{sec:tempModelError}
Unlike the velocity and concentration error maps of Fig.~\ref{fig:surrogateError}, the temperature error does not concentrate in a single, coherent region of the parameter space, reflecting the larger number of competing thermal mechanisms at play (surface shading, transpirative cooling, and the advected plume). Some trends remain visible nonetheless: the error increases at low $\kappa$, consistent with the velocity and concentration trends of Fig.~\ref{fig:surrogateError}, and mildly for $\theta>120^\circ$, coinciding with the onset of windward-wall heating and the resulting buoyant vortex reversal discussed in Section~\ref{sec:surrogatePerf}. A weaker increase is also discernible at both ends of the $U_\tau$ range relative to its central values, suggesting a comparatively lower accuracy of the temperature surrogate away from moderate inflow velocities.
\begin{figure}[h!]
	\centering
	\includegraphics[scale=.65]{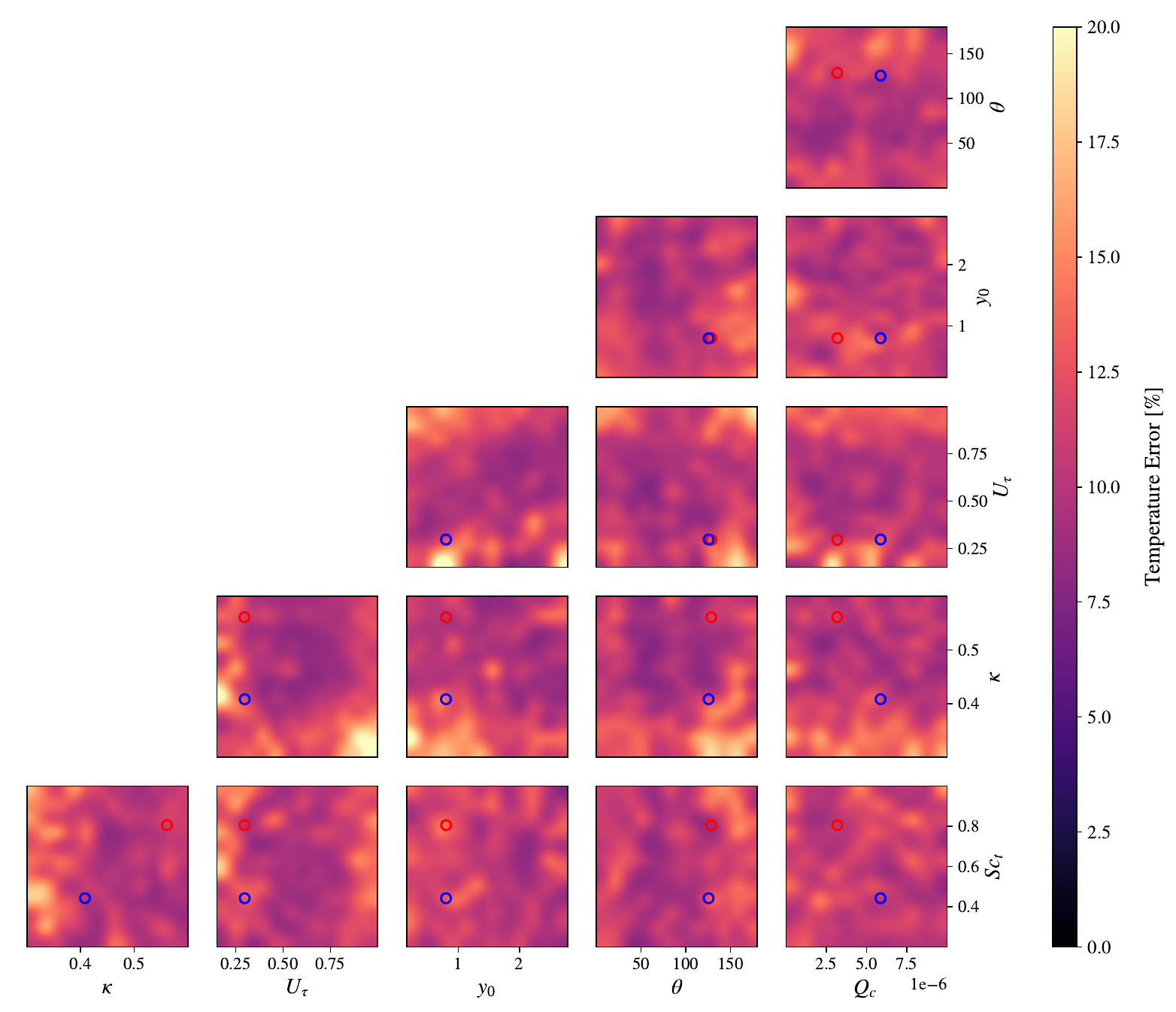}
	\caption{Error heatmap matrix illustrating the localized performance of the PCA+MLP surrogate model for the temperature field across pairwise parameter combinations. The maps display the mean relative $L_2$ error, computed by averaging the test errors within two-dimensional parameter bins. Red and blue circles mark the locations of the two representative test cases (Sample 1 and Sample 2) within the parametric space.}
	\label{fig:surrogateErrorTemp}
\end{figure}

\newpage
\section{Additional results}\label{app:AdditRes}
\begin{figure}[H]
    \centering
    \includegraphics[scale=0.88]{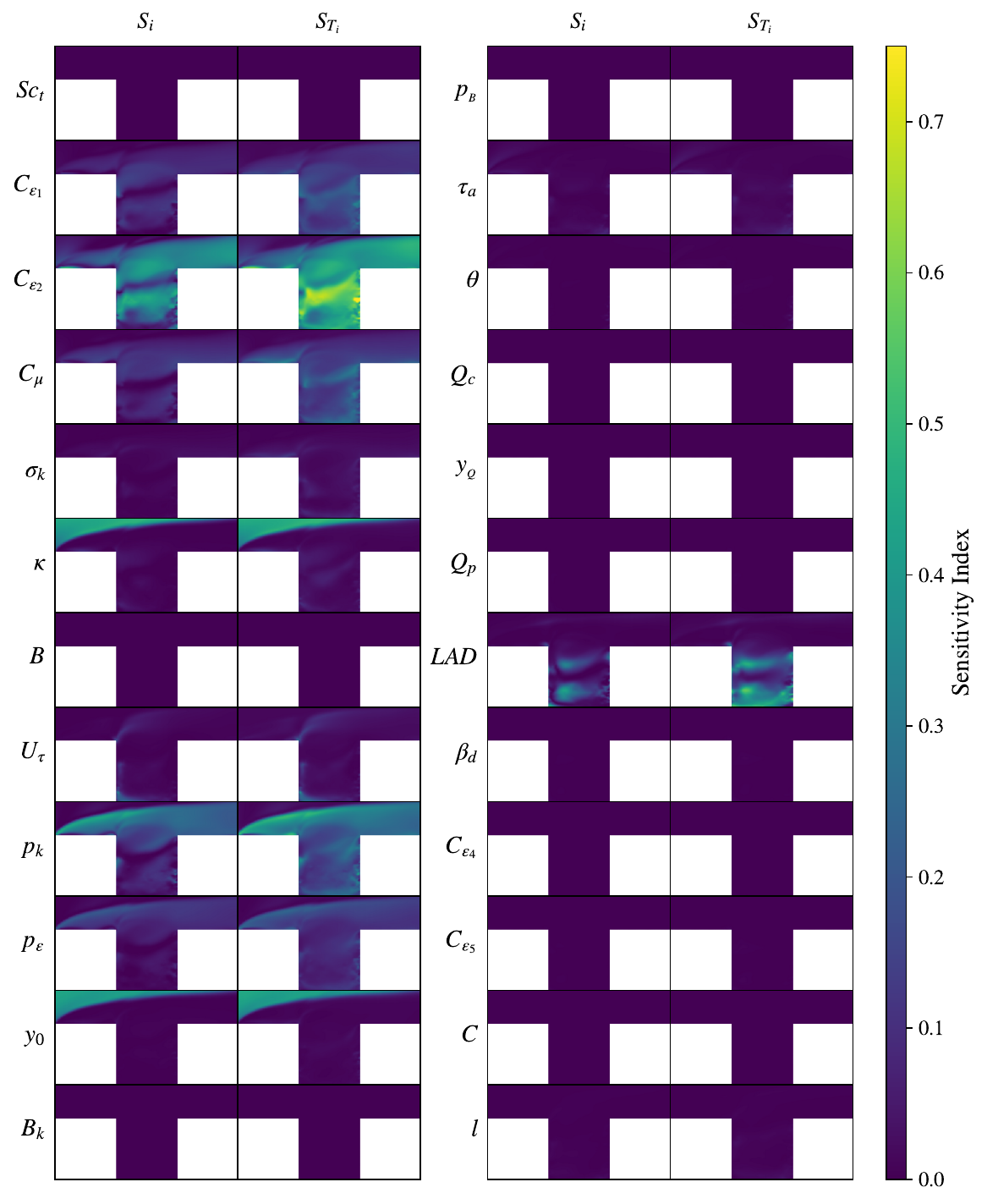}
    \caption{Spatial distribution of the first-order ($S_i$) and total-effect ($S_{T_i}$) Sobol sensitivity indices for the \textbf{velocity field} under the \textbf{thermal scenario}, considering the baseline uncertainty ($\gamma_{_0}$).}
    \label{fig:ThermV1}
\end{figure}
\begin{figure}[h]
    \centering
    \includegraphics[scale=0.88]{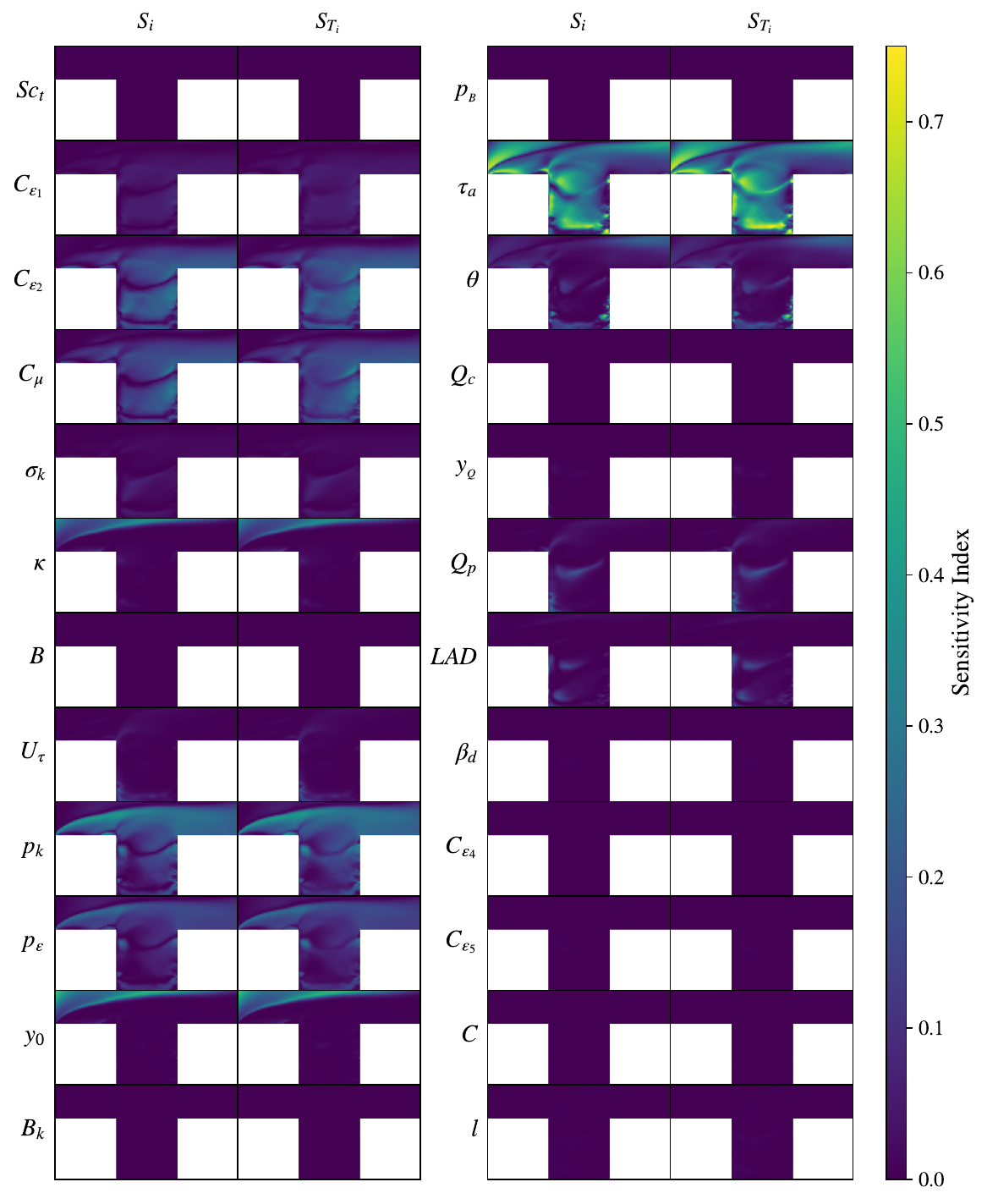}
    \caption{Spatial distribution of the first-order ($S_i$) and total-effect ($S_{T_i}$) Sobol sensitivity indices for the \textbf{velocity field} under the \textbf{thermal scenario}, considering a reduced uncertainty of $12.5\%\,\gamma_{_0}$.}
\end{figure}
\begin{figure}[h]
    \centering
    \includegraphics[scale=0.88]{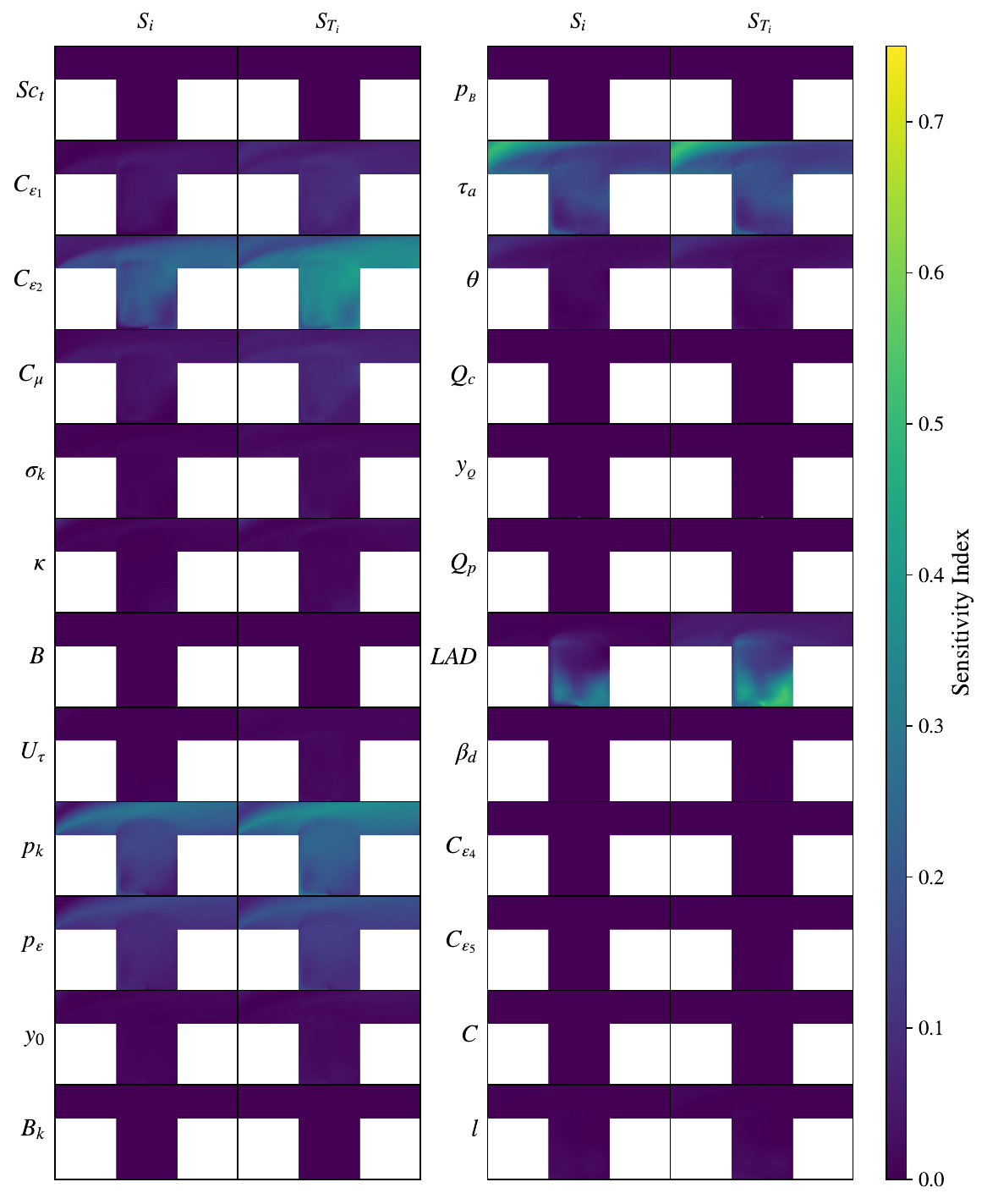}
    \caption{Spatial distribution of the first-order ($S_i$) and total-effect ($S_{T_i}$) Sobol sensitivity indices for the \textbf{temperature field} under the \textbf{thermal scenario}, considering the baseline uncertainty ($\gamma_{_0}$).}
\end{figure}
\begin{figure}[h]
    \centering
    \includegraphics[scale=0.88]{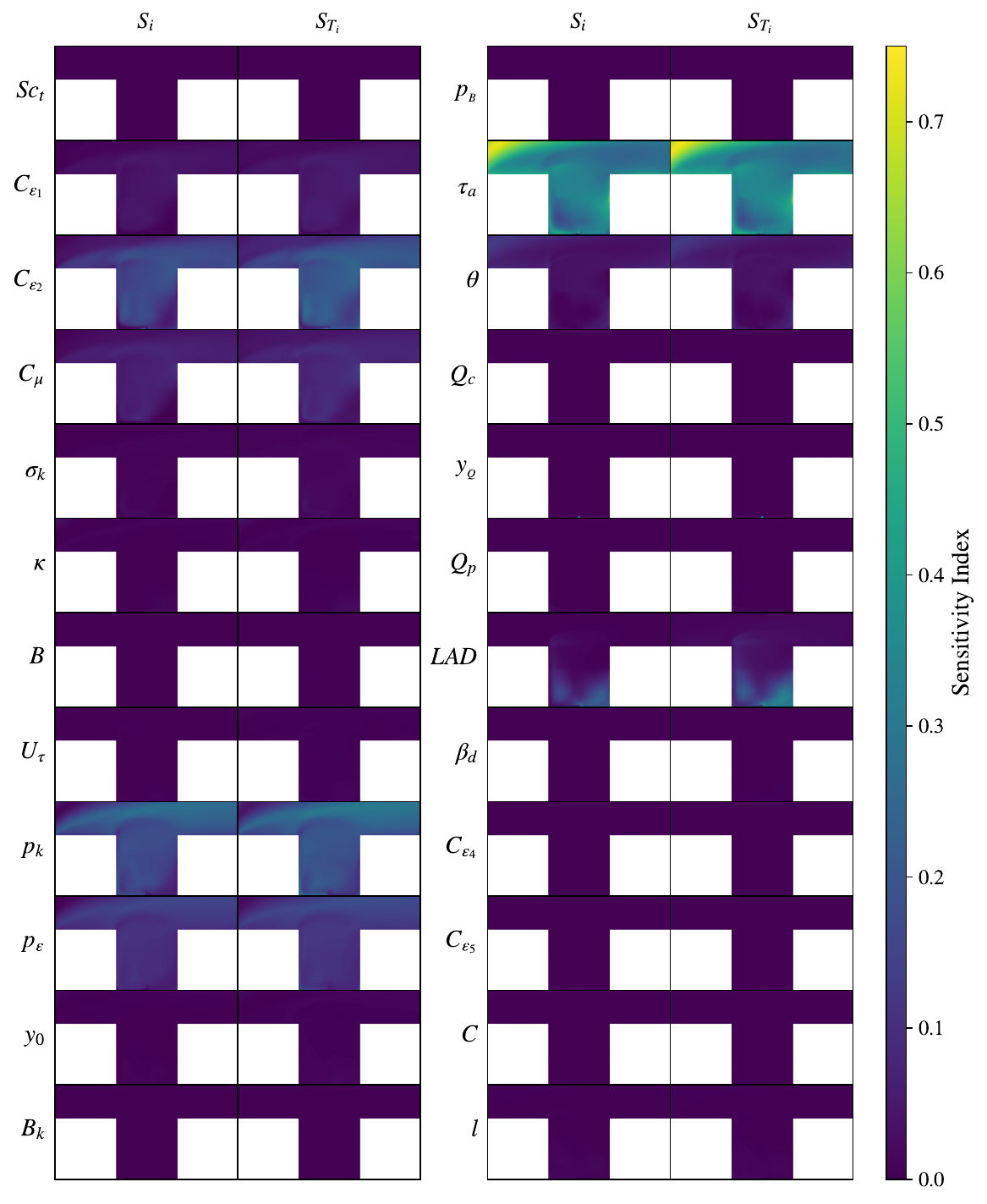}
    \caption{Spatial distribution of the first-order ($S_i$) and total-effect ($S_{T_i}$) Sobol sensitivity indices for the \textbf{temperature field} under the \textbf{thermal scenario}, considering a reduced uncertainty of $50\%\,\gamma_{_0}$.}
\end{figure}
\begin{figure}[h]
    \centering
    \includegraphics[scale=0.88]{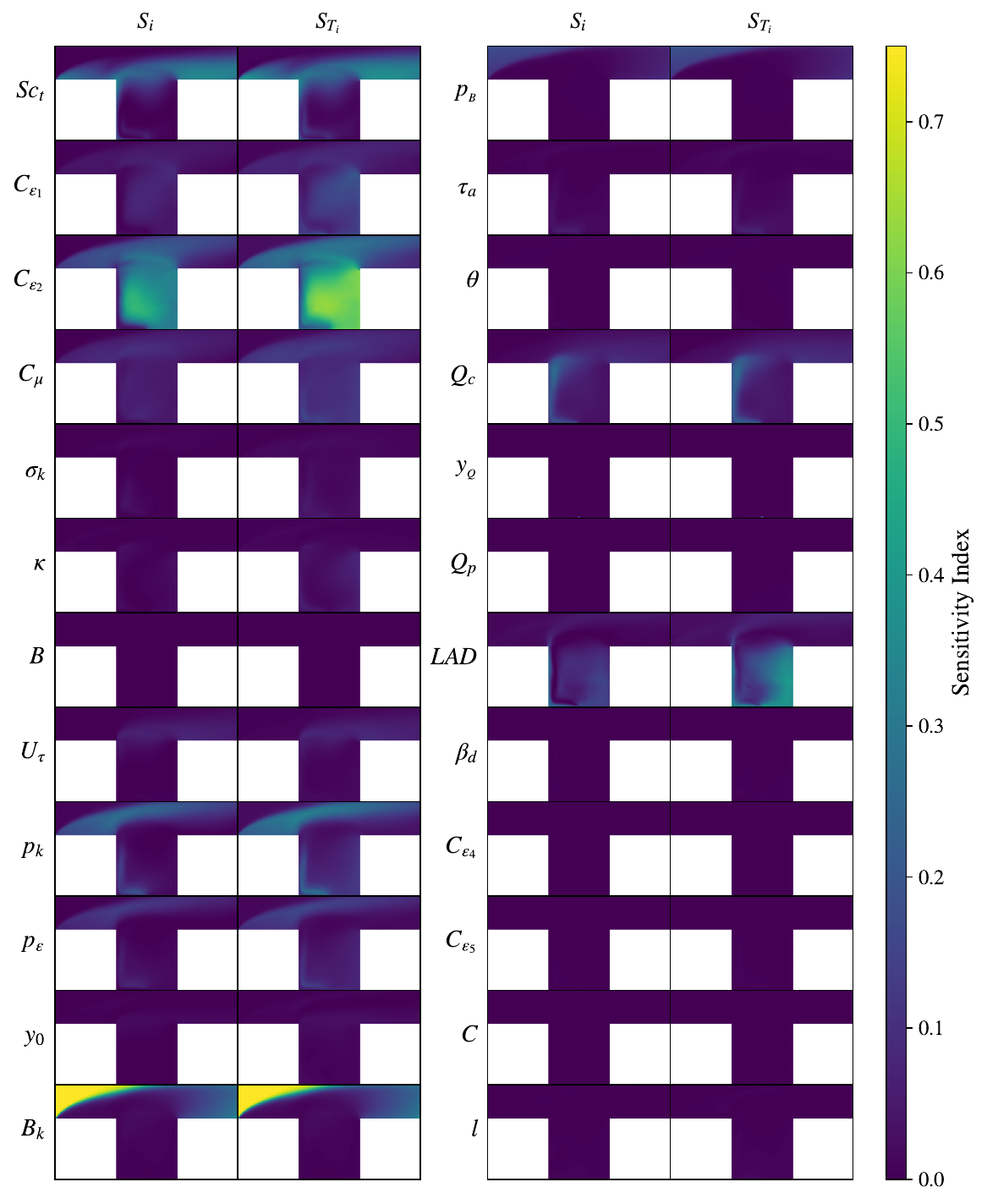}
    \caption{Spatial distribution of the first-order ($S_i$) and total-effect ($S_{T_i}$) Sobol sensitivity indices for the \textbf{concentration field} under the \textbf{thermal scenario}, considering the baseline uncertainty ($\gamma_{_0}$).}
    \label{fig:ThermC1}
\end{figure}
\begin{figure}[h]
    \centering
    \includegraphics[scale=0.88]{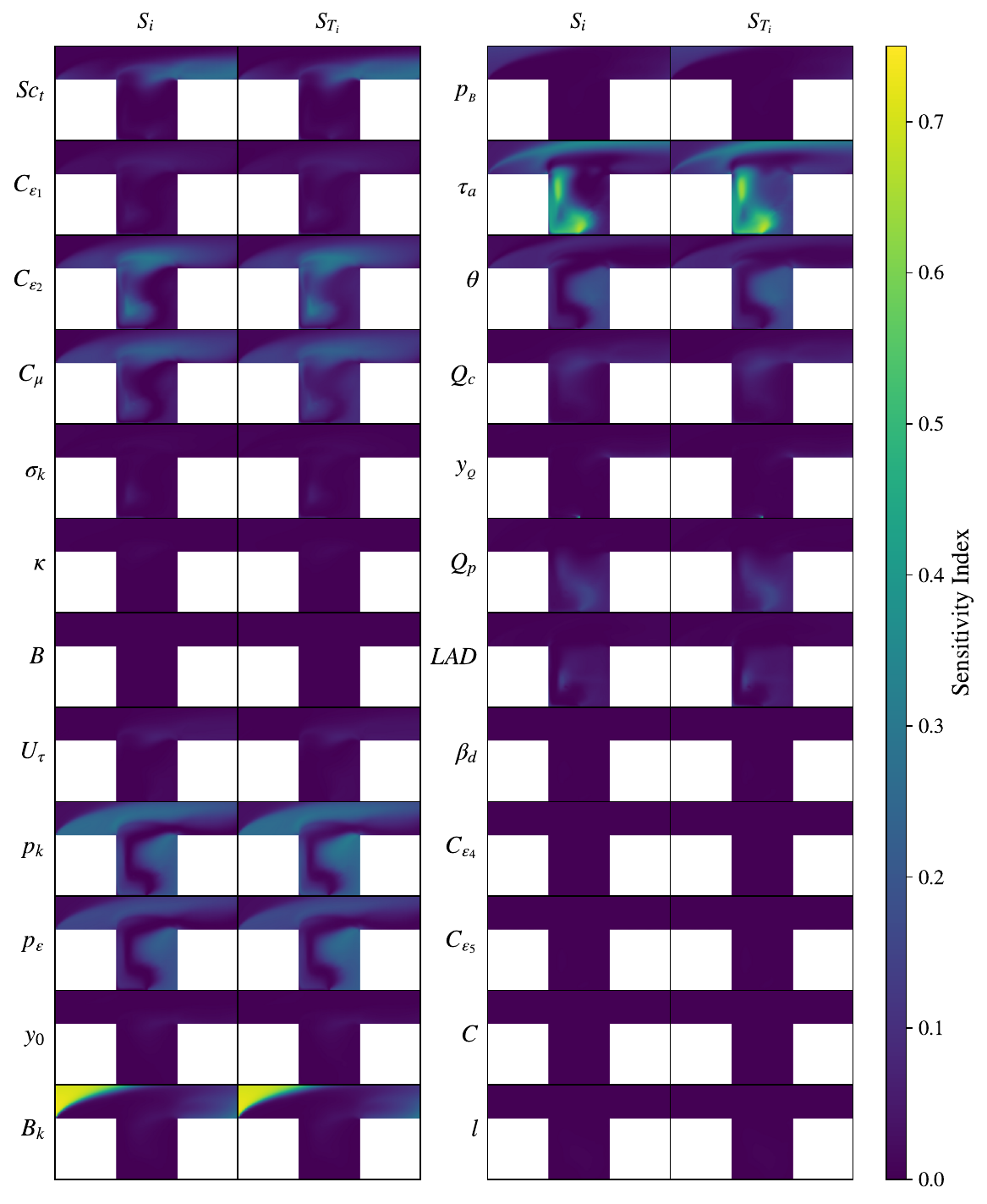}
    \caption{Spatial distribution of the first-order ($S_i$) and total-effect ($S_{T_i}$) Sobol sensitivity indices for the \textbf{concentration field} under the \textbf{thermal scenario}, considering a reduced uncertainty of $12.5\%\,\gamma_{_0}$.}
\end{figure}

\begin{figure}[h]
    \centering
    \includegraphics[scale=0.88]{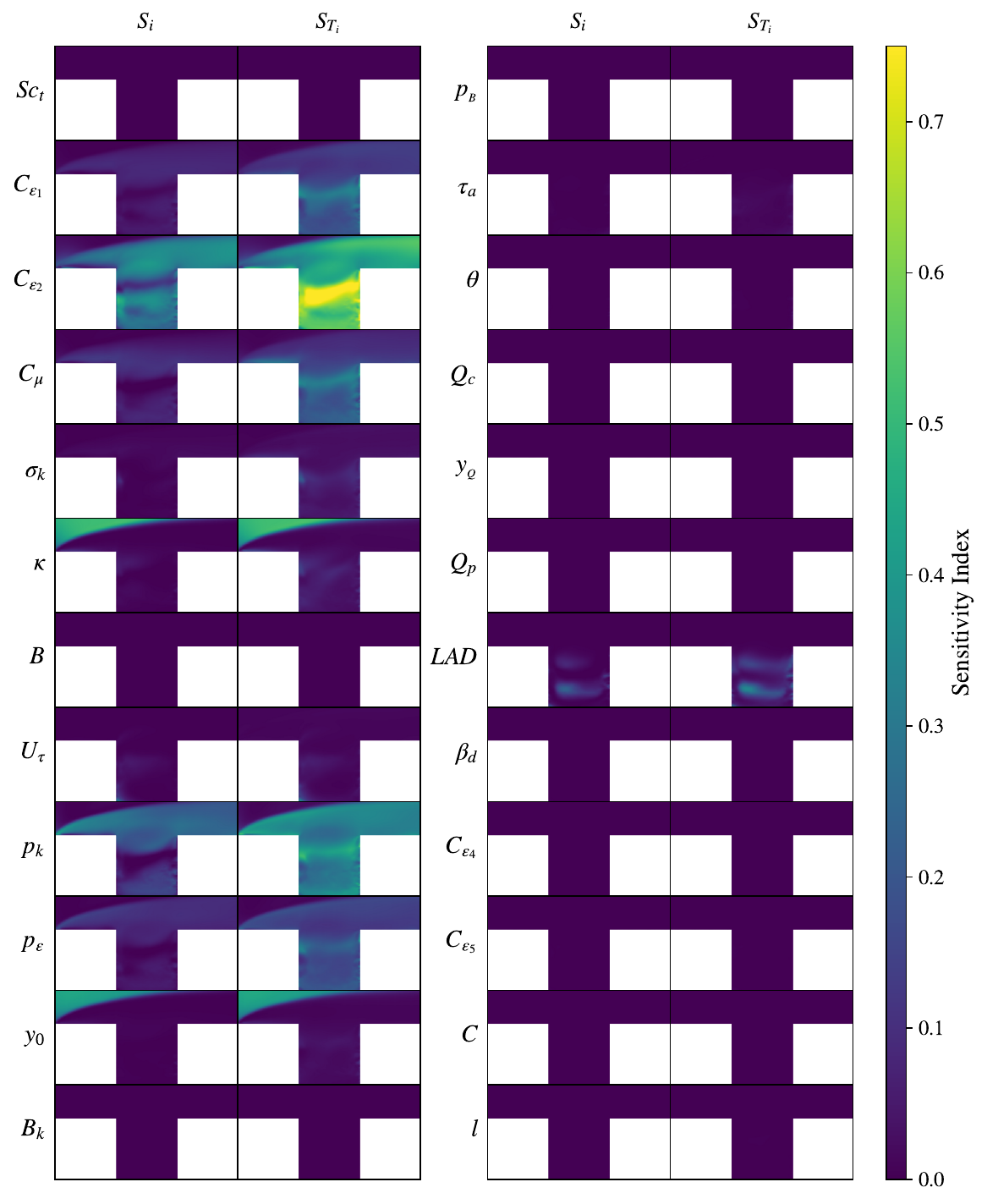}
    \caption{Spatial distribution of the first-order ($S_i$) and total-effect ($S_{T_i}$) Sobol sensitivity indices for the \textbf{velocity field} under the \textbf{vegetated scenario}, considering the baseline uncertainty ($\gamma_{_0}$).}
    \label{fig:VegU1}
\end{figure}
\begin{figure}[h]
    \centering
    \includegraphics[scale=0.88]{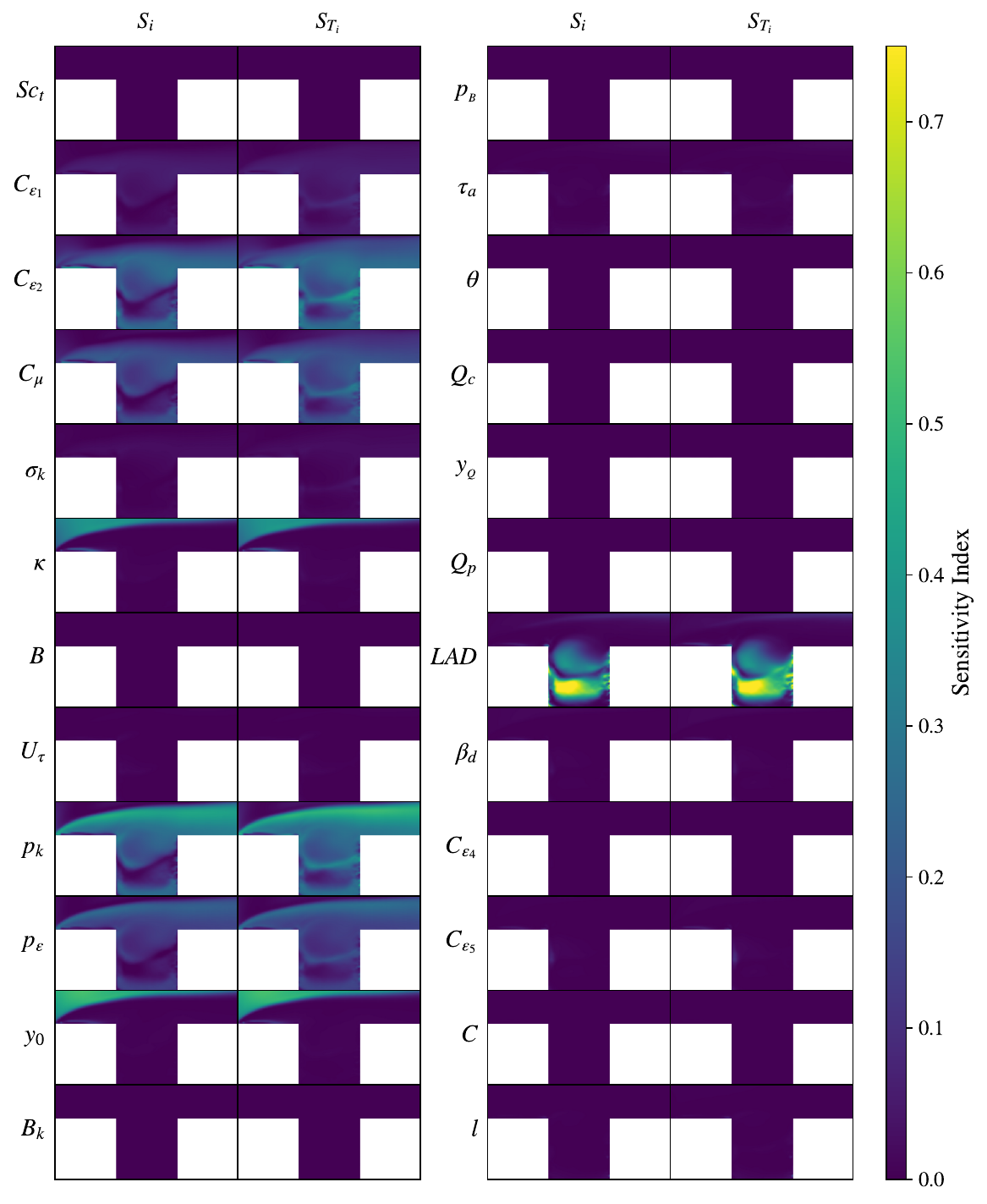}
    \caption{Spatial distribution of the first-order ($S_i$) and total-effect ($S_{T_i}$) Sobol sensitivity indices for the \textbf{velocity field} under the \textbf{vegetated scenario}, considering a reduced uncertainty of $12.5\%\,\gamma_{_0}$.}
    \label{fig:VegU0.125}
\end{figure}
\begin{figure}[h]
    \centering
    \includegraphics[scale=0.88]{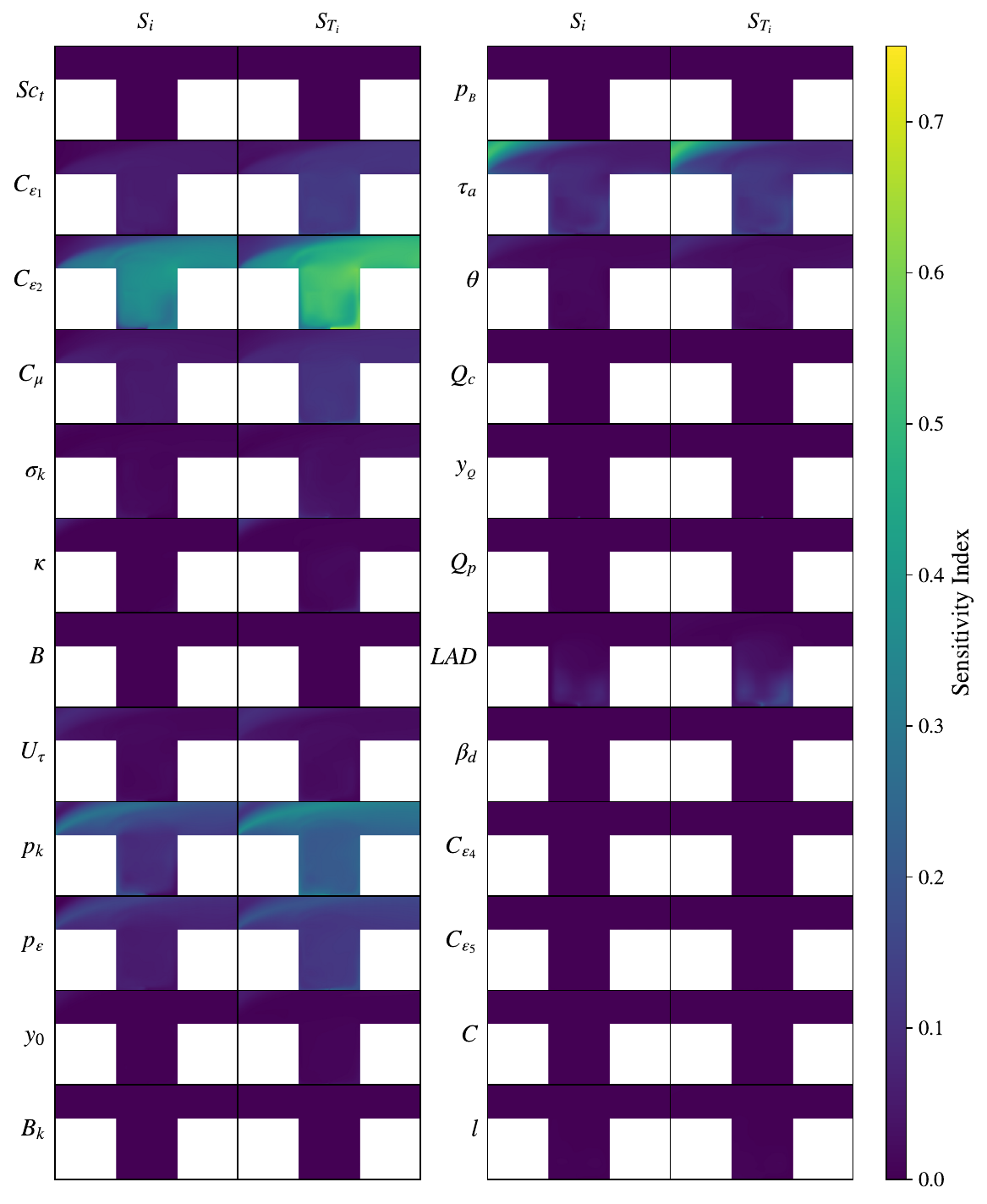}
    \caption{Spatial distribution of the first-order ($S_i$) and total-effect ($S_{T_i}$) Sobol sensitivity indices for the \textbf{temperature field} under the \textbf{vegetated scenario}, considering the baseline uncertainty ($\gamma_{_0}$).}
    \label{fig:VegT1}
\end{figure}
\begin{figure}[h]
    \centering
    \includegraphics[scale=0.88]{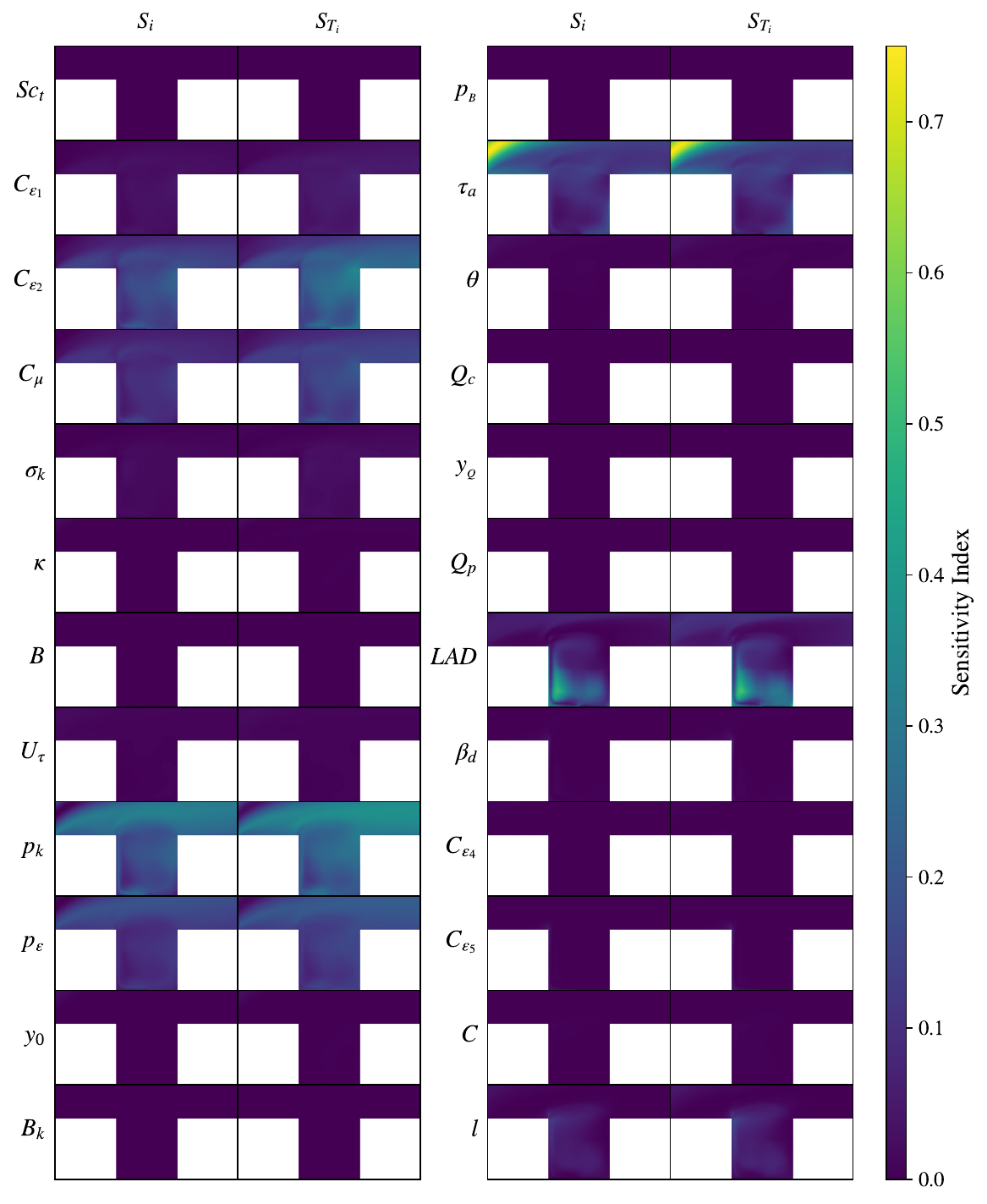}
    \caption{Spatial distribution of the first-order ($S_i$) and total-effect ($S_{T_i}$) Sobol sensitivity indices for the \textbf{temperature field} under the \textbf{vegetated scenario}, considering a reduced uncertainty of $12.5\%\,\gamma_{_0}$.}
    \label{fig:VegT0.125}
\end{figure}
\begin{figure}[h]
    \centering
    \includegraphics[scale=0.88]{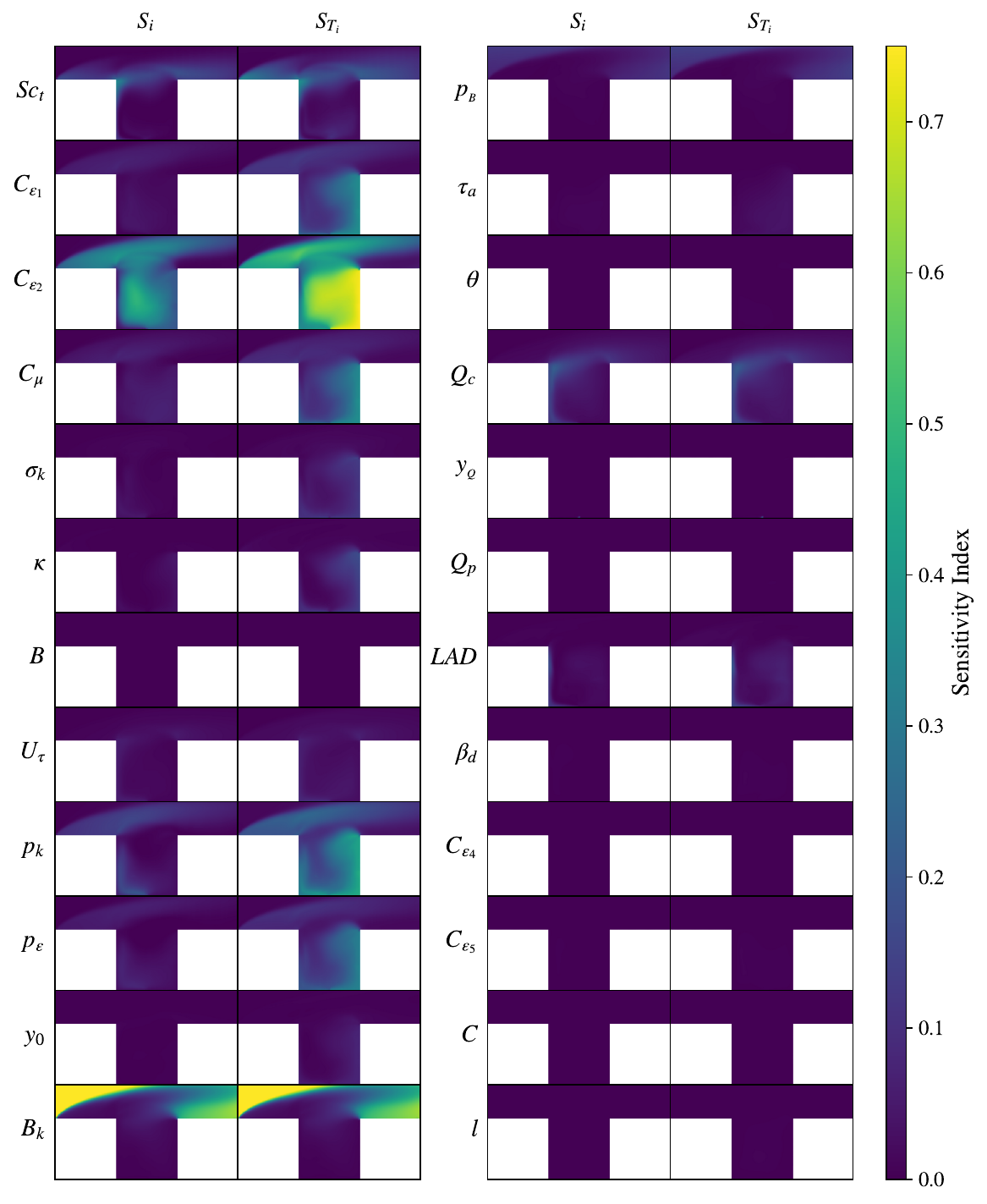}
    \caption{Spatial distribution of the first-order ($S_i$) and total-effect ($S_{T_i}$) Sobol sensitivity indices for the \textbf{concentration field} under the \textbf{vegetated scenario}, considering the baseline uncertainty ($\gamma_{_0}$).}
    \label{fig:VegC1}
\end{figure}
\begin{figure}[h]
    \centering
    \includegraphics[scale=0.88]{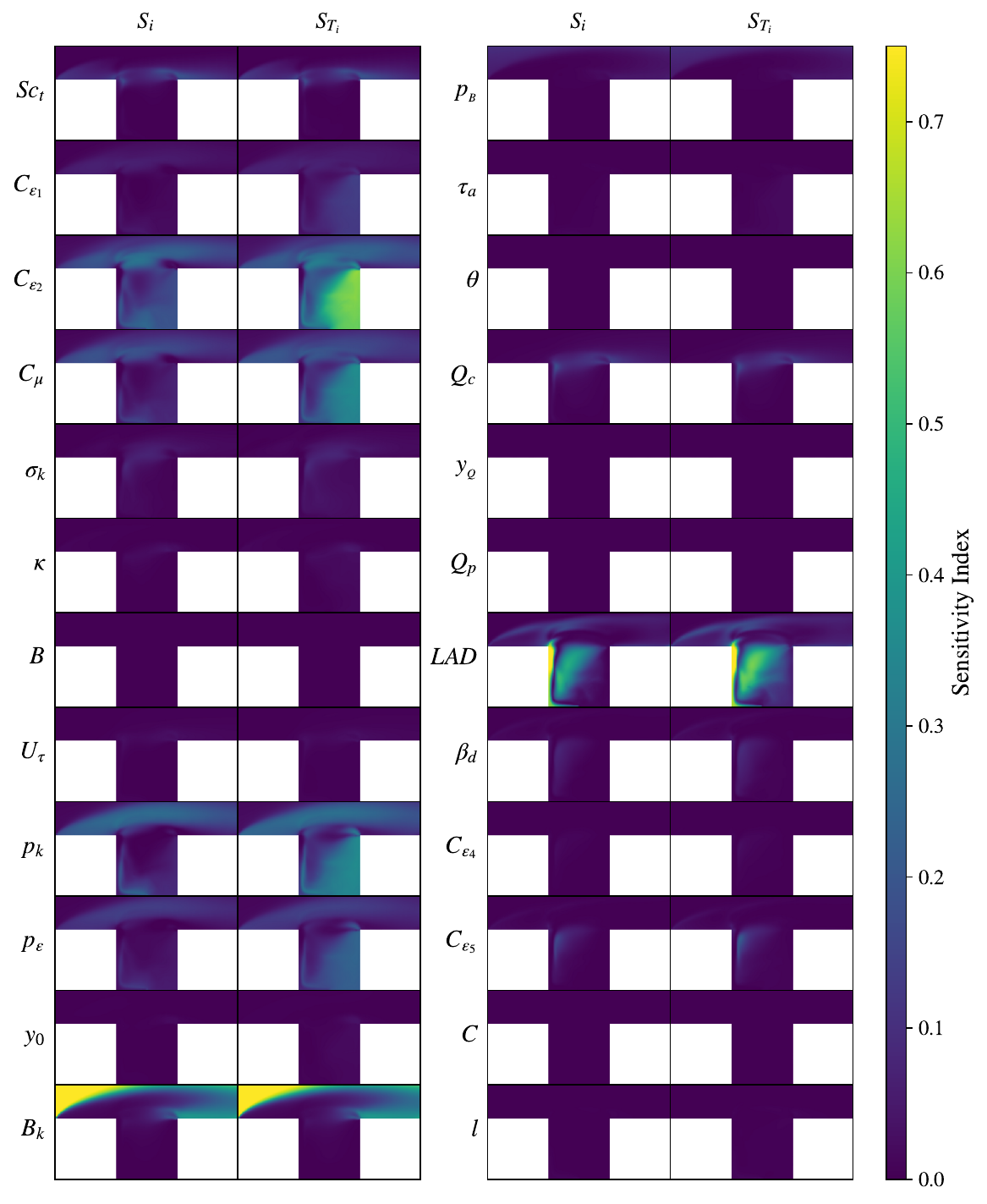}
    \caption{Spatial distribution of the first-order ($S_i$) and total-effect ($S_{T_i}$) Sobol sensitivity indices for the \textbf{concentration field} under the \textbf{vegetated scenario}, considering a reduced uncertainty of $12.5\%\,\gamma_{_0}$.}
    \label{fig:VegC0.125}
\end{figure}

\end{document}